\documentclass[]{interact}

\usepackage[caption=false]{subfig}% Support for small, `sub' figures and tables
\usepackage{amsmath}
\usepackage{amssymb}
\usepackage{hyperref}
\usepackage{xcolor}
\DeclareMathOperator{\arcsec}{^{\prime\prime}}
\usepackage[numbers,sort&compress]{natbib}% Citation support using natbib.sty
\bibpunct[, ]{[}{]}{,}{n}{,}{,}% Citation support using natbib.sty
\renewcommand\bibfont{\fontsize{10}{12}\selectfont}% Bibliography support using natbib.sty
\makeatletter% @ becomes a letter
\def\NAT@def@citea{\def\@citea{\NAT@separator}}% Suppress spaces between citations using natbib.sty
\makeatother% @ becomes a symbol again

\theoremstyle{plain}% Theorem-like structures provided by amsthm.sty

\theoremstyle{definition}

\theoremstyle{remark}

\begin{document}

%\articletype{ARTICLE TEMPLATE}% Specify the article type or omit as appropriate

\title{Strong Gravitational Lensing with the James Webb Space Telescope}

\author{
\name{Adi Zitrin\textsuperscript{a}\thanks{CONTACT Adi Zitrin. Email: zitrin@bgu.ac.il}}
\affil{\textsuperscript{a}Department of Physics, Ben-Gurion University of the Negev, P.O. Box 653, Be'er-Sheva 8410501, Israel}
}

\maketitle

\begin{abstract}
The theory of General Relativity predicts that, since massive bodies curve spacetime, light from a distant source would be deflected by a foreground massive object -- a phenomenon known as \emph{Gravitational Lensing}. Historically, the strength of deflection of light from background stars by the sun, during the 1919 solar eclipse, supplied one of the first proofs for the theory of General Relativity. However, it is only in the last few decades, with the advent of the Hubble Space Telescope and other large, ground-based facilities, that lensing has become a principal tool in modern astronomy. Lensing allows us to study both the matter content of the lensing bodies such as galaxies or clusters of galaxies, mainly dominated by the otherwise-invisible \emph{dark matter}, and the distant background sources that are being lensed by them. Strong gravitational lensing, where sources are substantially magnified and multiply imaged, is particularly useful to that end. The substantial magnification enables a high-resolution view of the sources and the detection of fainter and farther objects than would otherwise be possible; and image multiplicity helps in verifying the distance to them, and in studying variable or transient sources. Paired with the unprecedented capabilities of the James Webb Space Telescope (JWST), lensing now allows us to observe, detect, and study distant sources like never before. I summarise recent advances in strong-lensing applications and near-future prospects with JWST.
\end{abstract}


\section{Introduction}

\textit{From everyday life to the distant Universe~~} Sunsets, rainbows, stars, or the phases of the moon  -- the curious human has long been fascinated with optical phenomena. This is perhaps only natural, given that we perceive the world primarily through vision, and that these phenomena hold clues about our physical world. Indeed, optics is also a prominent field in physics, and light and its nature are fundamental to our physical Universe. 

A magnifying glass in the hands of a child opens a world of marvellous possibilities. Through the lens, objects appear magnified, and in the case of more complex lenses, often seem distorted and appearing multiple times. It can be used to concentrate sunlight onto a spot - causing small burns on wood or in dry leaves, to observe tiny creatures, or to read miniscule letters written on a single grain of rice. 

This review concentrates on a phenomenon which is often referred to as the astronomical parallel to the magnifying glass -- \emph{strong gravitational lensing}. I start with a brief history of gravitational lensing, followed by some qualitative description of the phenomenon and its different regimes. I then overview how, technically, (strong-) lensing is modelled and the range of science it facilitates; from studies of distant galaxies and black holes, through the nature of dark matter, to cosmology. I review how strong-lensing science has been established with the Hubble Space Telescope (HST, or \emph{Hubble}) over the past 2-3 decades, and the shifts taking place since the launch of the James Webb Space Telescope (JWST), in particular towards studying small, point-like (stars, star clusters), varying (quasars) or exploding transient (supernovae (SNe)) sources. Since they are the most massive gravitationally bound objects in the Universe, this review concentrates in particular on strong lensing by \emph{clusters of galaxies}, where most prominent lensing manifestations appear, although much of the science is also relevant for smaller lenses such as galaxies. I mention some recent advances from various cluster-lensing programs with JWST, together with some prospects for strong-lensing science in the near future with JWST, and for its interplay with recent or upcoming observatories from ground and space.

\subsection{A brief history of gravitational lensing}\label{history}

If light were assumed to consist of some test particles (corpuscles), then -- similar to other massive bodies -- when passing near a massive astronomical body the particles should be attracted to it, and thus deflected with respect to their original trajectory. The acceleration the particles would experience is independent of their mass, and the strength of deflection, the \emph{deflection angle}, can be calculated in the framework of Newtonian mechanics. This calculation is first attributed to Henry Cavendish in 1784, apparently following a letter by John Michell to him concerning the effect of gravity on the speed of light \citep[][]{ellis2010gravitational,meneghetti2021introduction}. The Newtonian deflection angle comes out to be $\hat\alpha=\frac{2GM}{c^2 r}$, where $G$ is the universal gravitational constant, $c$ the speed of light, $M$ the deflector's mass and $r$ the distance of closest approach to it. For example, the deflection angle from the sun for a light ray passing right next to its surface, becomes $\approx0".875$. Distant stars lying next to the line-of-sight to the limb of the sun at any given moment, if we could see them during daytime, would thus appear deflected by this much. 

Einstein's General Relativity (GR), published in 1915-1916 \cite{Einstein1915SPAW.......844E,Einstein1916AnP...354..769E}, has had a different concept of gravity and its interaction with space and time. In GR, massive bodies curve spacetime such that light, which -- contrary to the Newtonian treatment -- is now massless\footnote{Be it either that light is made of zero rest-mass particles -- photons, or simply a wave. It should be noted that also massive particles travel in geodesics, but not \emph{null} geodesics, so that their deflection would be stronger, depending on their velocity.} and travels in \emph{null} geodesics, seems deflected to a distant observer. The deflection angle in this framework, as calculated by Einstein, comes out to be $\hat\alpha=\frac{4GM}{c^2 r}$, i.e. twice the prediction from Newtonian physics and in the case of the sun for a ray near its surface, equals to about 1".75.

In an effort led by Arthur Eddington and Astronomer Royal Frank Dyson, the deflection of light from stars behind the sun was measured during the solar eclipse of 1919 \cite{Dyson1920RSPTA.220..291D}. The eclipse supplied a golden opportunity to test a key prediction of the theory of General Relativity. The measurements showed that the theory was successful in predicting the bending of light in a gravitational field -- the second test of GR that Einstein had proposed (the first being its prediction for the precession of the perihelion of Mercury, which had already been known at the time).

Despite its critical historical role in the early days of modern physics, lensing has begun shaping as a substantial astronomical tool, observationally at least, only over the past few decades, when increasingly larger telescopes and then the HST became available. 

The first observational instance of a strongly lensed, multiply imaged source was the doubly imaged quasar Q0957+561, published in 1979 \cite{Walsh1979Natur.279..381W}. Quasars are bright, quasi-stellar objects (QSOs) powered by accretion onto a supermassive black hole; they are a type of Active Galactic Nucleus (AGN). Several other multiply imaged quasars were then detected in the 1980's, including an ``Einstein cross" -- a quadruply imaged quasar, published in 1985 \citep{Huchra1985AJ.....90..691H}. Around the same time, the first gravitational arc was observed, lensed by the massive galaxy cluster Abell 370 \cite{Lynds1986BAAS...18R1014L,soucail87,LyndsPetrosian1989ApJ...336....1L}. Curiously, at the time, since lensing manifestations were not yet well established, it was not immediately clear that this was a gravitational arc. A spectroscopic measurement published about a year later confirmed that indeed, the source lay behind the cluster \cite{Soucail1988A&A...191L..19S,Paczynski1987Natur.325..572P}. This detection perhaps marks the beginning of strong-lensing science. During the 1990's, methodologies and analyses of gravitational lensing signatures by galaxy clusters started emerging, in the weak- and strong-lensing regimes \cite[e.g.][]{Tyson1990ApJ...349L...1T,Kaiser1993ApJ...404..441K,Kaiser1995ApJ...449..460K,Kneib1993Lenstool,kneib96}. The spatial resolution and imaging quality afforded by the HST first with e.g. the Wide Field and Planetary Camera 2 (WFPC2) and then with later-generation instruments installed during the first decade of the 2000's -- such as the Advanced Camera for Surveys (ACS) and the Wide-Field Camera 3 (WFC3) -- revealed more of both the weak- and strong-lensing power of galaxy clusters and motivated the development of new lensing-analysis methods \cite{kneib04,Smith2005MNRAS.359..417S,Broadhurst2005a,Broadhurst2005b,halkola06,jullo09,Liesenborgs2006,Zitrin2009_cl0024}. This recognition has sparked dedicated lensing campaigns to observe the most massive galaxy clusters and their magnified backgrounds \cite{EbelingMacsCat2001,richard10,PostmanCLASHoverview,Coe2019RELICS}, revealing some of the farthest known galaxies \cite[e.g.][]{Zheng2012NaturZ,Coe2012highz}, which supplied an additional motivation for such programs. These campaigns established that essentially every massive cluster acts as a giant strong lens in the sky -- a 'cosmic telescope'. One of the deepest surveys of lensing fields was the Director's Discretionary Time program \emph{Hubble Frontier Fields} \cite{Lotz2016HFF}, which targetted 6 massive galaxy clusters with over 800 HST orbits around earth. These observational efforts were accompanied by advancements in computational lens-modelling techniques (see section \ref{modeling}), which over the past couple of decades have reached a high level of accuracy and robustness, with greater grasp of the underlying uncertainties and systematics. In that sense, the JWST has arrived at exactly the right moment, with the lensing community already mature and prepared to exploit the new powers offered by the telescope.

In parallel to the above developments in strong-lensing science, other lensing regimes have also experienced much advancement over the past few decades. For example, various instruments, including wide-field cameras on ground-based telescopes, have been used to study weak-lensing on larger samples of clusters (e.g., \cite{Oguri2012SL,Merten2014CLASHcM,Umetsu2014CLASH_WL,Umetsu2020WLreview,Stern2019WL_SPT,Sereno2017WL,Hoekstra2015,Gruen2014MNRAS.442.1507G}), and dedicated missions and experiments have concentrated on micro-lensing, especially towards the galactic bulge or the Magellanic clouds \citep{Udalski1992AcA....42..253U,Sumi2003MOA,Tisserand2007A&A...469..387T}. The different types of lensing are described in Section \ref{types}.

\begin{figure}
\centering
    \begin{tabular}{c c}
\includegraphics[width=0.49\textwidth]{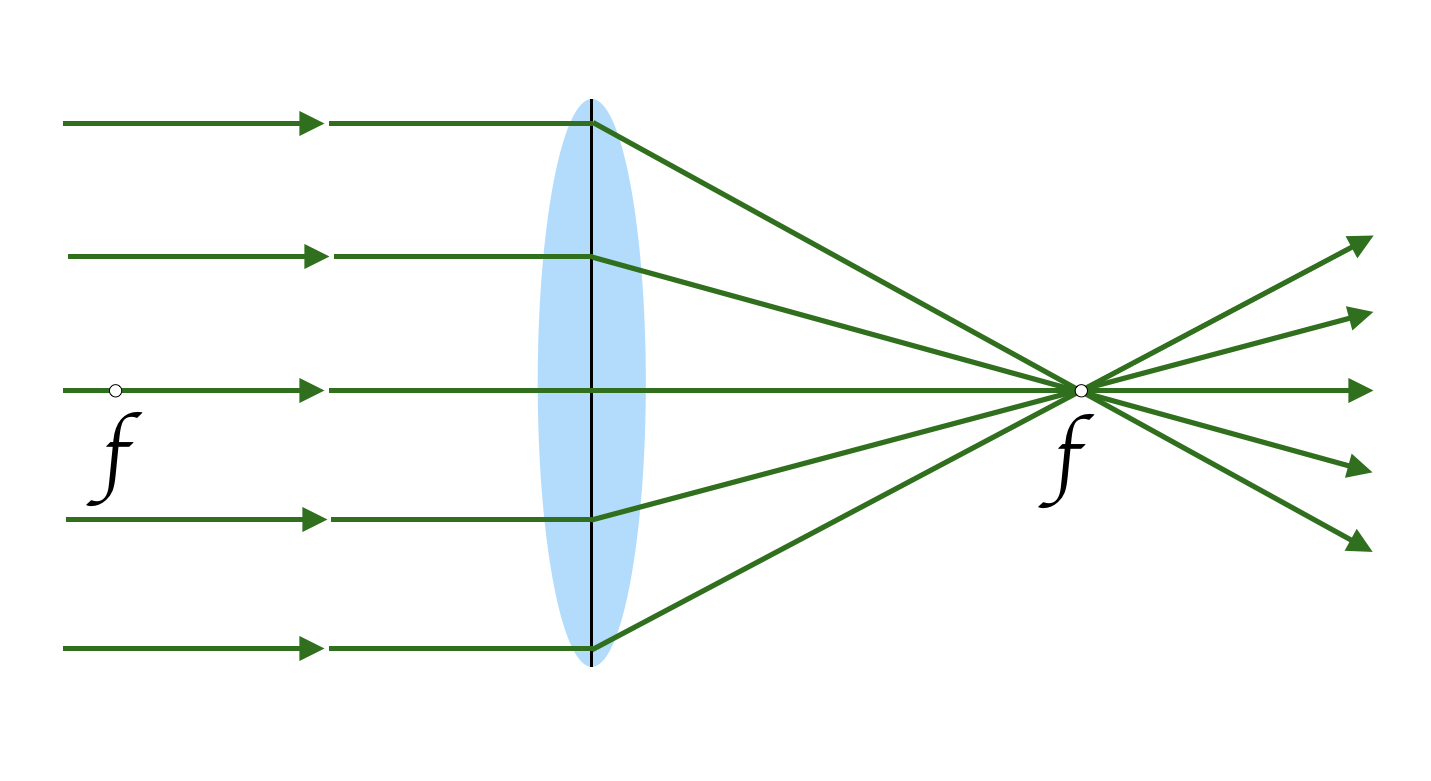}
 &
\includegraphics[width=0.49\textwidth]{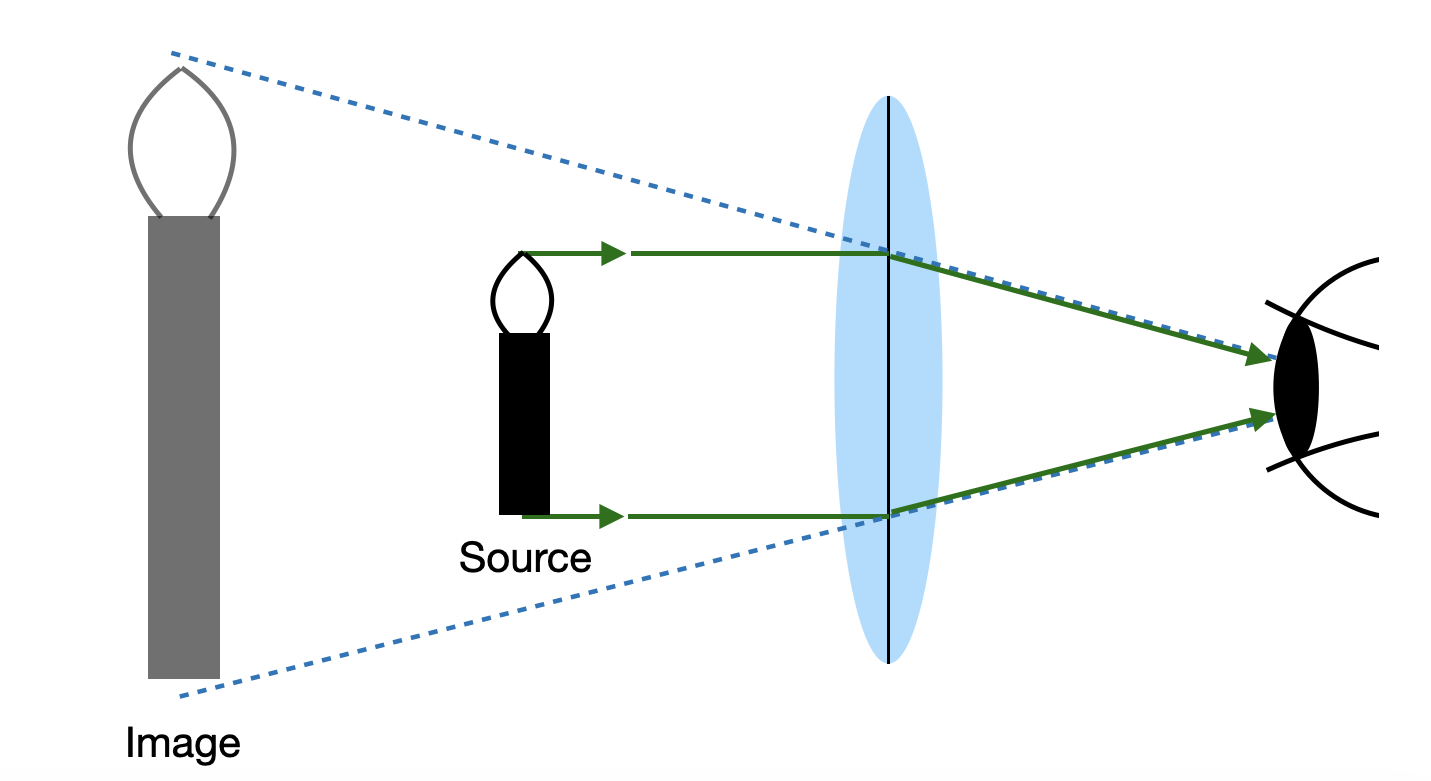}\\
{{(a)}} & {{(b)}}
\end{tabular}
\caption{Qualitative description of a magnifying glass. (a) Ray diagram for a convex lens. Parallel rays are focused onto a focal point after passing through the lens, marked here on each side of the lens. (b) Demonstration of why a magnifying glass magnifies. Light rays from the source to the observer are marked with green solid lines. To the observer -- the eye depicted in the figure -- the rays arrive from a wider angle than that spanned by the actual source, so that the object appears larger (Note: not to scale).} 
\label{fig:convex}
\end{figure} 

\subsection{A qualitative description of lensing}\label{background}

Imagine a ray of light travelling through air and hitting a thin, glass convex lens such as a magnifying glass. Due to the higher speed of light in air than in glass, the ray would generally bend inwards when entering the glass, and then bend inwards again, when exiting the lens, and continue in that final direction unperturbed. In total, for each ray reaching the lens, depending on its position, there is an effective angle by which the ray is bent with respect to its initial direction. Parallel rays reaching the lens would be focused onto a focal point after passing through the lens (see Figure \ref{fig:convex}, panel (a)). Now assume there is a source on one side of the lens, and an observer on the other side of the lens. The lens converges towards the observer -- which we can depict as a small aperture -- light rays that without the lens would not reach them, thus magnifying the object. A demonstration of lens magnification is shown in Figure \ref{fig:convex}, panel (b). Light from the source (shown here as a candle but any other object would do) arrives to the observer at a wider angle than would be subtended by the source in the absence of the lens, making it appear larger. While the physics causing the effect are somewhat different hence demanding caution, we can attempt to make an analogy with gravitational lenses: Similar to the bending light ray passing from air to glass and vice versa, to a distant observer the speed of light appears reduced near a gravitating mass, implying a similar 'refraction' should occur. Thus, similarly to a magnifying glass --  the lens converges light towards the observer and magnifies far away, background objects. 

\begin{figure}
\includegraphics[width=1\textwidth]{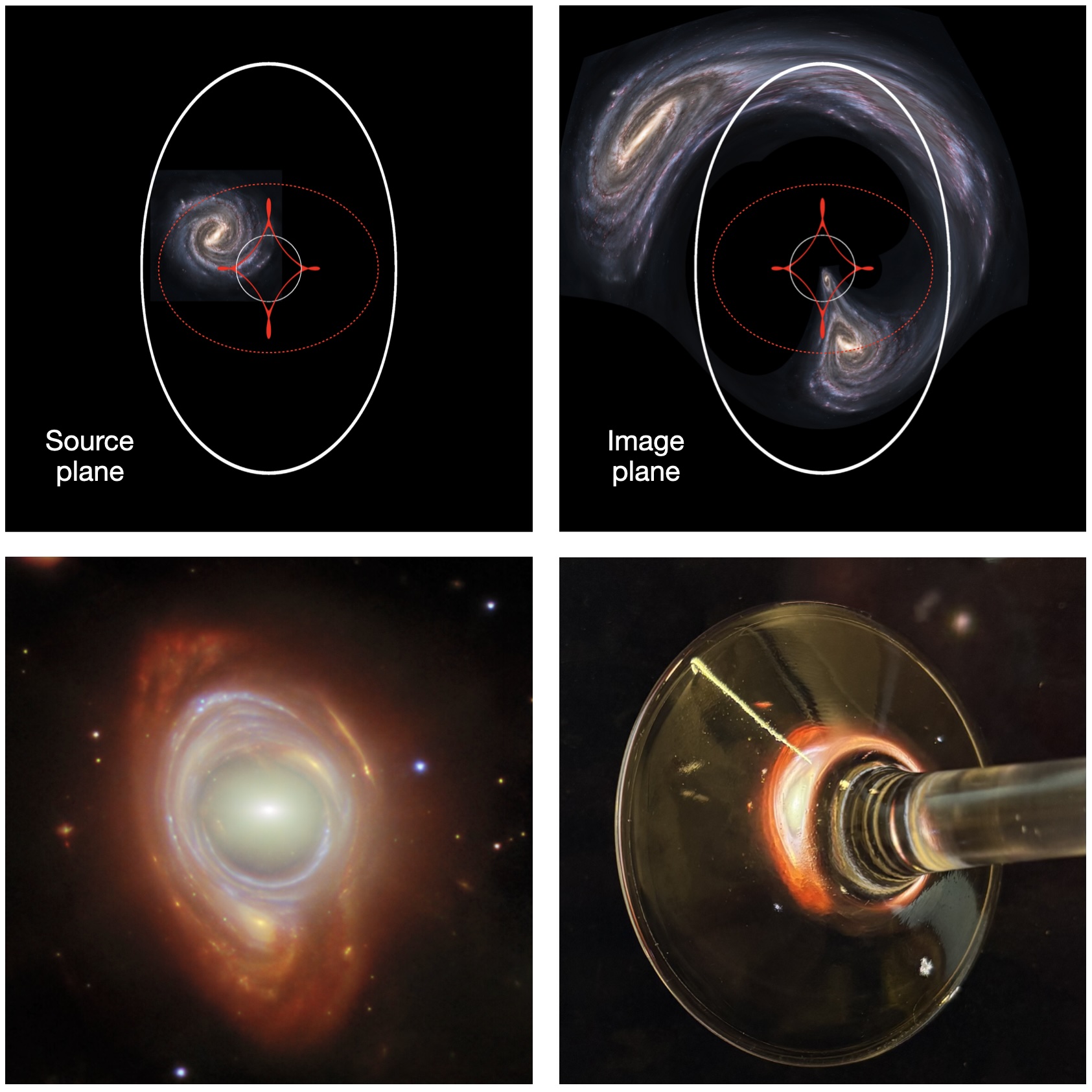} 
\caption{Examples of lensing. \emph{Upper row:} Image formation by an elliptical lens. The critical curves of the (invisible) lens are shown in white, and their corresponding caustics in red. We plant a source between the outer and inner caustics, as seen in the subfigure on the \emph{left}, which after lensing results in the image distribution seen in the subfigure on the \emph{right}. Note, only parts of the source sitting within the (outer) caustic are being multiply imaged. Parts of the source sitting within the outer caustic but outside the inner one are imaged three times, whereas parts of the source galaxy sitting also within the inner caustic are imaged five times. The source-galaxy image was adopted from the web -- it is an artist impression of the Milky Way based in part on NASA/ESA images (Credit: Nick Risinger). \emph{Bottom left:} JWST image of a background galaxy lensed by a foreground elliptical galaxy, forming a wide Einstein ring. Image is from the SLICE JWST survey (Credit: ESA/Webb, NASA \& CSA, G. Mahler and M.A. McDonald). \emph{Bottom right:} Lensing through the bottom of a wine glass, using the image of the lensing-system on the left as input.}
\label{fig:EinsteinRing}
\end{figure}

But there are some substantial differences between the two cases. For example, while lensing by a magnifying glass is weakly colour (i.e. wavelength) dependent, gravitational lensing in achromatic. Another important difference is that, while both cases deal with converging lenses, a simple magnifying glass deflects more strongly rays that are farther away from the lens centre (higher impact parameter), whereas in gravitational lensing the deflection usually decreases with distance from the lens' centre. This means that for the gravitational lens there is no focal point, and leads to a different image formation than in a regular magnifying glass. For example, in strong gravitational lensing images are multiplied, and while surface brightness is conserved, get highly distorted and stretched. As a matter of fact, a better resemblance to a gravitational lens is obtained by using the bottom of a wine glass, or what is sometimes called conical or axicon lenses. The glass' thickness profile causes light rays to refract in a similar way to a well-behaved gravitational lens, as demonstrated in Figure \ref{fig:EinsteinRing}. Such lenses would lens a light source sitting on the optical axis, for example, into a ring. There are various examples in the literature or online of lensing analogies with different glass lenses or with various phenomena from everyday life\footnote{The following website has a nice concentration of such examples: https://vela.astro.uliege.be/themes/extragal/gravlens/bibdat/engl/DE/didac.html} (e.g. \cite{Treu2010ARA&A..48...87T,Surdej1990LNP...360...57S,Surdej1993LIACo..31..199S}).

Another analogy to everyday lenses can come from applying Fermat's principle, known from geometric optics, to the case of a gravitational lens \cite{Schneider1985,BlandfordNarayan1986,NarayanBartelmann1996Lectures}. If we consider all possible light rays from a source to the observer, Fermat's principle dictates that the time of arrival of a photon, given by the path integral of each ray, should be extremised (this can be thought of as a generalisation of  the common notion that light takes the fastest, or shortest route). Only light rays whose path integral is stationary will thus reach the observer. Hence, one can in principle construct a time-arrival surface, whose extrema dictate where images of a certain source would form. This also means that there is a \emph{time delay} between different images of the same source; a delay which comes in handy for various cosmological tests as we shall see soon.

Before proceeding we need to define a couple of additional terms. We mentioned that the gravitational lens has no focal point. It has however a somewhat related characteristic called \emph{critical lines}. These lines define the symmetry points of the lens, and in principle, mark where the magnification diverges, i.e., becomes infinite. These lines, when mapped back to the source plane, are called \emph{caustics} \cite{Berry01011976,Schneider1992grle.book.....S}. Caustics are often seen on the bottom of a swimming pool on a sunny day: The water and ripples on top of it act as an unregulated lens, focusing light rays from the sun into the observed web of caustics. In gravitational lensing, sources sitting on the caustic would be very highly magnified, where the attained magnification is limited by the source size and by the interference of light.\footnote{The reader may notice a slight difference in terminology: In the swimming-pool scenario we referred to the caustics as the pattern forming on the observer's plane, whereas in lensing they are usually defined in the source plane.} Figure \ref{fig:EinsteinRing} shows an example of the critical curves for an elliptical strong lens. Sources that sit within the caustics are multiply imaged, where the resulting configuration depends on the source position. 

Finally, let us also define the \emph{Einstein radius}. When a source is located behind the centre of a circularly symmetric lens, a ring would appear, known as the \emph{Einstein ring}. The radius of the ring is directly related to the mass of the lens interior to the ring's angular radius, demonstrating that lensing allows us to 'weigh' astronomical objects.

More thorough coverage and useful formulae for lensing can be found in the various books and reviews about lensing already mentioned, such as \cite{Kneib2011review,Schneider1985,Schneider1992grle.book.....S,Treu2010ARA&A..48...87T,Umetsu2020WLreview}.

\subsubsection{Different regimes of lensing}\label{types}
Gravitational lensing can be practically divided into three regimes: strong, weak, and micro \cite{Meylan2006lensing}. While the underlying physics is similar -- namely the warping of spacetime by a massive body -- the different scales involved in the different regimes make the observables, methodologies, and applications distinct. 

When the projected mass density of a lens exceeds a critical density\footnote{The critical density depends on the relevant distances of and between lens and source and is typically of the order of a few kilograms per square meter.}, as is typically the case in the centres of galaxies or massive clusters of galaxies, multiple images of the background sources appear. This is \emph{the strong-lensing regime}, which is at the focus of this review. The mean separation between the images that form grows with the lens' mass, resulting in scales of one or a few arcseconds to a few tens of arcseconds, for galaxies and clusters of galaxies, respectively. The multiple images are typically magnified each by factors of a few to a few tens compared to the source\footnote{Smaller sources in principle can attain even a higher magnification next to the critical curve.}, and are often also significantly distorted, stretched or sheared, sometimes merging on the critical curve such that they form giant arcs, as the first giant arc discovered behind Abell 370.% The modelling of strong-lensing is described in Section \ref{modeling}. 

Farther away from a galaxy cluster's centre where the projected density is lower, is \emph{the weak-lensing regime}. In that regime, background galaxies get only slightly distorted, magnified and sheared - effects that, since the intrinsic shape of each galaxy is unknown, can only be measured statistically. In the weak-lensing regime, the averaged distortions in each location can be approximated by the measured ellipticites of background galaxies. Inversion techniques can be applied, to recover -- albeit with some known degeneracies -- the underlying mass distribution that led to the measured shear field. This can be done both directly by inverting the data, non-parametrically, or parametrically -- i.e. assuming certain parametric forms to describe the mass distribution and iterating for the best fitting result \citep{Kaiser1993ApJ...404..441K,Merritt2006AJ....132.2685M,Umetsu2014CLASH_WL,Niemiec2023MNRAS.524.2883N}. In the case of galaxy clusters the relevant scale to which the weak-lensing regime extends beyond the strong-lensing regime is typically arcminutes (or Mpc scales at the lens\footnote{A parcsec, pc, is equal to 3.086e+16 metres, or 3.26 light years.}). Note, also galaxies produce a weak-lensing effect around them, but given their lower mass compared to clusters and the smaller area of measurable influence they produce, analyses of the weak-lensing signal around galaxies are often done by stacking many galaxies together (e.g. \cite{Fischer2000AJ....120.1198F}). Weak coherent shearing can also be produced by large-scale structure in the Universe; this is known as Cosmic Shear. Mapping the Universe using cosmic shear is an ambitious goal set by some instruments such as the Euclid space mission \cite{Euclid2025A&A...697A...1E}, or the Rubin Observatory \citep{LSST2019ApJ...873..111I}. As will be further mentioned below, given its excellent spatial resolution and depth, JWST is also a very useful instrument for measuring weak-lensing by galaxy clusters.

As was already established, stars cause lensing, too. Imagine looking towards the galactic bulge, where the number density of stars is high. Recall also that there are vast numbers of stars spread between us and the galactic bulge, and that stars are constantly moving in the galaxy. Occasionally, a star between us and the galactic bulge would coincide temporarily with a background star from the bulge. The relative motion of background and foreground stars results in a temporal amplification of the light from the background star. This is the \emph{micro-lensing regime}: A star at a significant distance usually produces a lensing signal (or Einstein ring) of the order of a micro-arcsecond. Multiple images of the background star are therefore impossible to observe with current instruments, but the aggregated effect of magnification is observed, producing a typical light curve -- flux as a function of time -- whose properties depend on, e.g., the lensing mass, distances, and on the relative velocities. In fact, micro-lensing has also become a primary tool for detecting planets around the lensing stars, as these produce a secondary (smaller) peak on the light curve \citep{MaoPaczynski1991ApJ...374L..37M,GouldLoeb1992ApJ...396..104G}.

Some lensing scenarios can involve a mixture of the different regimes. For example, a cluster galaxy far away from the cluster's centre well in the weak-lensing regime can be chance-aligned with another background galaxy, resulting in strongly lensed images. A more scientifically important example is seen in instances of strongly lensed quasars. Since quasars are sufficiently small sources (say $10^{12}$ km in size), they are also prone to micro- or milli-lensing by objects in the lens such as stars, star clusters, small dark matter clumps or other substructure. What matters in determining whether a source would be susceptible to small-scale lensing is its size compared to the Einstein radius. For example, the Einstein radius of a solar-mass lens at a cosmological distance (say, billions of parsecs) strongly lensing a background source, is about 1 micro-arcsecond. One micro-arcsecond at the quasar's distance is indeed comparable to the size of its emitting region. The effect is seen when comparing the light curves of the quasar's multiple images,  where each light curve shows additional fluctuations not observed in the other images.  Larger or more massive substructure in the lens can also cause longer-term \emph{flux ratio anomalies}.

\subsection{Strong-lens modelling}\label{modeling}
It should become insightful to understand how lens modelling is carried out in practice. By 'lens modelling' we refer to constructing a model for the projected mass distribution of the lens. The potential, deflection angle, Fermat surface, and other relevant quantities all consistently relate to that mass distribution. Although the constraints we have are mainly from the positions of multiple images, i.e., on the deflection angle directly, it is often easier to model the potential, and from it to derive all other properties (deflection angle, mass distribution, etc.). 
I also remind the reader that we are concentrating here on strong lensing by galaxy clusters. In the case of galaxy-scale lenses, only one source typically, but sometimes two or three sources are seen lensed, forming multiple images around the galaxy and the symmetry is usually simple and clear (this however does not mean that obtaining a very precise solution is an easy task!). In contrast, galaxy clusters are much more complex bodies and can have many dozens, sometimes hundreds, of multiple images spread across their field. 

\begin{figure}[!t]
\includegraphics[width=1\textwidth]{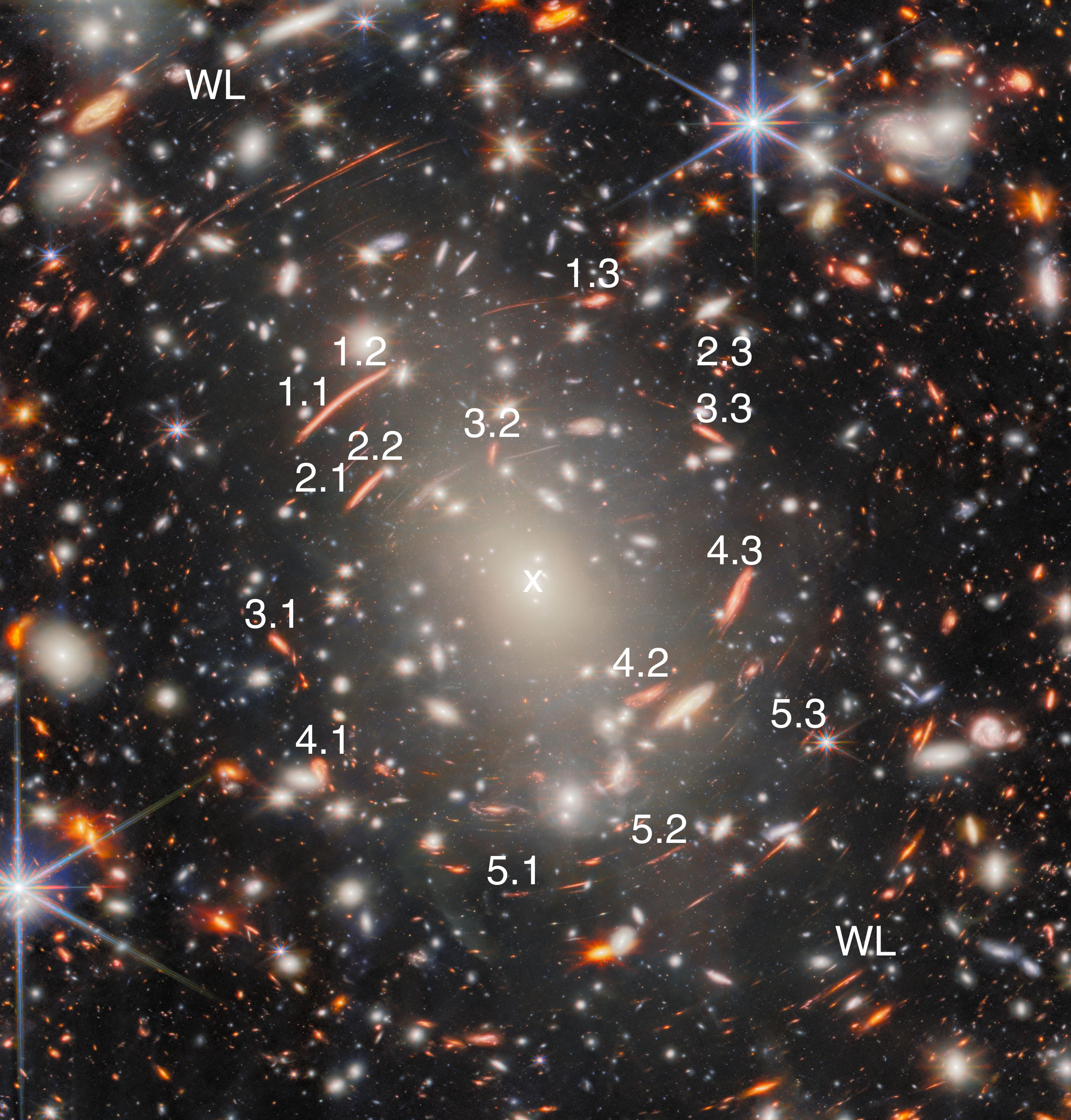} 
\caption{One of the deepest fields imaged with JWST/NIRCam to date, the lensing cluster AS1063. Image was taken as part of the GLIMPSE survey \cite{Atek2025arXiv251107542AOverview}, reaching about 31 AB magnitudes per band before accounting for lensing magnification. In the centre of the image lies the brightest cluster galaxy, marked with an "X", sitting at the heart of the cluster's potential well and thus lensing centre.  Around it are seen multiple images of strongly lensed background sources, some of which are indexed with a different leading number per family. Outwards beyond the strong-lensing regime, singly lensed tangentially sheared arcs are seen, where the weak-lensing regime roughly commences (marked with "WL"). Credit: ESA/Webb, NASA \& CSA, H. Atek, M. Zamani (ESA/Webb), R. Endsley.}
\label{fig:Glimpse}
\end{figure}

Our first mission when modelling a galaxy cluster strong-lens then is to identify the families of multiple images of the different background sources. The idea is to use the lens equation, which maps an image to the source plane, while remembering that all images of a certain source should trace back to the same source position. Although the trained eye can often associate a substantial part of the families based on symmetry and colour information, the identification often involves photometric or spectroscopic \emph{redshift} measurements, as well as an iterative process using the lens model under construction. \emph{Redshift}, denoted hereafter by $z$, defines the factor ($1+z$) by which the Universe has expanded since the observed light left the source. Generally, it is also a measure of distance: The higher the redshift, the larger the distance. Since the universe expands, objects farther away from us -- whose light also took longer to get to us --  are \emph{redshifted}. Photons emitted at a certain wavelength from a source at redshift $z$ will be received by us at a wavelength longer by a factor $\times(1+z)$. In practice, the redshift of a source can be measured spectroscopically, by comparing the observed and rest-frame wavelengths of known emission lines; or photometrically, by probing from a range of spectral templates which redshift and template best produce the observed broadband colours. Redshifts are important not only for associating multiple images, but also because they translate into the relevant distances needed for pinning down the lens model. For multiply imaged sources that do not have a spectroscopic measurement, a photometric redshift can be assigned, or alternatively, their redshift can be left free as one of the parameters of the model.   

Assuming at least some multiple images families have been identified, we can now construct a first model\footnote{There are in fact methods, such as Light-Traces-Mass \cite{Broadhurst2005a,Zitrin2009_macs0717} that are useful for the detection of multiple-image families following an initial well-guessed mass distribution.}. The basic idea, which is similar in nearly all methods, is to decide on a manner to describe the mass distribution, or on a set of assumptions, and then iterate on many realisations of the mass distribution under those assumptions, searching for the one that reproduces best the observations. For example, 'free-form' methods \cite{Diego2005Nonparam,WilliamsSaha2000AJ....119..439W,Liesenborgs2006} might assume that the larger-scale mass distribution can be modelled as a combination of some basis functions such as pixels or Gaussians, and are thus not limited to a certain pre-determined overall form \cite{Wagner2018A&A...612A..17W}. Parametric techniques, in contrast \cite{Jullo2007Lenstool,halkola06,oguri10a}, typically assume a pre-determined functional form for cluster galaxies, which are needed for an accurate reproduction of multiple images, and a large-scale functional form to represent the dark matter distribution (which is assumed to dominate the mass budget), and then iterate on the values of the parameters of these functions until the solution which best reproduces the observables (i.e. multiple images) is found.  Typically, only the positions of multiple images are being used, but it is also possible to use the shape and surface-brightness distribution, magnification ratios, or time delays (see e.g. Table 1 in \cite{SuyuEncore2026A&A...708A.291S}). The optimisation, i.e., minimisation of a cost function (typically $\chi^2$) quantifying the distance between observed and predicted multiple images, and exploration of the relevant parameter space can be done either via a grid, or more efficient methods such as gradient descent or Markov Chain Monte-Carlo. The typical accuracy for cluster lens models is of the order of a few tenths of an arcsecond. Such deviations are expected due e.g., to matter along the light-of-sight that is not independently modelled \cite{Host2012LOS}, or outliers in the several scaling relations assumed \cite{bergamini19} -- for example, in the scaling relations which determine the mass and extent of galaxies based on their luminosities. Thus contemporary lens modelling software is already capable of producing very high-fidelity results of the cluster's mass distribution. Such results have been vetted by examining the methodologies on a set of numerically simulated clusters, as well as by confronting their predictions for time delays and magnifications of lensed supernovae with observations and simulations \cite{Treu2016Refsdal,meneghetti17}. These tests have been imperative in the efforts to compare and improve high-end lens modelling techniques. 

\section{'Classical' applications of strong lensing}\label{applications}
We mentioned that over the past 2-3 decades lensing has become an important tool in astronomy. However, we did not yet explain to what end. Since we have a 'cosmic lens' or a 'looking glass' in the sky, it distorts and magnifies background objects that lie behind it. Hence, one immediate application of strong-lensing is that it allows us to study far-away galaxies in greater detail, as well as to observe faint and distant galaxies that without lensing may be too faint to be observed. Indeed, this advantage has led to cluster-lensing campaigns with the HST, constantly supplying record-breaking, high-redshift (and thus early) galaxies, thanks to lensing \cite{Franx1997,kneib04,Richard2011,Monna2014RXC2248,Zitrin2014highz,Bradley2008,Zheng2012NaturZ,Coe2012highz}.

In terms of numbers, however, it should not be obvious that lensing is advantageous for studying high-redshift galaxies. First, some other, record-breaking high-redshift galaxies have been detected also without the aid of lensing (mostly brighter ones; e.g., \cite{Ellis2013Highz,Bouwens2015LF,Oesch2016z11,Finkelstein2015}). Second, while lensing magnifies the background behind the lens and thus pushes more objects above the detection limit, since surface brightness has to be conserved, lensing effectively probes a smaller area in the source plane, per given field-of-view of a telescope. In terms of numbers, this trade-off between higher magnification and reduced source-plane area becomes advantageous only if the high-redshift luminosity function -- the density of galaxies per luminosity and redshift bin (see also section \ref{LF}) -- is sufficiently steep \citep{Broadhurst1995MagBias}. Indeed, thanks to lensing campaigns, it was established that the luminosity function was in fact as steep \citep[e.g.][]{atek18,ishigaki15a}. But regardless of the slope or net gain, lensing is in any case beneficial -- and maybe the only way -- for uncovering objects that would otherwise be simply too faint to be observed. 

Another advantage from strong lensing for high-redshift galaxies comes from image multiplicity. The deflection angle of an image scales as the ratio of the distance between lens and source and the distance to the source. This ratio saturates for high-redshift sources as the two distances grow more similar, but it is very useful for discriminating between lower-redshift and higher-redshift sources: Since the deflection depends on source redshift, the lens size or area within the critical curves increases for more distant, higher-redshift sources, increasing the angular separations between multiple images. Because most high-redshift galaxies have been identified photometrically based on their colours \citep{Bradley2008,Zheng2012NaturZ}, this 'nesting effect' could thus be used as a decisive, reinforcing evidence for the high-$z$ nature of the candidates \citep{Zitrin2014highz,Coe2012highz}. This was particularly useful in the era prior to JWST, when obtaining spectra for high-redshift candidates was mostly limited to observations from the ground, and was thus extremely challenging and often impossible. 

In order to analyse the lensed backgrounds we must construct a lens model, so we can obtain, for example, an estimate for the magnification by which background objects are magnified. The lens model essentially defines a mass distribution for the lens. Since most of the matter in the Universe, particularly in massive objects such as galaxy clusters, is believed to be dominated by an unseen component dubbed \emph{dark matter}, we thus have a means for mapping the unseen. Hence, a second immediate feat from lensing is the ability to map cosmic structures, and in particular the distribution of the otherwise-invisible dark matter. Such efforts are crucial for our understanding of dark matter, which continues to be one of astronomy's -- and in fact physics' -- greatest mysteries. Some unique examples, such as the Bullet Cluster \cite{Clowe2006Bullet,Bradac2006Bullet,Markevitch2004ApJ...606..819M} which exhibits an offset between dark matter -- probed by lensing and coincident with the galaxies -- and the gas lagging behind, imply that the dark matter self-interaction cross section is very small, consistent with zero (i.e. dark matter is referred to as collisionless). Furthermore, comparing properties of the resulting mass maps to analytical models or numerical simulations teaches us about dark matter, structure formation, and our physical universe through various cosmological parameters \cite{Voit2005RvMP...77..207V}. In addition, galaxy clusters, being the most massive gravitationally bound objects in the universe, are great laboratories for studying various astrophysical processes such as gas physics and dynamics. Coupling the gravitational potential inferred from lensing with a range of other observables such as X-ray or the Sunyaev-Zel'dovich effect teaches us about cluster physics and can also enable 3D mapping of the clusters \cite{Sigel2018ApJ...861...71S,Sereno2018ApJ...860L...4S}. 

Lensing also supplies additional probes of cosmology. The time delay between multiple images depends on the distances of and between the lens and source. The relevant combination of distances is often called the time-delay distance, providing an absolute distance scale which is inversely proportional to the Hubble constant, H$_{0}$. The observed time delay of multiply imaged variable sources such as quasars, combined with a lens model and redshift measurement, thus constitutes one of the few principal means to measure the expansion rate of the Universe \cite{Suyu2017MNRAS.468.2590SHolicow,Wong2020MNRAS.498.1420Wtension}. As such, it has been of great importance adding to the 'Hubble tension' between low-redshift distance ladder measurements of the Hubble constant \cite{Riess2021ApJ...908L...6Rtension,Freedman2021ApJ...919...16Ftension} and those obtained by \emph{Planck} observations of the \emph{Cosmic Microwave Background} \cite{Planck2020A&A...641A...6Pfinal}. In other instances, since lensing distances of background sources depend on the cosmological parameters such as matter density and on the dark energy equation of state, their ratios can be used to constrain these parameters \cite{Jullo2010}.

In summary, then, the main scientific themes that have been typically studied with lensing are dark matter, cosmology, and distant or high-redshift galaxies. Observations with the HST over the past few decades allowed for sufficiently high resolution data to realise these objectives and push these fields to previously unimaginable frontiers, revealing, for example in the high-redshift regime, galaxies as early as 500 Myr after the Big Bang.

\section{Strong lensing with JWST: A new era}\label{JWST}

A major leap in strong-lensing science followed the advent of JWST. After more than 20 years in development and much anticipation by the astronomical community and science aficionados alike, the JWST was launched into space on Dec 25th 2021. With an effective mirror size of 6.5m in diameter, JWST is the largest telescope in space. It operates in a wide wavelength range from red-optical to the mid-infrared (0.6–28 $\mu$m, roughly), with four advanced instruments for imaging and spectroscopy. Its size and wavelength coverage supply unprecedented sensitivity and resolution; it can reach to $\sim$30 AB magnitudes\footnote{Astronomers often express flux densities on a logarithmic scale called magnitudes. The AB magnitude system, for example, is defined as $m = -2.5 \log_{10}(f_\nu) - 48.60$, where the flux density $f_{\nu }$ is in units of $\text{erg s}^{-1}\text{ cm}^{-2}\text{ Hz}^{-1}$.} in a reasonable exposure time, with a spatial resolution (FWHM) of better than 0.03$\arcsec$ at short wavelengths, and 0.07$\arcsec$ around 2 $\mu$m. The observatory is thus ideal for studying distant sources -- whose UV and optical light, due to the expansion of the Universe, is redshifted towards longer wavelengths. Indeed, a primary goal of the JWST is to study the first stars and galaxies -- 'first light'.

\begin{figure}[!ht]
\includegraphics[width=1\textwidth]{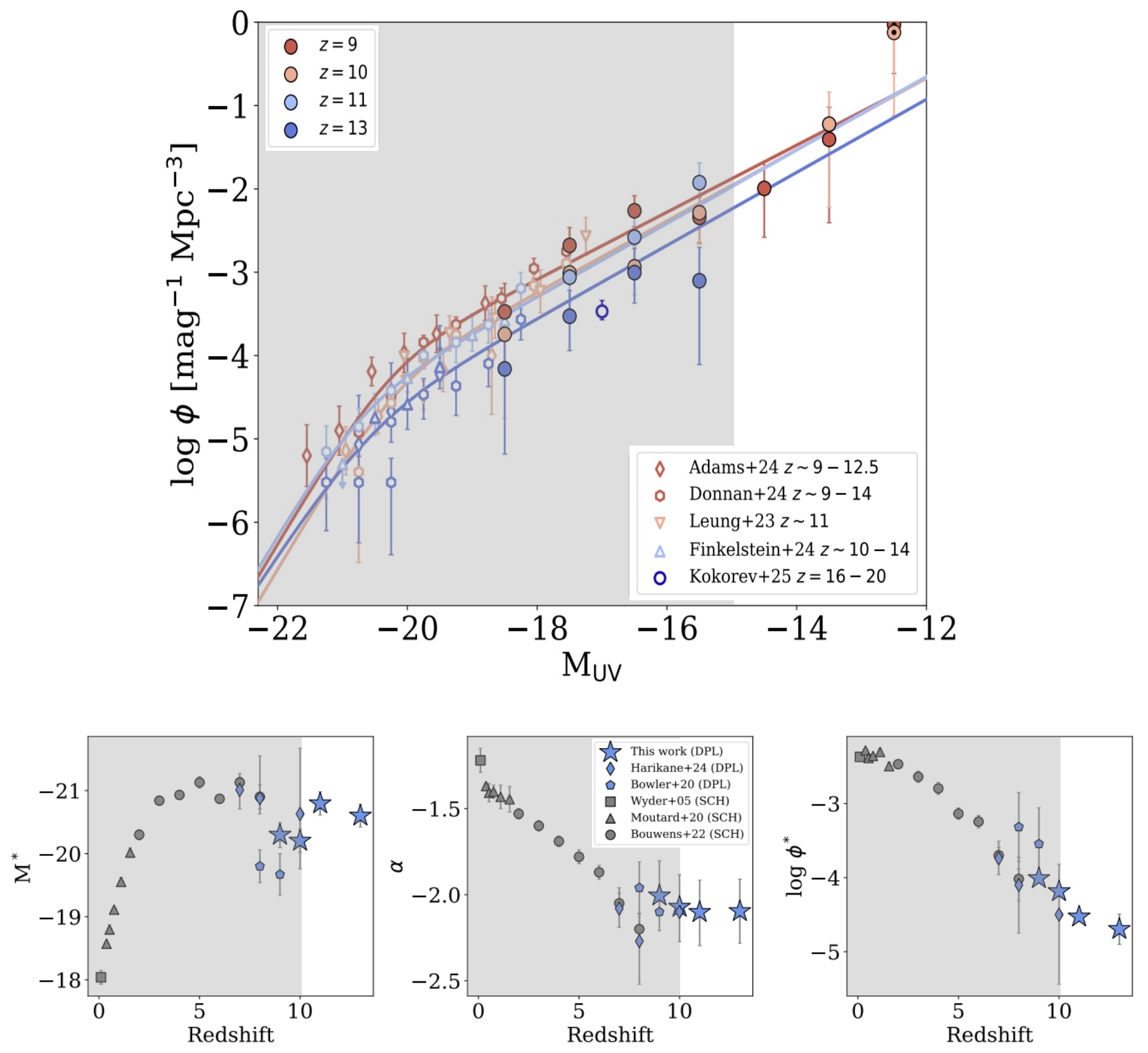} 
\caption{Deep view of the faint end of the luminosity function allowed by JWST+Lensing. The upper subfigure shows the luminosity function derived from very-deep observations of the lensing cluster AS1063 taken for the JWST/GLIMPSE program (seen in Figure \ref{fig:Glimpse}) \cite{Chemerynska2026MNRAS.546f2267C}, compared to several other works from the literature \cite{Adams2023MNRAS.518.4755A,Donnan2024MNRAS.533.3222D,Leung2023ApJ...954L..46L,Finkelstein2024ApJ...969L...2F,Kokorev2025ApJ...983L..22Kz16}. The bottom panels show the evolution of the luminosity function parameters with redshift, also compared to various works \cite{Harikane2024ApJ...960...56H,Bowler2020MNRAS.493.2059B,Wyder2005ApJ...619L..15W,Moutard2020MNRAS.494.1894M,Bouwens2022ApJ...940...55B}. Note also the steepening of the faint-end slope $\alpha$ at high-redshift, making lensing particularly beneficial. I overplot here, on the figures adapted from \cite{Chemerynska2026MNRAS.546f2267C}, the limits reachable with HST as shaded grey regions, taken here nominally as M$_{UV}$=-15 (at $z\sim8$), and $z=10$. As can be seen, the combination of JWST with lensing enables us to improve the constraints and probe far beyond these limits, to very faint and distant galaxies inaccessible otherwise.}
\label{fig:Chemerynska}
\end{figure}

\subsection{High-redshift galaxies with JWST: An uncharted territory}
\subsubsection{The UV luminosity function and reionisation of the Universe}\label{LF}
It has been well established that the inter-galactic Hydrogen was reionised during the first billion years of the Universe. One of the long standing questions regarding this important process, was whether galaxies drove reionisation thanks to UV radiation from massive stars, or whether it was AGN or other exotic sources that supplied the required radiation field. The key to answering this question is the number of galaxies and their ionising contribution. The 'counting' of galaxies in different wavelengths and luminosity bins forms what is known as the \emph{luminosity function}, whose shape and evolution are thus of great interest, as they help us characterise the shape and timeline of reionisation. Because there are far more fainter galaxies than brighter ones, lensing -- which can typically push observational limits for detecting distant galaxies by several magnitudes -- is bound to play a critical role in probing the faint and distant population in the early universe. 

A common functional form used to describe the luminosity function is the \emph{Schechter function} \cite{schechter76}. It comprises of a power-law with slope $\alpha$ on the faint end, and an exponential cutoff towards the brighter end where fewer galaxies are expected, beyond a characteristic luminosity $L^{*}$ (or absolute magnitude $M^{*}$). The normalisation of the function is given by $\Phi^{*}$. See for example Fig. \ref{fig:Chemerynska}.

Prior to JWST, pairing deep HST observations with lensing fields such as the Hubble Frontier Fields clusters allowed astronomers to construct robust, high-redshift luminosity functions to an ultraviolet absolute magnitude as faint as $M_{UV}\sim-15$
\footnote{
$M_{UV}$ refers to the \emph{absolute magnitude} (in the ultra-violet, in this case), defined as the apparent magnitude of the galaxy, $m$, were it placed 10 parsecs away, $M\approx m-5\log_{10}(\frac{d}{10})$, with $d$ the distance to the galaxy, in parsecs. $M_{UV}\sim-15$ at $z\sim7-8$ corresponds to an apparent magnitude of about $32-33$ AB.
}
towards $z\sim8$ \cite{Ishigaki2018HFF,atek18}, hinting that galaxies were indeed, likely responsible for reionisation; in particular, if the slope of the luminosity function continued unchanged to fainter magnitudes \cite{Livermore2017ApJ...835..113L}. But given this regime was largely unreachable with \emph{Hubble}, it was not clear whether this was indeed the case \cite{Bouwens2017ApJ...843..129Bturnover,atek18}. Some early programs with JWST, such as GLASS-JWST \cite{Treu2022ApJ...935..110TGlass}, the CAnadian NIRISS Unbiased Cluster Survey (CANUCS, \cite{Willott2022PASP..134b5002WCanucs,Sarrouh2026ApJS..282....3SCanucs}), Prime Extragalactic Areas for Reionization and Lensing Science (PEARLS, \cite{Windhorst2023AJ....165...13W}), and Ultradeep NIRSpec and NIRCam Observations before the Epoch of Reionization (UNCOVER, \cite{Bezanson2024ApJ...974...92B}), extended this previous success and targetted lensing fields to push observational limits deeper than ever before. Some of these programs reached already imaging depths of $\sim29-30$ magnitudes per band before additional $\sim2-3$ magnitudes from lensing, finding for example -- among many other important results -- that galaxies were likely responsible for reionisation \cite{Atek2024Natur.626..975Aionized}. The gravitational lensing \& NIRCam imaging to probe early galaxy formation and sources of reionisation (GLIMPSE, \cite{Atek2025arXiv251107542AOverview}) program, which observed with JWST one of the Hubble Frontier Fields clusters, AS1063, for 20-40 hours per band reaching 31 AB magnitudes before accounting for further boost from lensing, supplies one of the deepest fields ever imaged (Fig. \ref{fig:Glimpse}) and pushes the limits towards $M_{UV}\sim-12$, marking a major advancement over the limits reached with \emph{Hubble} and enabling better insight into the drivers of reionisation unachievable without the aid from lensing. The luminosity function from this survey is shown in Fig. \ref{fig:Chemerynska}. This additional insight also includes lensed galaxies exhibiting potential signatures of population III stars - the first generation of stars that was formed in the Universe \cite{FujimotoPopIII2025ApJ...989...46F}. The ongoing program Vast Exploration for Nascent Unexplored Sources (VENUS, e.g. \cite{Nakane2025arXiv251114483N}) is targetting a large number of clusters, which should both help to overcome cosmic variance and allow for substantial areas of high magnification, thus probing the $M_{UV}\simeq-12$ regime at $z\sim6–9$ as well.

\subsubsection{The most distant galaxies in the Universe}
\begin{figure}[!hb]
\centering 
\includegraphics[width=0.8\textwidth,trim={0.5cm 0 0 0},clip]{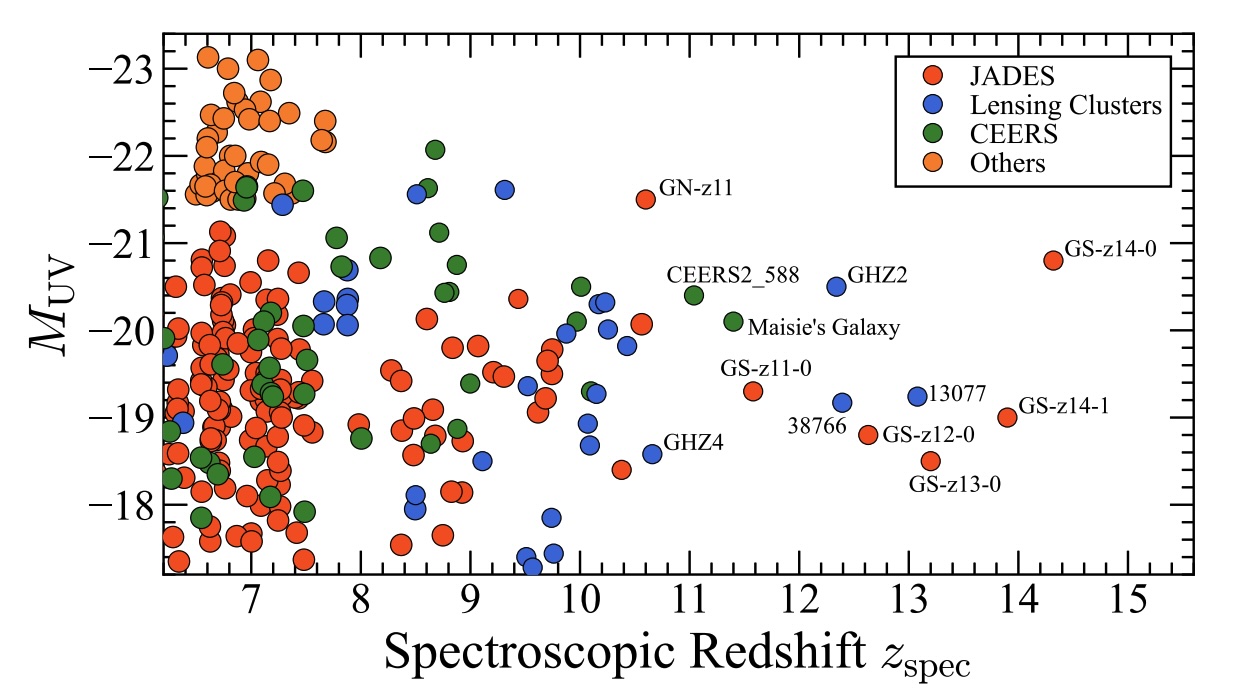} 
\caption{
Compilation of high-redshift galaxies from the literature. Figure is taken from \cite{Harikane2025ApJ...980..138H}; see references therein. Note the prominent contribution of lensing to high-redshift galaxy detection and the numerous galaxies now detected and spectroscopically confirmed with JWST beyond the limits allowed by HST ($z\sim11)$.}
\label{fig:Harikane2}
\end{figure}

The ability of JWST to observe fainter and farther than ever before naturally revealed new, record-breaking galaxies. Its spectroscopic capabilities now also allow us to confirm their redshifts, through emission lines or through their spectral break near the Lyman limit. While \emph{Hubble} could only observe galaxies out to redshift $\sim11-12$ \cite{Ellis2013Highz,Coe2012highz,Oesch2016z11}, at the time of writing there are already several galaxies spectroscopically confirmed with JWST around redshift $\sim14-14.5$  \citep{CurtisLake2023NatAs...7..622C,Carniani2024Natur.633..318C,Naidu2026OJAp....956033N1444} (see also Fig. \ref{fig:Harikane2} here), and some photometrically selected galaxies up to higher redshifts \cite{Hainline2026arXiv260115959H}. Interestingly, although $z\sim12-13$ galaxies have already been spectroscopically detected behind lensing clusters (e.g. \cite{Wang2023ApJ...957L..34Whighz,Castellano2024ApJ...972..143C,Roberts-Borsani2024ApJ...976..193R}), currently the 2-3 farthest galaxies at $z\sim14$ are relatively bright field galaxies chosen from wider field surveys with JWST, and not strongly lensed galaxies. This may be regarded as somewhat surprising since with \emph{Hubble}, individual, high-redshift lensed sources seemed to have been more prominent (e.g. \cite{Coe2012highz,Zitrin2014highz,Salmon2020RELICSHighz}), although the two farthest galaxies detected with \emph{Hubble}, GNz11 \cite{Oesch2016z11} or UDFj-39546284 \cite{Bouwens2011NaturZ10Gal,Ellis2013Highz,Robertson2023NatAs...7..611RHighzCOnf} are also bright field galaxies. We may thus expect that new record-breaking galaxies will also be discovered behind lensing clusters soon, but this prediction depends on various factors such as the shape of the luminosity function, contamination from intra-cluster light, and how JWST resources are divided between field surveys and cluster lensing programs. Note, lensed candidates have also been detected to redshift 16 and beyond \cite{Kokorev2025ApJ...983L..22Kz16}, albeit currently lacking spectroscopy. 

\newpage
\subsection{Strongly lensed Active Galactic Nuclei and Supermassive Black Holes}

Multiply imaged quasars and AGN have long been of great interest. They can be used to measure the expansion rate of the Universe through time delays \cite{Wong2020MNRAS.498.1420Wtension,Napier2023}, substructure in the lens through flux anomalies \cite{Hezaveh2016ApJ...823...37H}, the size of the source through micro- or small-scale lensing \cite{Mediavilla2011ApJ...730...16M}, or its mass through changes in brightness (and their echo from outer regions; a technique known as \emph{reverberation mapping}), which can be monitored more effectively thanks to lensing time delays \cite{Williams2021ApJ...911...64W,golubchik24}. Nevertheless, while many dozens of lensed quasars have been found to date, and many more are expected now with new large sky surveys such as Euclid, Roman, and Rubin/LSST, most of those are lensed by galaxies. Only several multiply imaged quasars have been found behind clusters to date (e.g. \cite{Inada2003Natur.426..810I,Inada2006,Dahle2013,Oguri2013MNRAS.429..482O,sharon2017,Shu2018,Bogdan2022,Martinez2022,Ducourant2026A&A...707A.345D}). 

It has been shown that bright, strongly lensed objects that maintain a point-like morphology may suggest an underlying supermassive black hole \cite{furtak23b}. Although JWST has been active for only four years, a few other multiply imaged AGN candidates have already been found following this rationale \cite{furtak23d,Allingham2025ApJ...990L..25A}. Thanks to its increased resolution and depth, it can be assumed that additional multiply imaged AGN will be similarly found with JWST soon. Another benefit from lensing comes from the magnification, which allows us to study smaller-mass black holes. Lensing surveys of large numbers of clusters with JWST such as the VENUS program, can now measure black hole masses down to $\sim10^5$ solar masses in lensed galaxies and shed light on their co-evolution with their host galaxies through cosmic history.

\begin{figure}
\includegraphics[width=0.5\textwidth]{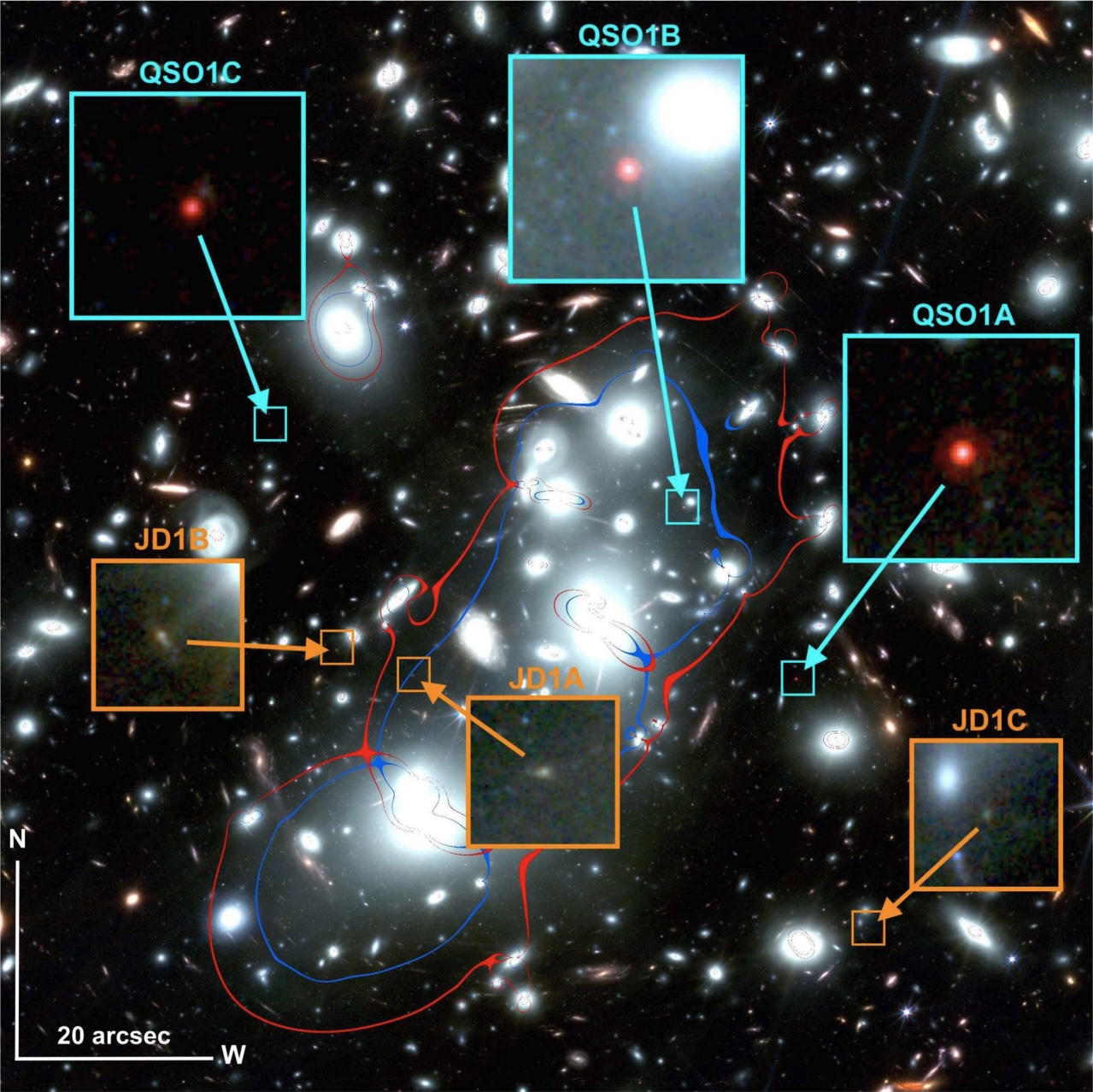} 
\includegraphics[width=0.5\textwidth]{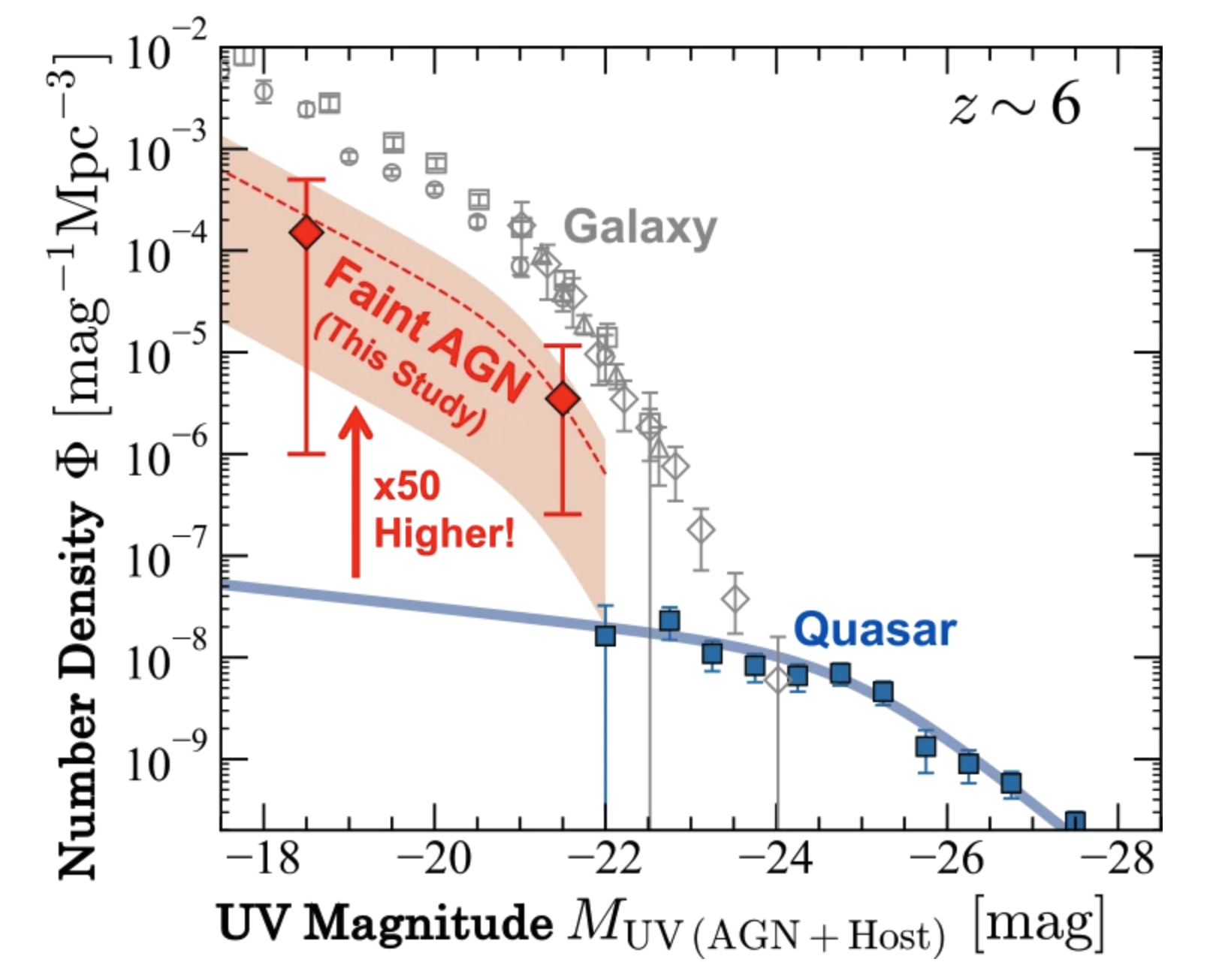} 
\caption{JWST revealed an unexpected population of faint AGN. Left: Image of the lensing cluster Abell 2744 taken as part of the UNCOVER JWST survey (Credit: NASA, ESA, CSA, I. Labbe, R. Bezanson, L. Furtak, A. Zitrin; figure from \cite{furtak23d}). A distinct, triply imaged red and compact source was detected, labelled QSO1. Despite large magnification and shear, it remained a point source over its three images suggesting a very small source size. Another high-redshift triply imaged galaxy, JD1, is highlighted for comparison. A2744-QSO1 became a proto-type of a population now known as LRDs. Together with its unique V-shaped SED, this suggested that LRDs could be powered by SMBH \cite{furtak24b}. Right: The population of faint AGN discovered by JWST is $\times50$ more abundant than extrapolations from previous, brighter quasars. Figure from \citep{Harikane2025Ap&SS.370...85HReview} (see also references therein). Lensing helps push the luminosity function fainter by about 2 additional magnitudes \citep{greene24}.}
\label{fig:Harikane1}
\end{figure}

In its first year of operations JWST detected an abundance of faint AGN at high redshifts \citep{Harikane2023,Kocevski2023}, with lensing supplying some additional, unexpected finds. In one of the first, deep cluster fields observed with JWST, Abell 2744, imaged for UNCOVER program, a puzzling source was discovered: A triply imaged, $z\sim7$ \emph{extremely red and compact} source (\cite{furtak23d}; see Figure \ref{fig:Harikane1} here). The source, despite being magnified and sheared by lensing, remained a point source across its three images, thus implying a size of less than $\sim30$ pc. Given its clear quasi-stellar appearance in JWST data it was given the name Abell2744-QSO1. It turned out this source had originally been found with \emph{Hubble} as a $z\sim7$ candidate \cite{Atek2014A2744}, but \emph{Hubble} lacked the resolution and wavelength coverage needed to reveal its uniqueness. Spectral energy distribution (SED) modelling using the extended wavelength coverage allowed by JWST suggested a high mass, similar to other, over-massive red objects previously reported by e.g. \cite{Labbe2023Natur.616..266L}. But now, thanks to the size constraint from lensing, it was obvious that it would be difficult to have so many stars packed in such a small region; the density would be unreasonably high. The most obvious explanation was, similar to brighter quasars or AGN, that the radiation was due to accretion onto a SMBH. Indeed, follow-up observations revealed broad Balmer-series emission lines of several thousand km/s widths, indicative of AGN, and further revealing a high black-hole to host mass ratio \cite{furtak24b}. But the object had a couple of other distinct features which were not typical of broad-line AGN: It had a blue UV spectral slope but red optical slope, forming a unique V-shape SED, and it lacked X-ray or radio detections, motivating also other alternative explanations. Several other such red and compact objects were quickly detected around the same time, in the field or more modestly magnified regions, and were given the appropriate name \emph{Little Red Dots} \cite{Matthee2024ApJ...963..129M,Greene2024ApJ...964...39G,Labbe2025ApJ...978...92L}. Many dozens of LRDs have been by now detected across various JWST fields at a range of redshifts \cite{deGraaf2025arXiv251121820D}.

\begin{figure}[!b]
\includegraphics[width=1\textwidth]{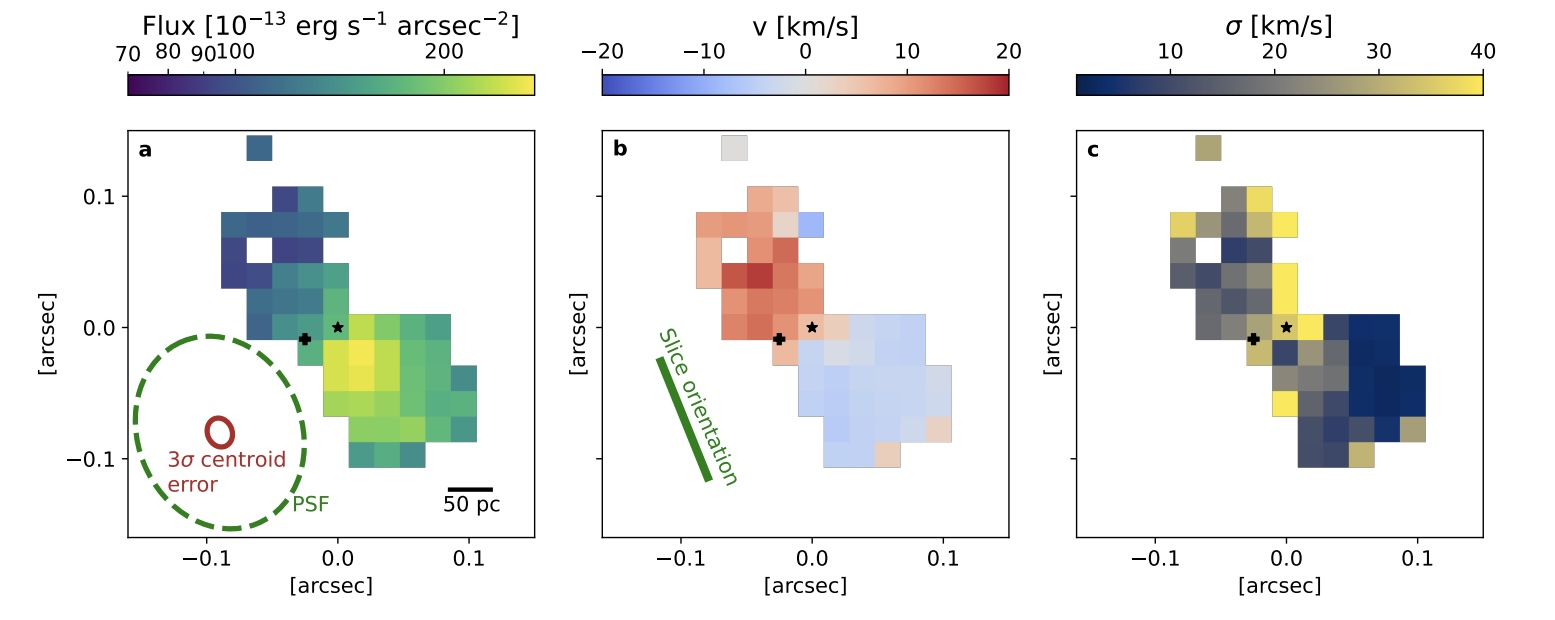} 
\caption{JWST allows to resolve the sphere of influence around strongly magnified SMBHs in the early universe, enabling a direct dynamical mass measurement. The Figure, taken from \cite{Juodzbalis2025arXiv250821748J}, shows the H$\alpha$ flux, velocity, and velocity dispersion fields around A2744-QSO1 using NIRSpec integral field observations.}
\label{fig:SphereOfInfulence}
\end{figure} 

Abell2744-QSO1 continued to play a pivotal role in understanding LRDs. Since there exists a substantial time delay between the images, it is possible to search for variability -- an expected signature of Type I AGN -- among the three images of QSO1, which effectively span about 20 years of observations. Signs for variability were found in the strengths (or equivalent widths) of broad Balmer line emission from QSO1 \cite{Furtak2025A&A...698A.227F,Ji2025MNRAS.544.3900J}, supporting the AGN scenario. 

Lensing also enables another important measurement for AGN: For the high magnifications typically seen for strongly lensed sources, and for high enough black hole masses, the sphere of influence of the BH becomes observable with JWST (e.g. \cite{NewmanBHmass2025arXiv250317478N}). NIRSpec/IFU observations supplied such direct dynamical mass measurement also for Abell2744-QSO1 \cite{Juodzbalis2025arXiv250821748J} (see Fig. \ref{fig:SphereOfInfulence} here).

\begin{figure}
\includegraphics[width=1\textwidth]{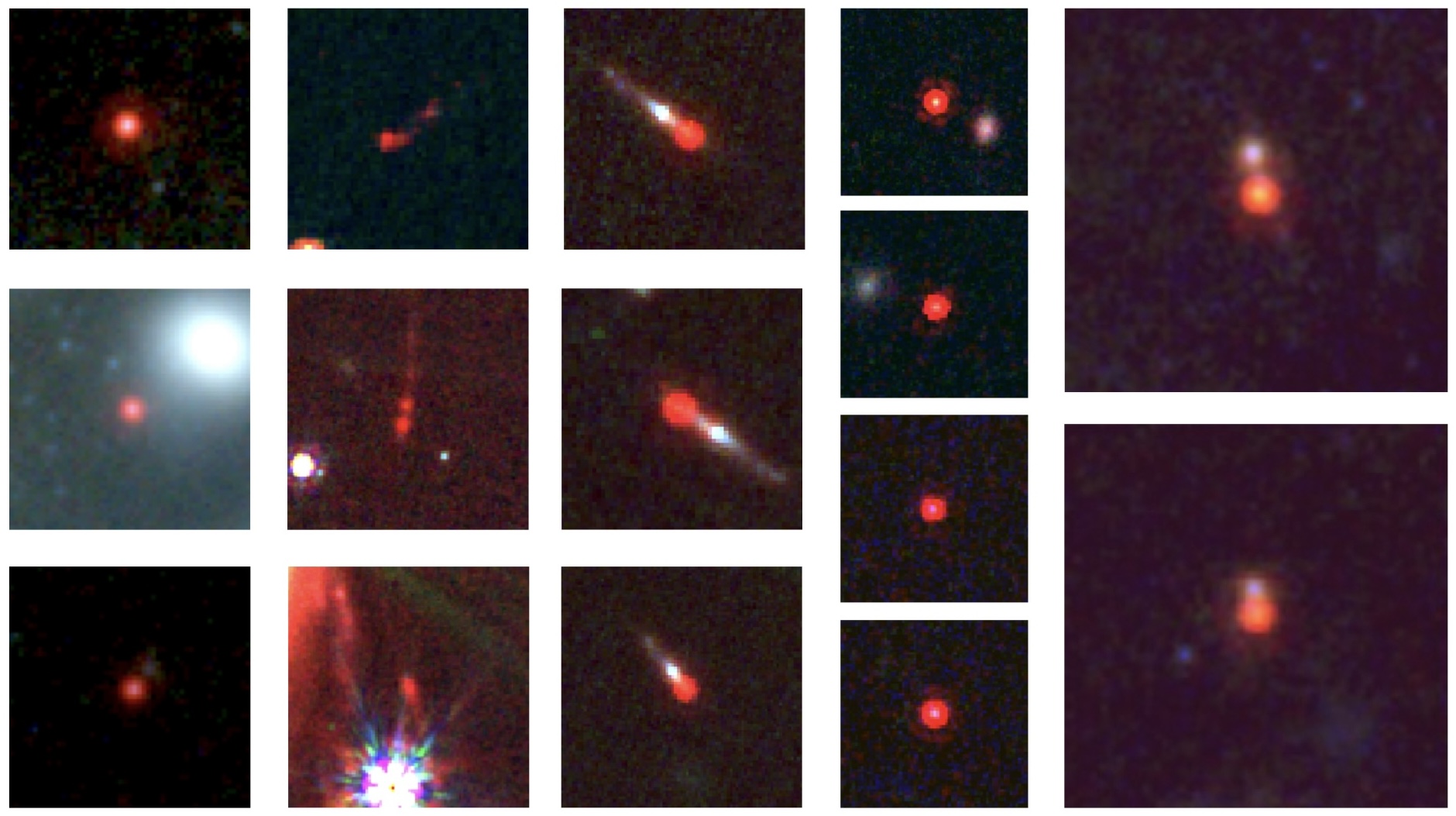} 
\caption{Compilation of strongly lensed LRDs. Each column shows a different system of a multiply imaged LRD (see \cite{furtak23d,Yanagisawa2026arXiv260106015Y_doubleLRD,Baggen2026arXiv260202702B,Zhang2025arXiv251205180ZRXC221LRDs,Golubchik2025arXiv251202117G_LRD}). The stamps were constructed here from public data and their sizes arbitrary. As can be seen, LRDs maintain a point-like appearance despite significant magnification, which places strong constraints on their physical size. Some of the LRDs have also a notable, nearby companion.}
\label{fig:LRDs}
\end{figure}

A few more multiply imaged LRDs have been by now discovered in recent lensing campaigns \cite{ZhangRXC221LRD2025arXiv251205180Z,Baggen2026arXiv260202702B}. Some of these show additional detail such as a close-by blue emission \citep{Golubchik2025arXiv251202117G_LRD,Baggen2026arXiv260202702B} or a LRD close pair \cite{YanagisawaLRDpair2026arXiv260106015Y}, thanks to lensing magnification (see Fig. \ref{fig:LRDs}). Investigations of the nature of some of these companions are ongoing.

\subsection{A magnified view of distant galaxies}
\begin{figure}
\includegraphics[width=1\textwidth]{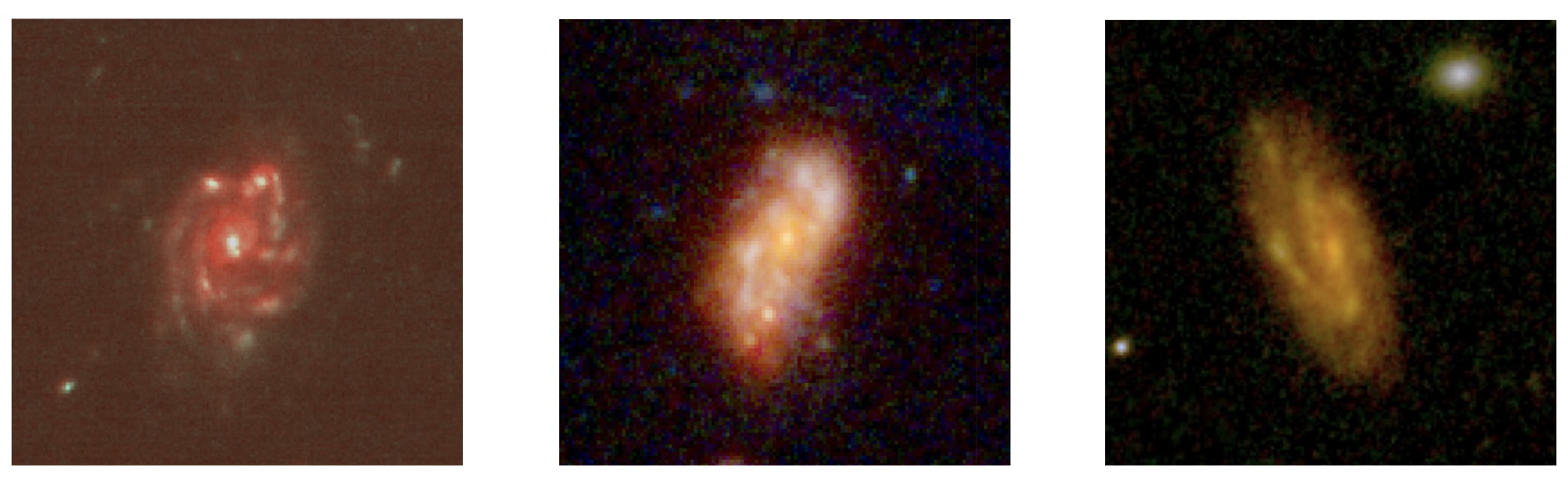}
\caption{JWST observes disk and spiral galaxies farther than ever before. From left to right, the subfigures show the Big Wheel galaxy at $z=3.25$ (\cite{Wang2025NatAs...9..710W}; not gravitationally lensed); galaxy Alaknanda, at $z\simeq4$ (\cite{Jain2025A&A...703A..96J}; moderately gravitationally magnified by the galaxy cluster Abell 2744);  galaxy Charybdis at $z\sim3.6$ -- an apparently spiral galaxy strongly lensed and triply imaged by the galaxy cluster MACS J1931.8-2635, enabling the detailed view \cite{Allingham2026arXiv260214074A_Eos}. Stamps are constructed from public data and their orientations and sizes are arbitrary.}
\label{fig:Spiral}
\end{figure}

The high magnification from lensing coupled with JWST allows us to obtain a more detailed view of distant galaxies than was previously possible, reaching a spatial resolution of about a few tens of pc in the source plane. \emph{Hubble}, coupled with lensing, supplied a magnified view of distant galaxies in some representative cases going down to 50-100 pc resolutions at redshifts $\sim2-5$, for example \cite{sharon12,Zitrin2011MS1358}. With JWST, this enhancement is further boosted and these numbers can be improved by factors of a few per object, simply given the ratio in mirror diameters. Together with the extended wavelength coverage, this also allows us to observe different types of galaxies in detail to higher redshifts. For example, spiral galaxies at increasing redshifts have long attracted attention, as these can shed light on when and how these spiral structures formed (e.g. \cite{Law2012Natur.487..338L}). The JWST, with its increased resolution now enables us to observe spiral galaxies to higher redshifts than before and indeed, various samples up to at least $z\sim5$ have been assembled \cite{Costantin2023Natur.623..499C,Kuhn2024ApJ...968L..15K,Xiao2025A&A...696A.156X}. Such galaxies could be seen with even further detail, were they lensed. For example, Figure \ref{fig:Spiral} shows a clear spiral structure in some distant galaxies, including a couple of lensed ones, at redshifts $z\sim3-4$. 

\subsection{Star clusters}
Understanding star formation and galaxy evolution across cosmic history is one of the key goals of modern astronomy. As fundamental building-blocks of galaxies and an excellent tracer of their star formation history \cite{Claeyssens2023MNRAS.520.2180C,PortegiesZwart2010ARA&A..48..431P,Claeyssens2026arXiv260116281C,Adamo2020MNRAS.499.3267A}, star clusters are thus of great interest. Their formation and evolution, feedback mechanisms and effects on the interstellar medium, as well as the survival mechanism of gravitationally bound clusters through redshift, have been largely unknown. Indeed, directly observing star clusters beyond the low-redshift universe has been very challenging: Even young star clusters with orders of $10^5$ stars remain faint ($>29$ AB in the blue \cite{Claeyssens2026arXiv260116281C}), and with physical scales of $\sim$pc to couple dozen pc they remained largely unresolved (for example, a 20 pc object at redshift 1 occupies an angular size of a few 0.001 arcseconds). 

\begin{figure}[!t]
\includegraphics[width=1\textwidth]{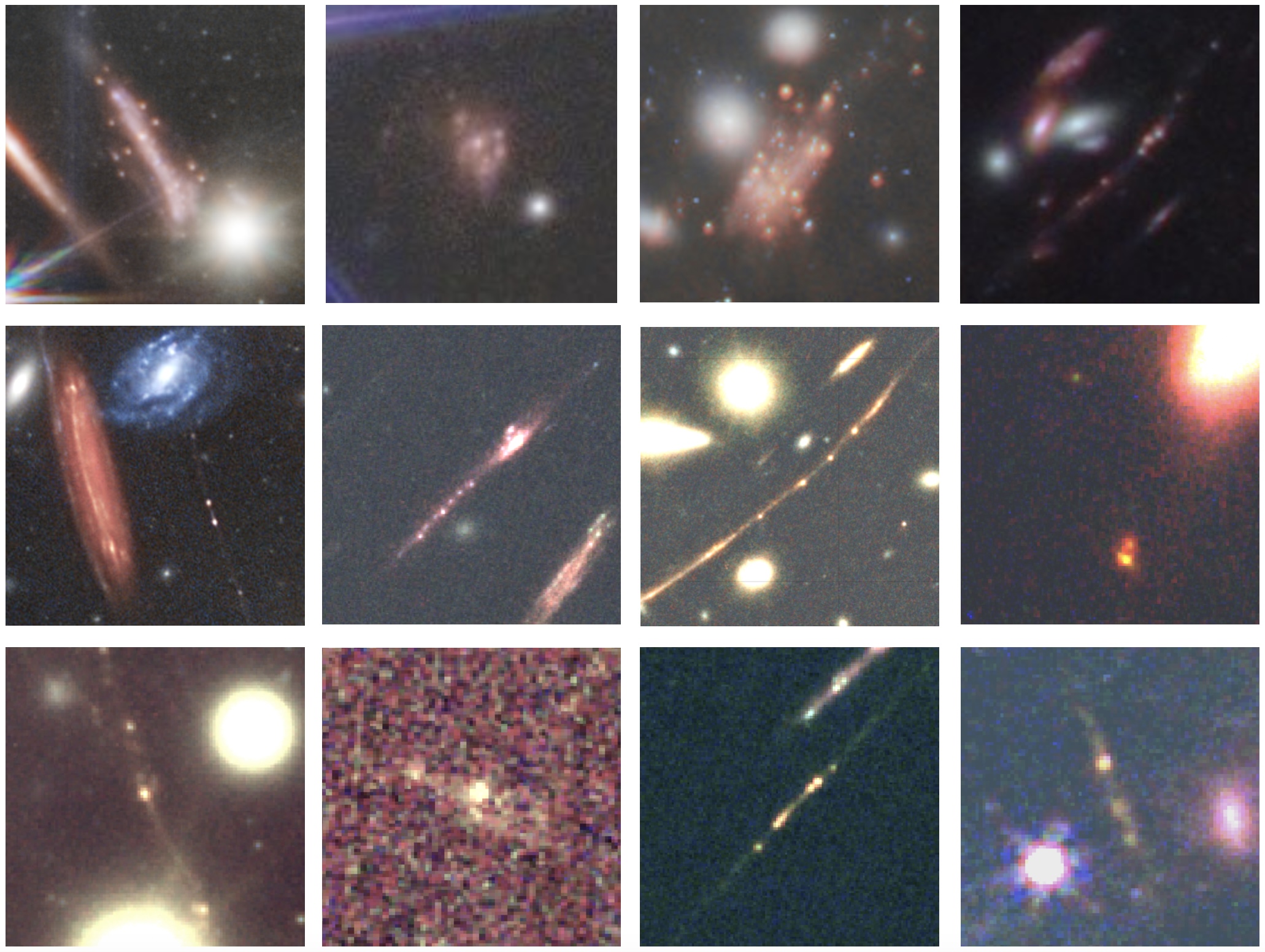} 
\caption{Collage of star-forming clumps, globular clusters, and highly magnified arcs seen with JWST throughout cosmic history. Shown here are example arcs and clumps up to redshift $z\sim11$ complied from the literature \cite{Mowla2022ApJ...937L..35M,
Abdurrouf2025arXiv251208054Az6clumps,
Bradac2025ApJ...995L..74Bz11SF,
Adamo2024Natur.632..513A,
Nakane2025arXiv251114483N,
Hsiao2023ApJ...949L..34Hz11,
Vanzella2023ApJ...945...53VSunrise,
Fujimoto2025NatAs...9.1553F} or from various lensing surveys \cite{Atek2025arXiv251107542AOverview,
Bezanson2024ApJ...974...92B,
Windhorst2023AJ....165...13W}, and constructed here from publicly available data (Credit: ESA/Webb, NASA \& CSA). Many of these clumps and point sources could not be resolved prior to JWST. Especially in highly magnified arcs, these are now resolved down to $\sim1$ pc. The orientations and stamp sizes are arbitrary.}
\label{fig:Clumps}
\end{figure}

One of the first images by JWST that was released to the public was of a galaxy cluster lens, SMACS J0723.3–7327, revealing an unprecedented amount of detail and specifically, many previously unobserved point sources; either star clusters in the intracluster light, or lensed star clusters in the background, for example in or around the halos of magnified distant galaxies. The first striking example for this was likely the Sparkler galaxy \cite{Mowla2022ApJ...937L..35M}, seen lensed by the cluster (upper-left stamp in Figure \ref{fig:Clumps} here). The JWST image of the galaxy revealed various star clusters surrounding it; objects we could simply not see before. Various similar galaxies have been detected since with JWST (see Figure \ref{fig:Clumps}), suggesting that, when combined with lensing, JWST allows us to study star formation at the smallest scales. 

But it is also JWST's wavelength coverage that plays a role in pushing the limits, enabling stellar population synthesis in star clusters at growing redshifts. In that respect it is worth noting that JWST has managed to study several gravitational arcs at high redshifts \citep{Vanzella2023ApJ...945...53VSunrise,Adamo2024Natur.632..513A}. To form a gravitational arc the source has to sit on the caustic, and this scenario is thus rarer. But this configuration is interesting in another aspect, too: A source sitting on the caustic would be significantly stretched and some parts of it very highly magnified. The turn-over in angular-diameter distance towards high-redshifts\footnote{The angular-diameter distance is a cosmological distance measure defined as the ratio of an object's physical size to its observed angular size. The angular-diameter distance increases with redshift since more distant sources appear smaller on the sky, but at some point turns over and decreases with source redshift. This is because in an expanding universe, very distant sources were much closer to us when they emitted the light that is now reaching us.} aids in increasing the probed physical resolution as well. Indeed, JWST enabled the detection of star clusters as small as 1 pc, revealing when, where, and how the first stellar populations and proto–globular clusters formed. At high redshifts these may also be the sites from which ionising radiation is coming, and is thus important also for understanding the role of early galaxies in reionisation.

\subsection{Lensing of stars at cosmological distances}
Lensing of distant stars marks a remarkable advancement in strong-lensing science over the past decade. In the previous subsection we mentioned that a gravitational arc forms when the source -- typically a galaxy -- sits on the caustic, and some parts of it get highly magnified. Because objects move in space, there is a relative transverse motion between the source, lens and observer, so that effectively, the source moves transversely with respect to the line-of-sight to the lens. Stars in that source galaxy will thus occasionally sweep across the caustic, and get extremely magnified as they do -- producing a typical light curve, as was already predicted three and a half decades ago \cite{miralda-escude91}. The maximum magnification that a source can attain is limited by the size of the source and thus, for stars, can in principle reach even $\sim10^5-10^6$. This means that \emph{individual stars} at cosmological distances can be temporarily observed with lensing. These events are often called \emph{caustic crossing events}. The caustic crossing time, given by the star's radius divided by the velocity (say, hundreds of km/s), is of timescale of hours to days, typically.  

The first instance of an extremely magnified star crossing the caustic of a cluster lens was serendipitously found with the HST already a decade ago, when monitoring the galaxy cluster MACS J1149.5+2223 \citep{Kelly2018NatAsCCE}. Once it was realised that \emph{Hubble} could see individual stars, an effort was made to characterise such events: What type of stars or other sources (for example pristine population III stars and their stellar-mass black hole accretion disks \cite{Windhorst2018ApJSCCE}) are more likely to be found, the expected rate, or dependency on various parameters such as microlens density \cite{Venumadhav2017ApJCCE,Diego2018ApJCCE,Oguri2018PhRvDCCE} or Initial Mass Function \cite{meena25,Li2025ApJ...988..178L} -- the distribution of masses of newly born stars. In parallel, more events of lensed stars were found in different clusters \citep{Chen2019CCE,Kaurov2019CCE,meena23a}, in random or dedicated lensing campaigns. Some systematic searches supplied important empirical limits suggesting that although CCEs can be seen with \emph{Hubble}, the rate of events is quite low simply because typical HST observations are not deep enough \cite{golubchik23}. The first event in MACS J1149.5+2223, it turned out, was at the relatively bright end of the expected distribution of events.

The first caustic-crossing events supplied another important insight. In practice, the lens itself is also composed of stars and other massive substructure. This means that the (macro) caustic is surrounded by smaller micro-caustics, or that it is not smooth, but breaks into a net of smaller micro-caustics \cite{Venumadhav2017ApJCCE,Diego2018ApJCCE}.  In fact, the amount of micro-lensing can also teach us about the composition of the lens and thus also important for studying the stellar population and compact dark matter content in the lens. Correspondingly, the light-curve during a caustic-crossing event, as was seen in MACS J1149.5+2223 \cite{Kelly2018NatAsCCE}, would break into multiple smaller peaks, each typically with magnification of 'only' a few hundreds to a few thousands. This, nonetheless, is still sufficient to observe single stars close to the caustic \cite{Meena2022MNRAS.514.2545M}. Bright, super-giant stars can be naturally observed as (micro-)caustic crossing transients also somewhat farther away from the caustic (even up to about $\sim1$ arcsecond or more; \cite{meena23b}). Very close to the position of the macro-caustic, when in the net, stars will appear semi persistent -- they can remain highly magnified for decades -- showing only minor flux fluctuations. Earendel, the highest-redshift star candidate to date  \cite[][]{Welch2022EarendelHST,Welch2022EarendelJWST}, is one of those cases, and thus in principle can also be a star cluster \cite{Pascale2025ApJ...988L..76PEarendel}.

\begin{figure}
\hspace{0.3cm}
\includegraphics[width=0.96\textwidth]{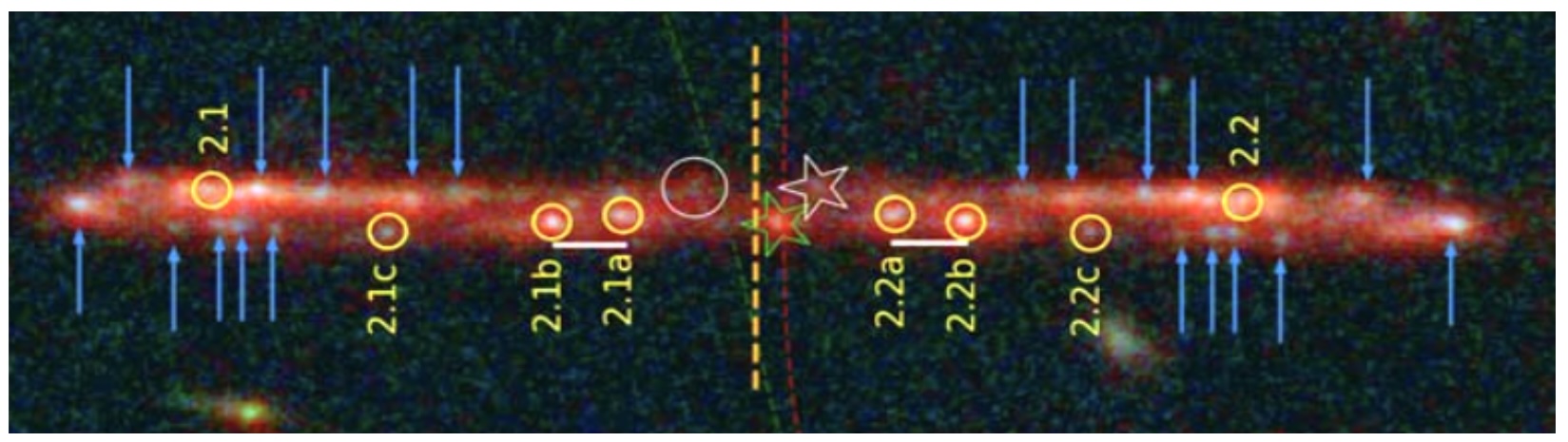}
\caption{Lensed star candidates in an arc at $z\sim4.8$, detected in JWST/NIRCam observations \cite{meena23a}. The figure shows two lensed star candidates (marked with a star shape) next to positions of the critical curves (dashed lines). The faint green and purple dashed lines show the critical curves from two different lens models, and the yellow dashed line marks the symmetry point of the arc. Counter images of various knots are marked with blue arrows, some of them also circled and numbered. Figure from \cite{meena23a}.}
\label{fig:starspectrum1}
\vspace{2cm}
\includegraphics[width=1\textwidth,trim={0.1cm 0cm 0cm 0cm},clip]{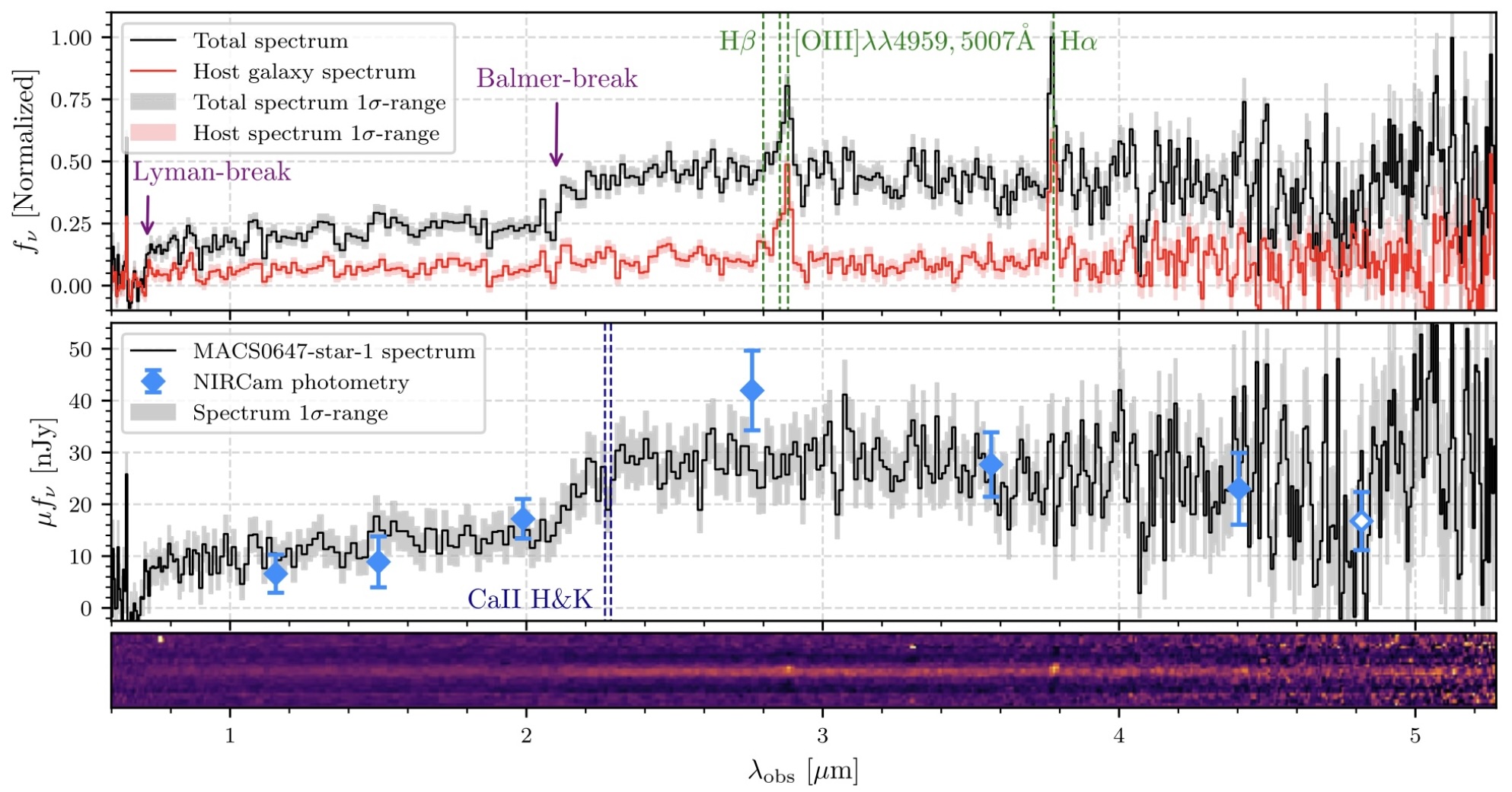} 
\caption{JWST can obtain spectra of stars at large cosmological distances. Figure shows the spectrum taken for the $z\simeq4.8$ star marked in green in Fig. \ref{fig:starspectrum1}. The upper panel shows the total and the host's JWST/NIRSpec spectrum. The clean JWST/NIRSpec spectrum of the star is seen in the middle panel. The bottom panel shows the 2D spectrum of through the slit. The star's spectrum, while of relatively low resolution, constitutes a remarkable proof of concept of JWST capabilities: Only JWST coupled with lensing can perform a similar observation. The spectrum yielded a temperature of $\sim15,000$ K for the star candidate in agreement with the photometric estimates, but a deeper, higher-resolution spectrum is needed to distinguish it from, e.g. a binary star system or a star cluster. Figure from \cite{furtak24a}.}
\label{fig:starspectrum2}
\end{figure} 

While Earendel was originally detected with HST, early results from the first year of JWST operations already revealed a couple of other high redshift $z\sim5$ stars \cite{meena23c} (Fig. \ref{fig:starspectrum1}, upper panel). Remarkably, follow-up observations with NIRSpec captured the spectrum of one of the stars \cite{furtak24a} (see Fig. \ref{fig:starspectrum2} here). The spectrum fitted well the expectations from a $\sim15,000$ K star -- although a double star or even a star cluster could not be ruled out. The spectrum is of relatively low resolution and not very deep but nevertheless a remarkable proof-of-concept: JWST can observe spectroscopically very distant stars. This added to the observed spectrum of another very intriguing lensed star candidate, \emph{Godzilla}, at $z\sim2.37$, which was measured both from the ground and with JWST (e.g. \cite{Rivera-Thorsen2024A&A...690A.269R, Pascale2024ApJ...976..166PGodzilla}), and a later JWST spectrum of Earendel \citep{Pascale2025ApJ...988L..76PEarendel}.

\begin{figure}
\includegraphics[width=1\textwidth]{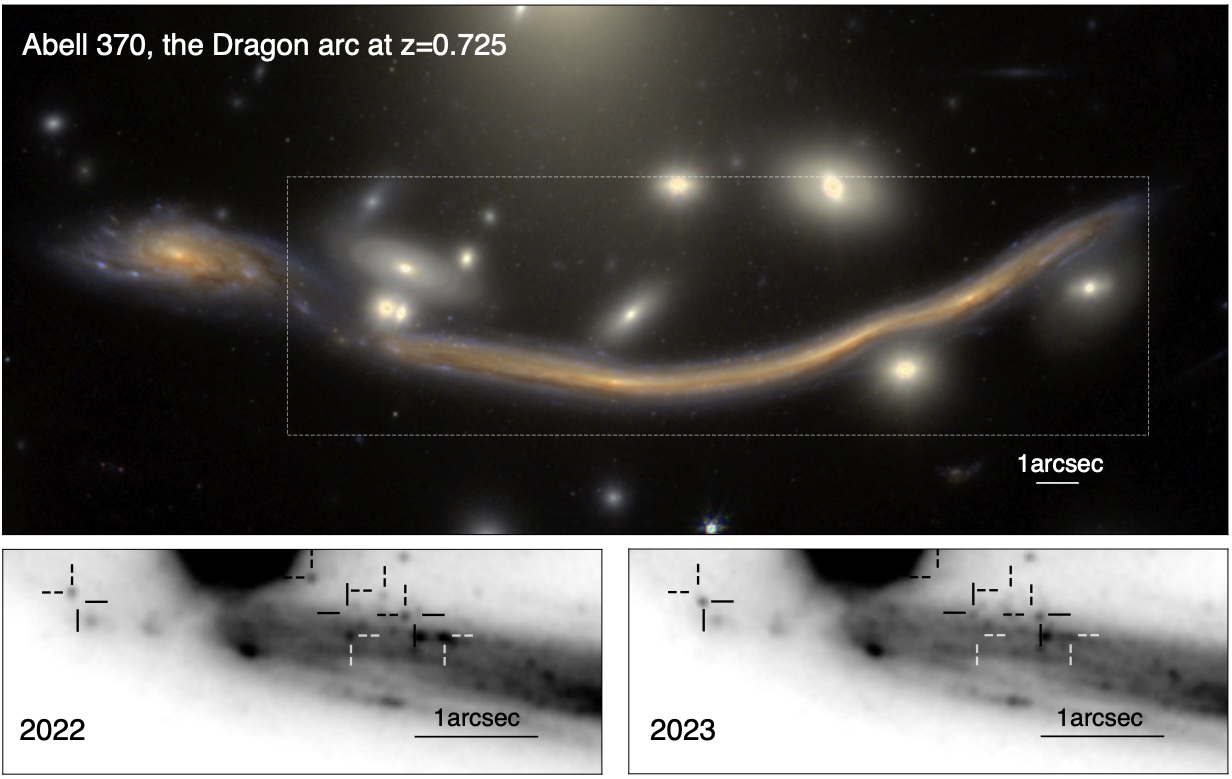} 
\caption{The first gravitational arc, seen in Abell 370 and now dubbed the Dragon, imaged with JWST. Imaging at two different epochs revealed over 40 transient point sources -- stars in the background spiral galaxy sitting near the caustics and getting temporarily, highly magnified. The bottom panels show a portion of the arc at the two different epochs, and marked in (dashed) half-crosses are the locations of sources (dis-)appearing between the two epochs. Figure from \cite{fudamoto25}.}
\label{fig:LensedStars}
\end{figure}

Another game changer for lensed stars from JWST came with the observation of the first gravitational arc in Abell 370 - which has since earned the name 'the Dragon arc' given it resembles a fire-spitting dragon. Observations of the cluster revealed more than 40 caustic-crossing events in one pair of repeat observations (Fig. \ref{fig:LensedStars}), superseding the total number of caustic-crossing events that had been detected to that date. Indeed, several programs continue with JWST to explore caustic crossing events and monitor the Dragon Arc as well, and the numbers of stars at cosmological distances should continue to be rapidly rising. 

The fact that individual stars can be now seen to cosmological distances may at first glance be hard to fathom, given the extreme magnification needed. However, when doing the numbers, it becomes clear that not only can we detect such stars, but that in fact we are bound to detect them, and in abundance; a prediction JWST now realises.

\subsection{Lensed supernovae and other exploding transients}\label{sn}
Strongly lensed supernovae have been a long sought-after observation. Already 60 years ago it was proposed that strongly lensed supernovae could, through time delays, constrain the expansion rate of the Universe \cite{Refsdal1964MNRAS}. While the same idea has been now applied to lensed quasars \cite{Suyu2017MNRAS.468.2590SHolicow,Wong2020MNRAS.498.1420Wtension}, for decades, no multiply imaged supernova was found. The first multiply imaged supernova that was clearly detected, lensed by a cluster, was found in 2014 with \emph{Hubble} behind the galaxy cluster MACS J1149.5+2223 \cite{Kelly2015Sci}. The supernova exploded in a spiral arm of a multiply imaged galaxy at a redshift of $z\simeq1.5$, further lensed by a local cluster galaxy into an Einstein cross. The supernova was named SN Refsdal, and spurred much excitement. Lens models predicted that another image of the supernova should appear about a year later in a different image of the spiral, a prediction that follow-up HST observations confirmed. SN Refsdal was then used to examine and improve lens models \cite{Rodney2016Refsdal,Treu2016Refsdal}, and to put constraints on the expansion rate of the universe \citep{Kelly2023Sci...380.1322K}.   

The fact that the lens models could predict where and when another image of the supernova would appear suggested that supernovae could thus be captured in their first moments of explosions. This is very important because the early light-curve evolution teaches us about the progenitor of the supernova and its environment \cite{Ofek2010ApJ...724.1396O,WaxmanKatz2017hsn..book..967W}. Only a couple other multiply lensed supernovae have been detected with \emph{Hubble} (e.g. \cite{Rodney2021NatAs...5.1118R}) and indeed, in at least one other instance, thanks to time delay, one of the images captured the supernova only hours after explosion, leading to constraints on the progenitor \cite{Chen2022Natur.611..256C}.

Lensed supernovae are also advantageous in other aspects. Lensing magnification allows us to see farther, and thus more distant supernovae could be observed through lensing. This should enable us to trace and understand stellar evolution through cosmic time, and potentially also observe rare explosions from massive, low-metallicity stars. Lensed supernovae of Type Ia could also then help in populating the Hubble diagram of brightness or distance versus redshift at the high redshift end, for stronger constraints on cosmology. Lensing of Type Ia supernovae, being standardisable candles, also supplies a precious test for the magnification estimates from different lens models \citep{Rodney2018Spock,Dhanasingham2026arXiv260411882D}. 

JWST's increased resolution and depth turned out to be a key factor for detecting more multiply imaged supernovae. In just three years of operation JWST has detected more lensed SNe by galaxy clusters than \emph{Hubble} \cite{Pierel2024ApJ...967L..37PEncore,Frye2024ApJ...961..171FSN_Hope,Larison2026AAS...24715606L}, suggesting that many more are to come with ongoing and future surveys. The increased depth coupled with lensing magnification essentially opens up significantly more volume in which lensed supernovae can be found per observation. While upcoming sky surveys from space such as Euclid and the Roman Space Telescope, or from the ground with large telescopes such as the Vera Rubin Observatory (LSST) are predicted to find many multiply imaged supernovae based on the much larger area probed and constant repeat visits, it was somewhat surprising -- at least for those who got used to the detection rates from \emph{Hubble} -- that JWST essentially turned into a transient detection machine: The current rate implies that just a few visits to a massive, strong-lensing galaxy cluster with JWST are sufficient to yield a multiply imaged supernova.

\begin{figure}[!ht]
\centering
\includegraphics[width=0.95\textwidth]{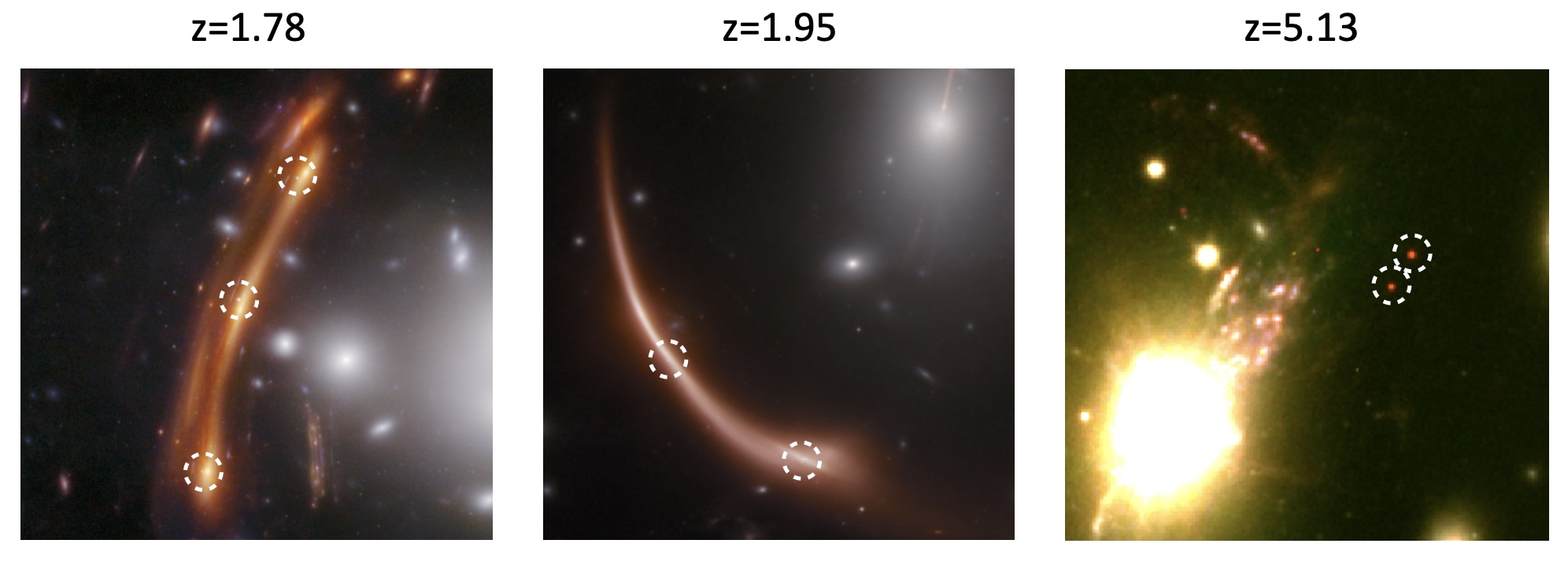}\vspace{0.6cm}
\includegraphics[width=0.96\textwidth,trim={0cm 1cm 1cm 1cm},clip]{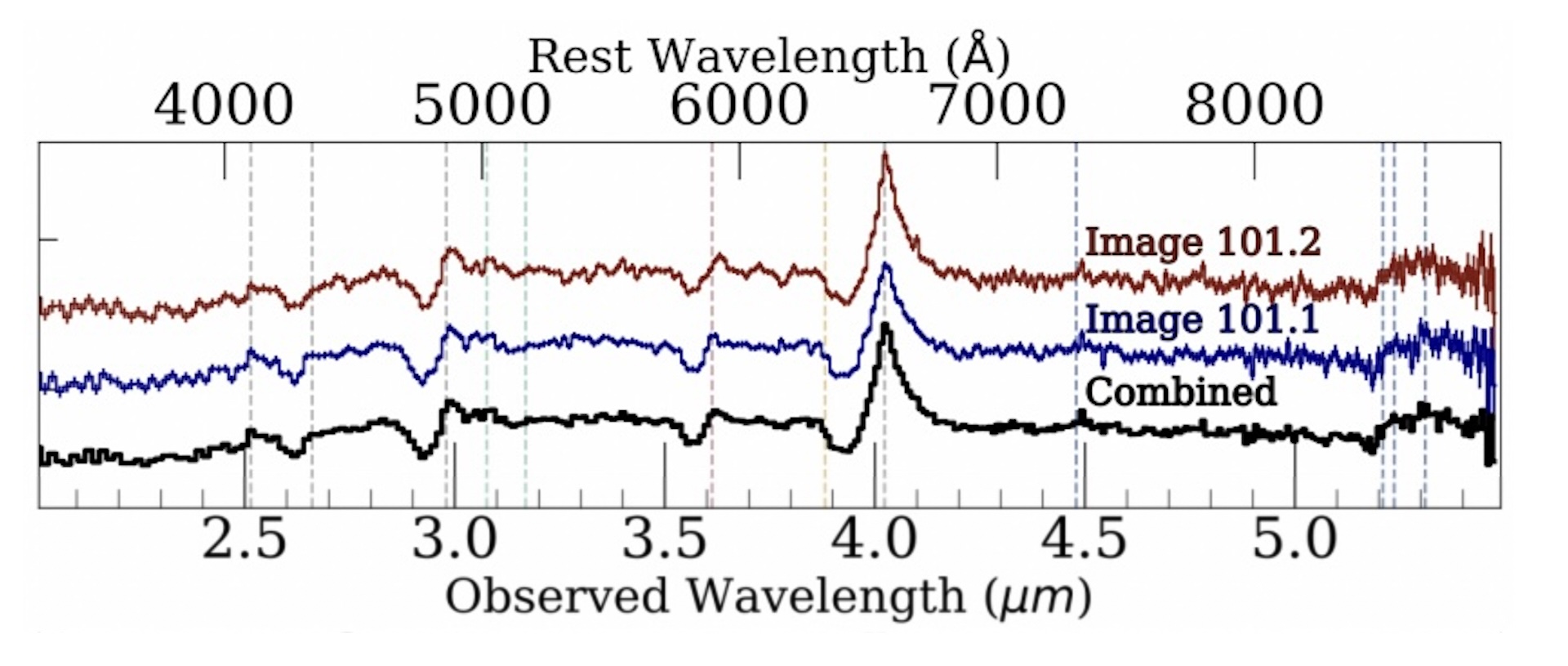} 
\caption{Multiply imaged supernovae detected with JWST. Upper row, Left: Supernova H0PE at $z=1.78$ \cite{Frye2024ApJ...961..171FSN_Hope}. Middle: Supernova Encore at $z=1.95$ \cite{Pierel2024ApJ...967L..37PEncore}. Right: Supernova Eos -- the farthest directly observed supernova to date, at $z=5.13$ \cite{Coulter2026arXiv260104156C_SNEos}. A spectrum of the supernova is seen in the lower panel, showing the relative flux density as a function of wavelength for the two supernova images, as well as their combined spectrum. The spectra were shifted vertically for better visibility. The spectra confirm that both images are counter-images of the same source -- a metal-poor Type II supernova (subfigure adapted from \cite{Coulter2026arXiv260104156C_SNEos}, further cut and edited here). JWST is the only instrument capable of measuring the optical spectrum of this supernova, which was originally detected with JWST thanks to lensing \cite{Allingham2026arXiv260214074A_Eos}.}
\label{fig:snEos}
\end{figure} 

JWST has recently observed the farthest, multiply imaged supernova at a redshift of $z=5.13$, SN Eos (\cite{Coulter2026arXiv260104156C_SNEos}; see Figure \ref{fig:snEos} here).
While most supernovae are detected through difference imaging showing the appearance or disappearance of the SN compared to previous data, SN Eos was detected thanks to lensing: Two red point sources were noted in recent imaging of the galaxy cluster MACS J1931.8-2635. The position of the two red point-sources suggested they were likely counter images of the same background source, but the lack of counter images where the lens model predicted them suggested it was a transient source, a redshift $z\sim5$ supernova \cite{Allingham2026arXiv260214074A_Eos}. Indeed, photometric and spectroscopic observations, and the absence of the SN images in previous \emph{Hubble} data, confirmed this prediction \cite{Coulter2026arXiv260104156C_SNEos}, demonstrating the power of JWST in finding and studying distant and lensed supernovae. 

It is worth mentioning that supernovae are not the only type of explosive transient sources that can be multiply imaged. The recent decade has seen increased interest in lensing of other transient sources such as Fast Radio Bursts, Gamma Ray Bursts, Tidal Disruption Events, or Gravitational Waves, and the range of science that can be done with them. While some of these cannot be directly observed with JWST, their progenitors or counterparts, could. It is thus safe to gamble that it is only a matter of time until more exciting supernovae or other transient results from JWST, will be delivered.

\subsection{Galaxy-Cluster Evolution}
Galaxy clusters are the most massive gravitationally bound structures in the Universe and are believed to form around $z\sim3$. Clusters contain hundreds to thousands of galaxies, and as a result of interactions within the dense cluster which remove the gas from the constituent galaxies and thus the fuel for forming new stars, most of these galaxies are red ellipticals, brighter redwards of $\sim$4000-5000 \AA\ rest-frame. JWST is thus an ideal observatory to investigate cluster evolution. The SLICE survey \cite{Mahler2024jwst.prop.5594M,Cerny2026ApJ..1001...60C}, for example, uses NIRCam to observe clusters out to nearly $z\sim2$, tracking 8 Gyrs of formation history of massive clusters, chosen from Sunyaev-Zel'dovich and X-ray catalogues. The cluster mass-redshift evolution can address many aspects of cluster evolution, from the formation of Brightest Cluster Galaxies, through infalling cluster galaxies and the build-up of stellar, dark matter and intra-cluster light content, to the large population of globular clusters now visible with JWST \cite{Diego2026arXiv260212332Dtight,Keatley2025ApJ...990...67Kglobulars,Hinrichs2026ApJ..1001...91HGlobulars}.

Most lensing studies to date, due to a combination of cluster lensing efficiency and observational capabilities, have concentrated on clusters at redshifts around $\sim0.2-0.5$, with some examples at somewhat higher redshifts towards $z\sim1$. In recent years, however, clusters have also been detected to cosmic noon ($z\sim2-3$), for example through their Sunyaev-Zel'dovich effect which does not experience redshift dimming, but their lensing signatures have been harder to capture. JWST now enables not only a deeper, longer-wavelength window into these clusters, but also into their magnified backgrounds, allowing us to derive reliable masses for them through strong (where apparent) and weak lensing. In fact, it turns out that JWST is an excellent instrument for weak-lensing studies \cite{Finner2023ApJ...953..102F,Harvey2024MNRAS.529..802H,Cha2024ApJ...961..186C}. While the field-of-view in JWST imaging is not very wide, its sensitivity and high spatial resolution reveal a high density of background sources and enable robust measurement of their ellipticities, and thus the shear signal.  

In addition to massive clusters at cosmic noon, early overdensities, which are believed to be the seeds of these later-time massive clusters of galaxies, are visible with JWST to high redshifts. Indeed, such proto-clusters have been observed with JWST \cite{Foo2025ApJ...995..219F,Bogdan2026Natur.649.1134B}, including spectroscopically, with some lensed galaxy overdensities seen out to cosmic dawn where the Universe was less than a Gyr old \citep{Morishita2023ApJ...947L..24M,Laporte2022A&A...667L...3L,Wu2025PASA...42..141WProto,Fudamoto2025arXiv250315597FHighzcluster}. The JWST, therefore, opens a precious window for studying the formation and evolution of clusters of galaxies through almost 14 Gyr of cosmic history!

\section{Conclusion}

One of the first images from JWST, released to the public in July 2022, was of a lensing cluster, SMACS J0723.3–7327, allowing a preview of JWST's capabilities. The depth and amount of apparent detail, compared to previous images of the same cluster field, were astonishing: Stellar clumps and a variety of point sources in the cluster and in-and-around background galaxies were seen in great abundances, for the first time; high-redshift galaxies were observed and spectroscopically measured beyond the limit of our capabilities from the ground; high-redshift overdensities were discovered; and a variety of new lensed features have surfaced.  

Over the next couple of years the exciting capabilities exhibited by the first JWST cluster image have continued to establish, deepen, and expand. Lensing campaigns in JWST's first few years of operations such as GLASS, UNCOVER, CANUCS, PEARLS, GLIMPSE, MAGNIF, SLICE, and VENUS, included both very deep observations of single clusters as well as shallower observations of growing samples of clusters, often with a component of follow-up spectroscopy or repeat observations for the study of potential transients. The number of clusters now covered with JWST in just a few cycles - by now many dozens of clusters have been observed -- does not fall shy of that covered with \emph{Hubble} in over 3 decades of operations. This is perhaps a testimony to the growing importance of lensing for studying dark matter, structure formation and cosmology and of course -- the distant universe.

Although a large number of clusters has been covered with JWST, the rate at which new lensing related studies and discoveries emerge remains surprising. High-redshift galaxies are being detected to higher redshifts or fainter magnitudes beyond those that have been reachable with \emph{Hubble}, often accompanied by UV and optical spectroscopy, and a detailed view of star-forming clumps, globular clusters, and galaxies such as spirals, can be seen to higher redshifts than ever before. Combined with lensing, JWST has opened new realms to study and discover -- in particular, of small, point-like sources. Lensed SMBH and AGN are easier to spot with JWST and are seen to higher redshifts and smaller masses; star clusters that we could not resolve before are now seen within galaxy clusters at $z<<1$, around galaxies in cosmic noon, and down to 1 pc scales at highly magnified arcs at redshifts beyond $\sim10-11$; redder and fainter sources previously elusive or completely unfamiliar have been detected, such as Little Red Dots; population III candidates have been found; and multiply imaged supernovae and extremely magnified lensed stars are being detected in greater numbers and towards higher redshifts than has been possible with \emph{Hubble}. 

The unique capabilities of JWST in terms of wavelength coverage, depth, and detail, also place it in excellent synergy with a range of instruments, such as the Atacama Large Millimeter/submillimeter Array (ALMA), Euclid mission, Vera Rubin observatory/LSST, the Nancy Grace Roman Space Telescope, and the upcoming 30-40m class of extremely large telescopes. For example, wide-field and survey instruments can scan the sky for intriguing objects and phenomena in the optical and infrared, with JWST then able to zoom-in on these objects and probe them in a unique range of wavelengths and to greater detail and depths. Further enhanced by the powers of gravitational lensing pushing JWST to see fainter and farther, we can anecdotally say: It is a bright future for faint sources.

\section*{Acknowledgement(s)}

This work uses observations made with the NASA/ESA/CSA James Webb Space Telescope. The data were obtained from the Mikulski Archive for Space Telescopes (MAST) at the Space Telescope Science Institute (STScI), which is operated by the Association of Universities for Research in Astronomy (AURA), Inc., under NASA contract NAS 5-03127 for JWST.
Some of stamp images presented herein were constructed using reduced data retrieved from the Dawn JWST Archive (DJA; see also \cite{Grizli} for the \texttt{grizli} data-reduction software). DJA is an initiative of the Cosmic Dawn Center (DAWN), which is funded by the Danish National Research Foundation under grant DNRF140. Other stamp images include reduced data from the UNCOVER, PEARLS, or VENUS repositories, to which the author has internal access. I would like to kindly thank the respective teams and acknowledge the hard work put into obtaining and reducing these data and making them available.
The author is also in debt to his research students, postdocs, and many collaborators, for many enriching discussions over the years. Finally, I would like to dedicate this review to Stella Seitz, Avishai Dekel, Mario Nonino, and Yannick Mellier; dear colleagues whom the lensing, high-redshift, and galaxy-evolution communities have lost over the past few years.

% \section*{Disclosure statement}

% An unnumbered section, e.g.\ \verb"\section*{Disclosure statement}", may be used to declare any potential conflict of interest and included \emph{in the non-anonymous version} before any Notes or References, after any Acknowledgements and before any Funding information.

\section*{Funding}
This work is supported by the Israel Science Foundation Grant No. 864/23.

\bibliography{interactnlmsample}

@ARTICLE{Rodney2018Spock,
author = {{Rodney}, S.~A. and {Balestra}, I. and {Bradac}, M. and {Brammer}, G. and
{Broadhurst}, T. and {Caminha}, G.~B. and {Chiriv{\i}}, G. and
{Diego}, J.~M. and {Filippenko}, A.~V. and {Foley}, R.~J. and
{Graur}, O. and {Grillo}, C. and {Hemmati}, S. and {Hjorth}, J. and
{Hoag}, A. and {Jauzac}, M. and {Jha}, S.~W. and {Kawamata}, R. and
{Kelly}, P.~L. and {McCully}, C. and {Mobasher}, B. and {Molino}, A. and
{Oguri}, M. and {Richard}, J. and {Riess}, A.~G. and {Rosati}, P. and
{Schmidt}, K.~B. and {Selsing}, J. and {Sharon}, K. and {Strolger}, L.-G. and
{Suyu}, S.~H. and {Treu}, T. and {Weiner}, B.~J. and {Williams}, L.~L.~R. and
{Zitrin}, A.},
title = "{Two peculiar fast transients in a strongly lensed host galaxy}",
journal = {Nature Astronomy},
archivePrefix = "arXiv",
eprint = {1707.02434},
year = 2018,
month = apr,
volume = 2,
pages = {324-333},
doi = {10.1038/s41550-018-0405-4},
adsurl = {http://adsabs.harvard.edu/abs/2018NatAs...2..324R}
}

@ARTICLE{Welch2022EarendelHST,
       author = {{Welch}, Brian and {Coe}, Dan and {Diego}, Jose M. and {Zitrin}, Adi and {Zackrisson}, Erik and {Dimauro}, Paola and {Jim{\'e}nez-Teja}, Yolanda and {Kelly}, Patrick and {Mahler}, Guillaume and {Oguri}, Masamune and {Timmes}, F.~X. and {Windhorst}, Rogier and {Florian}, Michael and {de Mink}, S.~E. and {Avila}, Roberto J. and {Anderson}, Jay and {Bradley}, Larry and {Sharon}, Keren and {Vikaeus}, Anton and {McCandliss}, Stephan and {Brada{\v{c}}}, Maru{\v{s}}a and {Rigby}, Jane and {Frye}, Brenda and {Toft}, Sune and {Strait}, Victoria and {Trenti}, Michele and {Sharma}, Soniya and {Andrade-Santos}, Felipe and {Broadhurst}, Tom},
        title = "{A highly magnified star at redshift 6.2}",
      journal = {\nat},
         year = 2022,
        month = mar,
       volume = {603},
       number = {7903},
        pages = {815-818},
          doi = {10.1038/s41586-022-04449-y},
archivePrefix = {arXiv},
       eprint = {2209.14866},
 primaryClass = {astro-ph.GA},
       adsurl = {https://ui.adsabs.harvard.edu/abs/2022Natur.603..815W}
}

@ARTICLE{Bouwens2015LF,
   author = {{Bouwens}, R.~J. and {Illingworth}, G.~D. and {Oesch}, P.~A. and 
	{Trenti}, M. and {Labb{\'e}}, I. and {Bradley}, L. and {Carollo}, M. and 
	{van Dokkum}, P.~G. and {Gonzalez}, V. and {Holwerda}, B. and 
	{Franx}, M. and {Spitler}, L. and {Smit}, R. and {Magee}, D.
	},
    title = "{UV Luminosity Functions at Redshifts z $\sim$ 4 to z $\sim$ 10: 10,000 Galaxies from HST Legacy Fields}",
  journal = {\apj},
archivePrefix = "arXiv",
   eprint = {1403.4295},
     year = 2015,
    month = apr,
   volume = 803,
      eid = {34},
    pages = {34},
      doi = {10.1088/0004-637X/803/1/34},
   adsurl = {http://adsabs.harvard.edu/abs/2015ApJ...803...34B}
}

@ARTICLE{Finkelstein2015,
   author = {{Finkelstein}, S.~L. and {Ryan}, Jr., R.~E. and {Papovich}, C. and 
	{Dickinson}, M. and {Song}, M. and {Somerville}, R.~S. and {Ferguson}, H.~C. and 
	{Salmon}, B. and {Giavalisco}, M. and {Koekemoer}, A.~M. and 
	{Ashby}, M.~L.~N. and {Behroozi}, P. and {Castellano}, M. and 
	{Dunlop}, J.~S. and {Faber}, S.~M. and {Fazio}, G.~G. and {Fontana}, A. and 
	{Grogin}, N.~A. and {Hathi}, N. and {Jaacks}, J. and {Kocevski}, D.~D. and 
	{Livermore}, R. and {McLure}, R.~J. and {Merlin}, E. and {Mobasher}, B. and 
	{Newman}, J.~A. and {Rafelski}, M. and {Tilvi}, V. and {Willner}, S.~P.
	},
    title = "{The Evolution of the Galaxy Rest-frame Ultraviolet Luminosity Function over the First Two Billion Years}",
  journal = {\apj},
archivePrefix = "arXiv",
   eprint = {1410.5439},
     year = 2015,
    month = sep,
   volume = 810,
      eid = {71},
    pages = {71},
      doi = {10.1088/0004-637X/810/1/71},
   adsurl = {http://adsabs.harvard.edu/abs/2015ApJ...810...71F}
}

@article{Berry01011976,
author = {M.V. Berry},
title = {Waves and Thom's theorem},
journal = {Advances in Physics},
volume = {25},
number = {1},
pages = {1--26},
year = {1976},
publisher = {Taylor \& Francis},
doi = {10.1080/00018737600101342},


URL = { 
    
        https://doi.org/10.1080/00018737600101342
    
    

},
eprint = { 
    
        https://doi.org/10.1080/00018737600101342
    
    

}

}

@ARTICLE{Ishigaki2018HFF,
   author = {{Ishigaki}, M. and {Kawamata}, R. and {Ouchi}, M. and {Oguri}, M. and 
	{Shimasaku}, K. and {Ono}, Y.},
    title = "{Full-data Results of Hubble Frontier Fields: UV Luminosity Functions at z $\sim$ 6-10 and a Consistent Picture of Cosmic Reionization}",
  journal = {\apj},
archivePrefix = "arXiv",
   eprint = {1702.04867},
     year = 2018,
    month = feb,
   volume = 854,
      eid = {73},
    pages = {73},
      doi = {10.3847/1538-4357/aaa544},
   adsurl = {http://adsabs.harvard.edu/abs/2018ApJ...854...73I}
}

@ARTICLE{Atek2014A2744,
   author = {{Atek}, H. and {Richard}, J. and {Kneib}, J.-P. and {Clement}, B. and
	{Egami}, E. and {Ebeling}, H. and {Jauzac}, M. and {Jullo}, E. and
	{Laporte}, N. and {Limousin}, M. and {Natarajan}, P.},
    title = "{Probing the z {\gt} 6 Universe with the First Hubble Frontier Fields Cluster A2744}",
  journal = {\apj},
archivePrefix = "arXiv",
   eprint = {1311.7670},
 primaryClass = "astro-ph.CO",
     year = 2014,
    month = may,
   volume = 786,
      eid = {60},
    pages = {60},
      doi = {10.1088/0004-637X/786/1/60},
   adsurl = {http://adsabs.harvard.edu/abs/2014ApJ...786...60A}
}

@ARTICLE{Coe2012highz,
       author = {{Coe}, Dan and {Zitrin}, Adi and {Carrasco}, Mauricio and {Shu}, Xinwen and {Zheng}, Wei and {Postman}, Marc and {Bradley}, Larry and {Koekemoer}, Anton and {Bouwens}, Rychard and {Broadhurst}, Tom and {Monna}, Anna and {Host}, Ole and {Moustakas}, Leonidas A. and {Ford}, Holland and {Moustakas}, John and {van der Wel}, Arjen and {Donahue}, Megan and {Rodney}, Steven A. and {Ben{\'\i}tez}, Narciso and {Jouvel}, Stephanie and {Seitz}, Stella and {Kelson}, Daniel D. and {Rosati}, Piero},
        title = "{CLASH: Three Strongly Lensed Images of a Candidate z {\ensuremath{\approx}} 11 Galaxy}",
      journal = {\apj},
         year = 2013,
        month = jan,
       volume = {762},
       number = {1},
          eid = {32},
        pages = {32},
          doi = {10.1088/0004-637X/762/1/32},
archivePrefix = {arXiv},
       eprint = {1211.3663},
 primaryClass = {astro-ph.CO},
       adsurl = {https://ui.adsabs.harvard.edu/abs/2013ApJ...762...32C}
}

@ARTICLE{Monna2014RXC2248,
   author = {{Monna}, A. and {Seitz}, S. and {Greisel}, N. and {Eichner}, T. and
	{Drory}, N. and {Postman}, M. and {Zitrin}, A. and {Coe}, D. and
	{Halkola}, A. and {Suyu}, S.~H. and {Grillo}, C. and {Rosati}, P. and
	{Lemze}, D. and {Balestra}, I. and {Snigula}, J. and {Bradley}, L. and
	{Umetsu}, K. and {Koekemoer}, A. and {Kuchner}, U. and {Moustakas}, L. and
	{Bartelmann}, M. and {Ben{\'{\i}}tez}, N. and {Bouwens}, R. and
	{Broadhurst}, T. and {Donahue}, M. and {Ford}, H. and {Host}, O. and
	{Infante}, L. and {Jimenez-Teja}, Y. and {Jouvel}, S. and {Kelson}, D. and
	{Lahav}, O. and {Medezinski}, E. and {Melchior}, P. and {Meneghetti}, M. and
	{Merten}, J. and {Molino}, A. and {Moustakas}, J. and {Nonino}, M. and
	{Zheng}, W.},
    title = "{CLASH: z $\sim$ 6 young galaxy candidate quintuply lensed by the frontier field cluster RXC J2248.7-4431}",
  journal = {\mnras},
archivePrefix = "arXiv",
   eprint = {1308.6280},
 primaryClass = "astro-ph.CO",
     year = 2014,
    month = feb,
   volume = 438,
    pages = {1417-1434},
      doi = {10.1093/mnras/stt2284},
   adsurl = {http://adsabs.harvard.edu/abs/2014MNRAS.438.1417M}
}

@ARTICLE{Zitrin2011MS1358,
   author = {{Zitrin}, A. and {Broadhurst}, T. and {Coe}, D. and {Liesenborgs}, J. and
	{Ben{\'{\i}}tez}, N. and {Rephaeli}, Y. and {Ford}, H. and {Umetsu}, K.
	},
    title = "{Strong-lensing analysis of MS 1358.4+6245: New multiple images and implications for the well-resolved z= 4.92 galaxy}",
  journal = {\mnras},
archivePrefix = "arXiv",
   eprint = {1009.3936},
 primaryClass = "astro-ph.CO",
     year = 2011,
    month = may,
   volume = 413,
    pages = {1753-1763},
      doi = {10.1111/j.1365-2966.2011.18252.x},
   adsurl = {http://adsabs.harvard.edu/abs/2011MNRAS.413.1753Z}
}

@ARTICLE{Zitrin2009_macs0717,
   author = {{Zitrin}, A. and {Broadhurst}, T. and {Rephaeli}, Y. and {Sadeh}, S.
	},
    title = "{The Largest Gravitational Lens: MACS J0717.5+3745 (z = 0.546)}",
  journal = {\apjl},
archivePrefix = "arXiv",
   eprint = {0907.4232},
 primaryClass = "astro-ph.CO",
     year = 2009,
    month = dec,
   volume = 707,
    pages = {L102-L106},
      doi = {10.1088/0004-637X/707/1/L102},
   adsurl = {http://adsabs.harvard.edu/abs/2009ApJ...707L.102Z}
}

@ARTICLE{Zitrin2009_cl0024,
   author = {{Zitrin}, A. and {Broadhurst}, T. and {Umetsu}, K. and {Coe}, D. and
	{Ben{\'{\i}}tez}, N. and {Ascaso}, B. and {Bradley}, L. and
	{Ford}, H. and {Jee}, J. and {Medezinski}, E. and {Rephaeli}, Y. and
	{Zheng}, W.},
    title = "{New multiply-lensed galaxies identified in ACS/NIC3 observations of Cl0024+1654 using an improved mass model}",
  journal = {\mnras},
archivePrefix = "arXiv",
   eprint = {0902.3971},
 primaryClass = "astro-ph.CO",
     year = 2009,
    month = jul,
   volume = 396,
    pages = {1985-2002},
      doi = {10.1111/j.1365-2966.2009.14899.x},
   adsurl = {http://adsabs.harvard.edu/abs/2009MNRAS.396.1985Z}
}

@ARTICLE{Kelly2018NatAsCCE,
   author = {{Kelly}, P.~L. and {Diego}, J.~M. and {Rodney}, S. and {Kaiser}, N. and 
	{Broadhurst}, T. and {Zitrin}, A. and {Treu}, T. and {P{\'e}rez-Gonz{\'a}lez}, P.~G. and 
	{Morishita}, T. and {Jauzac}, M. and {Selsing}, J. and {Oguri}, M. and 
	{Pueyo}, L. and {Ross}, T.~W. and {Filippenko}, A.~V. and {Smith}, N. and 
	{Hjorth}, J. and {Cenko}, S.~B. and {Wang}, X. and {Howell}, D.~A. and 
	{Richard}, J. and {Frye}, B.~L. and {Jha}, S.~W. and {Foley}, R.~J. and 
	{Norman}, C. and {Bradac}, M. and {Zheng}, W. and {Brammer}, G. and 
	{Benito}, A.~M. and {Cava}, A. and {Christensen}, L. and {de Mink}, S.~E. and 
	{Graur}, O. and {Grillo}, C. and {Kawamata}, R. and {Kneib}, J.-P. and 
	{Matheson}, T. and {McCully}, C. and {Nonino}, M. and {P{\'e}rez-Fournon}, I. and 
	{Riess}, A.~G. and {Rosati}, P. and {Schmidt}, K.~B. and {Sharon}, K. and 
	{Weiner}, B.~J.},
    title = "{Extreme magnification of an individual star at redshift 1.5 by a galaxy-cluster lens}",
  journal = {Nature Astronomy},
archivePrefix = "arXiv",
   eprint = {1706.10279},
     year = 2018,
    month = apr,
   volume = 2,
    pages = {334-342},
      doi = {10.1038/s41550-018-0430-3},
   adsurl = {http://adsabs.harvard.edu/abs/2018NatAs...2..334K}
}

@ARTICLE{Kelly2015Sci,
   author = {{Kelly}, P.~L. and {Rodney}, S.~A. and {Treu}, T. and {Foley}, R.~J. and 
	{Brammer}, G. and {Schmidt}, K.~B. and {Zitrin}, A. and {Sonnenfeld}, A. and 
	{Strolger}, L.-G. and {Graur}, O. and {Filippenko}, A.~V. and 
	{Jha}, S.~W. and {Riess}, A.~G. and {Bradac}, M. and {Weiner}, B.~J. and 
	{Scolnic}, D. and {Malkan}, M.~A. and {von der Linden}, A. and 
	{Trenti}, M. and {Hjorth}, J. and {Gavazzi}, R. and {Fontana}, A. and 
	{Merten}, J.~C. and {McCully}, C. and {Jones}, T. and {Postman}, M. and 
	{Dressler}, A. and {Patel}, B. and {Cenko}, S.~B. and {Graham}, M.~L. and 
	{Tucker}, B.~E.},
    title = "{Multiple images of a highly magnified supernova formed by an early-type cluster galaxy lens}",
  journal = {Science},
archivePrefix = "arXiv",
   eprint = {1411.6009},
     year = 2015,
    month = mar,
   volume = 347,
    pages = {1123-1126},
      doi = {10.1126/science.aaa3350},
   adsurl = {http://adsabs.harvard.edu/abs/2015Sci...347.1123K}
}

@ARTICLE{Chen2019CCE,
       author = {{Chen}, Wenlei and {Kelly}, Patrick L. and {Diego}, Jose M. and
         {Oguri}, Masamune and {Williams}, Liliya L.~R. and {Zitrin}, Adi and
         {Treu}, Tommaso L. and {Smith}, Nathan and {Broadhurst}, Thomas J. and
         {Kaiser}, Nick and {Foley}, Ryan J. and {Filippenko}, Alexei V. and
         {Salo}, Laura and {Hjorth}, Jens and {Selsing}, Jonatan},
        title = "{Searching for Highly Magnified Stars at Cosmological Distances: Discovery of a Redshift 0.94 Blue Supergiant in Archival Images of the Galaxy Cluster MACS J0416.1-2403}",
      journal = {\apj},
         year = 2019,
        month = aug,
       volume = {881},
       number = {1},
          eid = {8},
        pages = {8},
          doi = {10.3847/1538-4357/ab297d},
archivePrefix = {arXiv},
       eprint = {1902.05510},
 primaryClass = {astro-ph.GA},
       adsurl = {https://ui.adsabs.harvard.edu/abs/2019ApJ...881....8C}
}

@ARTICLE{Kaurov2019CCE,
       author = {{Kaurov}, Alexander A. and {Dai}, Liang and {Venumadhav}, Tejaswi and
         {Miralda-Escud{\'e}}, Jordi and {Frye}, Brenda},
        title = "{Highly Magnified Stars in Lensing Clusters: New Evidence in a Galaxy Lensed by MACS J0416.1-2403}",
      journal = {\apj},
         year = 2019,
        month = jul,
       volume = {880},
       number = {1},
          eid = {58},
        pages = {58},
          doi = {10.3847/1538-4357/ab2888},
archivePrefix = {arXiv},
       eprint = {1902.10090},
 primaryClass = {astro-ph.GA},
       adsurl = {https://ui.adsabs.harvard.edu/abs/2019ApJ...880...58K}
}

@ARTICLE{Oguri2018PhRvDCCE,
   author = {{Oguri}, M. and {Diego}, J.~M. and {Kaiser}, N. and {Kelly}, P.~L. and 
	{Broadhurst}, T.},
    title = "{Understanding caustic crossings in giant arcs: Characteristic scales, event rates, and constraints on compact dark matter}",
  journal = {\prd},
archivePrefix = "arXiv",
   eprint = {1710.00148},
     year = 2018,
    month = jan,
   volume = 97,
   number = 2,
      eid = {023518},
    pages = {023518},
      doi = {10.1103/PhysRevD.97.023518},
   adsurl = {http://adsabs.harvard.edu/abs/2018PhRvD..97b3518O}
}

@ARTICLE{Windhorst2018ApJSCCE,
   author = {{Windhorst}, R.~A. and {Timmes}, F.~X. and {Wyithe}, J.~S.~B. and 
	{Alpaslan}, M. and {Andrews}, S.~K. and {Coe}, D. and {Diego}, J.~M. and 
	{Dijkstra}, M. and {Driver}, S.~P. and {Kelly}, P.~L. and {Kim}, D.
	},
    title = "{On the Observability of Individual Population III Stars and Their Stellar-mass Black Hole Accretion Disks through Cluster Caustic Transits}",
  journal = {\apjs},
archivePrefix = "arXiv",
   eprint = {1801.03584},
     year = 2018,
    month = feb,
   volume = 234,
      eid = {41},
    pages = {41},
      doi = {10.3847/1538-4365/aaa760},
   adsurl = {http://adsabs.harvard.edu/abs/2018ApJS..234...41W}
}

@ARTICLE{Venumadhav2017ApJCCE,
   author = {{Venumadhav}, T. and {Dai}, L. and {Miralda-Escud{\'e}}, J.},
    title = "{Microlensing of Extremely Magnified Stars near Caustics of Galaxy Clusters}",
  journal = {\apj},
archivePrefix = "arXiv",
   eprint = {1707.00003},
     year = 2017,
    month = nov,
   volume = 850,
      eid = {49},
    pages = {49},
      doi = {10.3847/1538-4357/aa9575},
   adsurl = {http://adsabs.harvard.edu/abs/2017ApJ...850...49V}
}

@ARTICLE{Diego2018ApJCCE,
   author = {{Diego}, J.~M. and {Kaiser}, N. and {Broadhurst}, T. and {Kelly}, P.~L. and 
	{Rodney}, S. and {Morishita}, T. and {Oguri}, M. and {Ross}, T.~W. and 
	{Zitrin}, A. and {Jauzac}, M. and {Richard}, J. and {Williams}, L. and 
	{Vega-Ferrero}, J. and {Frye}, B. and {Filippenko}, A.~V.},
    title = "{Dark Matter under the Microscope: Constraining Compact Dark Matter with Caustic Crossing Events}",
  journal = {\apj},
archivePrefix = "arXiv",
   eprint = {1706.10281},
     year = 2018,
    month = apr,
   volume = 857,
      eid = {25},
    pages = {25},
      doi = {10.3847/1538-4357/aab617},
   adsurl = {http://adsabs.harvard.edu/abs/2018ApJ...857...25D}
}

@ARTICLE{Rodney2016Refsdal,
   author = {{Rodney}, S.~A. and {Strolger}, L.-G. and {Kelly}, P.~L. and 
	{Brada{\v c}}, M. and {Brammer}, G. and {Filippenko}, A.~V. and 
	{Foley}, R.~J. and {Graur}, O. and {Hjorth}, J. and {Jha}, S.~W. and 
	{McCully}, C. and {Molino}, A. and {Riess}, A.~G. and {Schmidt}, K.~B. and 
	{Selsing}, J. and {Sharon}, K. and {Treu}, T. and {Weiner}, B.~J. and 
	{Zitrin}, A.},
    title = "{SN Refsdal: Photometry and Time Delay Measurements of the First Einstein Cross Supernova}",
  journal = {\apj},
archivePrefix = "arXiv",
   eprint = {1512.05734},
     year = 2016,
    month = mar,
   volume = 820,
      eid = {50},
    pages = {50},
      doi = {10.3847/0004-637X/820/1/50},
   adsurl = {http://adsabs.harvard.edu/abs/2016ApJ...820...50R}
}

@ARTICLE{Treu2016Refsdal,
   author = {{Treu}, T. and {Brammer}, G. and {Diego}, J.~M. and {Grillo}, C. and 
	{Kelly}, P.~L. and {Oguri}, M. and {Rodney}, S.~A. and {Rosati}, P. and 
	{Sharon}, K. and {Zitrin}, A. and {Balestra}, I. and {Brada{\v c}}, M. and 
	{Broadhurst}, T. and {Caminha}, G.~B. and {Halkola}, A. and 
	{Hoag}, A. and {Ishigaki}, M. and {Johnson}, T.~L. and {Karman}, W. and 
	{Kawamata}, R. and {Mercurio}, A. and {Schmidt}, K.~B. and {Strolger}, L.-G. and 
	{Suyu}, S.~H. and {Filippenko}, A.~V. and {Foley}, R.~J. and 
	{Jha}, S.~W. and {Patel}, B.},
    title = "{''Refsdal'' Meets Popper: Comparing Predictions of the Re-appearance of the Multiply Imaged Supernova Behind MACSJ1149.5+2223}",
  journal = {\apj},
archivePrefix = "arXiv",
   eprint = {1510.05750},
     year = 2016,
    month = jan,
   volume = 817,
      eid = {60},
    pages = {60},
      doi = {10.3847/0004-637X/817/1/60},
   adsurl = {http://adsabs.harvard.edu/abs/2016ApJ...817...60T}
}

@ARTICLE{Zheng2012NaturZ,
   author = {{Zheng}, W. and {Postman}, M. and {Zitrin}, A. and {Moustakas}, J. and
	{Shu}, X. and {Jouvel}, S. and {H{\o}st}, O. and {Molino}, A. and
	{Bradley}, L. and {Coe}, D. and {Moustakas}, L.~A. and {Carrasco}, M. and
	{Ford}, H. and {Ben{\'{\i}}tez}, N. and {Lauer}, T.~R. and {Seitz}, S. and
	{Bouwens}, R. and {Koekemoer}, A. and {Medezinski}, E. and {Bartelmann}, M. and
	{Broadhurst}, T. and {Donahue}, M. and {Grillo}, C. and {Infante}, L. and
	{Jha}, S.~W. and {Kelson}, D.~D. and {Lahav}, O. and {Lemze}, D. and
	{Melchior}, P. and {Meneghetti}, M. and {Merten}, J. and {Nonino}, M. and
	{Ogaz}, S. and {Rosati}, P. and {Umetsu}, K. and {van der Wel}, A.
	},
    title = "{A magnified young galaxy from about 500 million years after the Big Bang}",
  journal = {\nat},
archivePrefix = "arXiv",
   eprint = {1204.2305},
 primaryClass = "astro-ph.CO",
     year = 2012,
    month = sep,
   volume = 489,
    pages = {406-408},
      doi = {10.1038/nature11446},
   adsurl = {http://adsabs.harvard.edu/abs/2012Natur.489..406Z}
}

@ARTICLE{Oguri2012SL,
       author = {{Oguri}, Masamune and {Bayliss}, Matthew B. and {Dahle}, H{\^a}kon and {Sharon}, Keren and {Gladders}, Michael D. and {Natarajan}, Priyamvada and {Hennawi}, Joseph F. and {Koester}, Benjamin P.},
        title = "{Combined strong and weak lensing analysis of 28 clusters from the Sloan Giant Arcs Survey}",
      journal = {\mnras},
         year = 2012,
        month = mar,
       volume = {420},
       number = {4},
        pages = {3213-3239},
          doi = {10.1111/j.1365-2966.2011.20248.x},
archivePrefix = {arXiv},
       eprint = {1109.2594},
 primaryClass = {astro-ph.CO},
       adsurl = {https://ui.adsabs.harvard.edu/abs/2012MNRAS.420.3213O}
}

@ARTICLE{Oesch2016z11,
   author = {{Oesch}, P.~A. and {Brammer}, G. and {van Dokkum}, P.~G. and 
	{Illingworth}, G.~D. and {Bouwens}, R.~J. and {Labb{\'e}}, I. and 
	{Franx}, M. and {Momcheva}, I. and {Ashby}, M.~L.~N. and {Fazio}, G.~G. and 
	{Gonzalez}, V. and {Holden}, B. and {Magee}, D. and {Skelton}, R.~E. and 
	{Smit}, R. and {Spitler}, L.~R. and {Trenti}, M. and {Willner}, S.~P.
	},
    title = "{A Remarkably Luminous Galaxy at z=11.1 Measured with Hubble Space Telescope Grism Spectroscopy}",
  journal = {\apj},
archivePrefix = "arXiv",
   eprint = {1603.00461},
     year = 2016,
    month = mar,
   volume = 819,
      eid = {129},
    pages = {129},
      doi = {10.3847/0004-637X/819/2/129},
   adsurl = {http://adsabs.harvard.edu/abs/2016ApJ...819..129O}
}

@ARTICLE{Umetsu2020WLreview,
       author = {{Umetsu}, Keiichi},
        title = "{Cluster-galaxy weak lensing}",
      journal = {\aapr},
         year = 2020,
        month = dec,
       volume = {28},
       number = {1},
          eid = {7},
        pages = {7},
          doi = {10.1007/s00159-020-00129-w},
archivePrefix = {arXiv},
       eprint = {2007.00506},
 primaryClass = {astro-ph.CO},
       adsurl = {https://ui.adsabs.harvard.edu/abs/2020A&ARv..28....7U}
}

@ARTICLE{Salmon2020RELICSHighz,
       author = {{Salmon}, Brett and {Coe}, Dan and {Bradley}, Larry and
         {Bouwens}, Rychard and {Brada{\v{c}}}, Marusa and {Huang}, Kuang-Han and
         {Oesch}, Pascal A. and {Stark}, Daniel and {Sharon}, Keren and
         {Trenti}, Michele and {Avila}, Roberto J. and {Ogaz}, Sara and
         {Andrade-Santos}, Felipe and {Carrasco}, Daniela and
         {Cerny}, Catherine and {Dawson}, William and {Frye}, Brenda L. and
         {Hoag}, Austin and {Johnson}, Traci Lin and {Jones}, Christine and
         {Lam}, Daniel and {Lovisari}, Lorenzo and {Mainali}, Ramesh and
         {Past}, Matt and {Paterno-Mahler}, Rachel and {Peterson}, Avery and
         {Riess}, Adam G. and {Rodney}, Steven A. and {Ryan}, Russel E. and
         {Sendra-Server}, Irene and {Strait}, Victoria and
         {Strolger}, Louis-Gregory and {Umetsu}, Keiichi and
         {Vulcani}, Benedetta and {Zitrin}, Adi},
        title = "{RELICS: The Reionization Lensing Cluster Survey and the Brightest High-z Galaxies}",
      journal = {\apj},
         year = 2020,
        month = feb,
       volume = {889},
       number = {2},
          eid = {189},
        pages = {189},
          doi = {10.3847/1538-4357/ab5a8b},
archivePrefix = {arXiv},
       eprint = {1710.08930},
 primaryClass = {astro-ph.GA},
       adsurl = {https://ui.adsabs.harvard.edu/abs/2020ApJ...889..189S}
}

@ARTICLE{Bouwens2011NaturZ10Gal,
   author = {{Bouwens}, R.~J. and {Illingworth}, G.~D. and {Labbe}, I. and
	{Oesch}, P.~A. and {Trenti}, M. and {Carollo}, C.~M. and {van Dokkum}, P.~G. and
	{Franx}, M. and {Stiavelli}, M. and {Gonz{\'a}lez}, V. and {Magee}, D. and
	{Bradley}, L.},
    title = "{A candidate redshift z\~{}10 galaxy and rapid changes in that population at an age of 500Myr}",
  journal = {\nat},
archivePrefix = "arXiv",
   eprint = {0912.4263},
 primaryClass = "astro-ph.CO",
     year = 2011,
    month = jan,
   volume = 469,
    pages = {504-507},
      doi = {10.1038/nature09717},
   adsurl = {http://adsabs.harvard.edu/abs/2011Natur.469..504B}
}

@ARTICLE{Bradley2008,
	author = {{Bradley}, L.~D. and {Bouwens}, R.~J. and {Ford}, H.~C. and
	{Illingworth}, G.~D. and {Jee}, M.~J. and {Ben{\'{\i}}tez}, N. and
	{Broadhurst}, T.~J. and {Franx}, M. and {Frye}, B.~L. and {Infante}, L. and
	{Motta}, V. and others},
   authorfull = {{Bradley}, L.~D. and {Bouwens}, R.~J. and {Ford}, H.~C. and
	{Illingworth}, G.~D. and {Jee}, M.~J. and {Ben{\'{\i}}tez}, N. and
	{Broadhurst}, T.~J. and {Franx}, M. and {Frye}, B.~L. and {Infante}, L. and
	{Motta}, V. and {Rosati}, P. and {White}, R.~L. and {Zheng}, W.
	},
    title = "{Discovery of a Very Bright Strongly Lensed Galaxy Candidate at z \~{} 7.6}",
  journal = {\apj},
archivePrefix = "arXiv",
   eprint = {0802.2506},
     year = 2008,
    month = may,
   volume = 678,
    pages = {647-654},
      doi = {10.1086/533519},
   adsurl = {http://adsabs.harvard.edu/abs/2008ApJ...678..647B}
}

@ARTICLE{Broadhurst2005a,
	author = {{Broadhurst}, T. and {Ben{\'{\i}}tez}, N. and {Coe}, D. and
	{Sharon}, K. and {Zekser}, K. and {White}, R. and {Ford}, H. and
	{Bouwens}, R. and {Blakeslee}, J. and {Clampin}, M. and others},
   authorfull = {{Broadhurst}, T. and {Ben{\'{\i}}tez}, N. and {Coe}, D. and
	{Sharon}, K. and {Zekser}, K. and {White}, R. and {Ford}, H. and
	{Bouwens}, R. and {Blakeslee}, J. and {Clampin}, M. and {Cross}, N. and
	{Franx}, M. and {Frye}, B. and {Hartig}, G. and {Illingworth}, G. and
	{Infante}, L. and {Menanteau}, F. and {Meurer}, G. and {Postman}, M. and
	{Ardila}, D.~R. and {Bartko}, F. and {Brown}, R.~A. and {Burrows}, C.~J. and
	{Cheng}, E.~S. and {Feldman}, P.~D. and {Golimowski}, D.~A. and
	{Goto}, T. and {Gronwall}, C. and {Herranz}, D. and {Holden}, B. and
	{Homeier}, N. and {Krist}, J.~E. and {Lesser}, M.~P. and {Martel}, A.~R. and
	{Miley}, G.~K. and {Rosati}, P. and {Sirianni}, M. and {Sparks}, W.~B. and
	{Steindling}, S. and {Tran}, H.~D. and {Tsvetanov}, Z.~I. and
	{Zheng}, W.},
    title = "{Strong-Lensing Analysis of A1689 from Deep Advanced Camera Images}",
  journal = {\apj},
   eprint = {arXiv:astro-ph/0409132},
     year = 2005,
    month = mar,
   volume = 621,
    pages = {53-88},
      doi = {10.1086/426494},
   adsurl = {http://adsabs.harvard.edu/abs/2005ApJ...621...53B}
}

@ARTICLE{Broadhurst2005b,
   author = {{Broadhurst}, T. and {Takada}, M. and {Umetsu}, K. and {Kong}, X. and
	{Arimoto}, N. and {Chiba}, M. and {Futamase}, T.},
    title = "{The Surprisingly Steep Mass Profile of A1689, from a Lensing Analysis of Subaru Images}",
  journal = {\apjl},
   eprint = {arXiv:astro-ph/0412192},
     year = 2005,
    month = feb,
   volume = 619,
    pages = {L143-L146},
      doi = {10.1086/428122},
   adsurl = {http://adsabs.harvard.edu/abs/2005ApJ...619L.143B}
}

@ARTICLE{EbelingMacsCat2001,
   author = {{Ebeling}, H. and {Edge}, A.~C. and {Henry}, J.~P.},
    title = "{MACS: A Quest for the Most Massive Galaxy Clusters in the Universe}",
  journal = {\apj},
   eprint = {arXiv:astro-ph/0009101},
     year = 2001,
    month = jun,
   volume = 553,
    pages = {668-676},
      doi = {10.1086/320958},
   adsurl = {http://adsabs.harvard.edu/abs/2001ApJ...553..668E}
}

@ARTICLE{Jullo2010,
   author = {{Jullo}, E. and {Natarajan}, P. and {Kneib}, {J.-P.} and {D'Aloisio}, A. and
	{Limousin}, M. and {Richard}, J. and {Schimd}, C.},
    title = "{Cosmological Constraints from Strong Gravitational Lensing in Clusters of Galaxies}",
  journal = {Science},
archivePrefix = "arXiv",
   eprint = {1008.4802},
 primaryClass = "astro-ph.CO",
     year = 2010,
    month = aug,
   volume = 329,
    pages = {924-927},
      doi = {10.1126/science.1185759},
   adsurl = {http://adsabs.harvard.edu/abs/2010Sci...329..924J}
}

@ARTICLE{Liesenborgs2006,
   author = {{Liesenborgs}, J. and {De Rijcke}, S. and {Dejonghe}, H.},
    title = "{A genetic algorithm for the non-parametric inversion of strong lensing systems}",
  journal = {\mnras},
   eprint = {arXiv:astro-ph/0601124},
     year = 2006,
    month = apr,
   volume = 367,
    pages = {1209-1216},
      doi = {10.1111/j.1365-2966.2006.10040.x},
   adsurl = {http://adsabs.harvard.edu/abs/2006MNRAS.367.1209L}
}

@ARTICLE{Zitrin2014highz,
   author = {{Zitrin}, A. and {Zheng}, W. and {Broadhurst}, T. and {Moustakas}, J. and
	{Lam}, D. and {Shu}, X. and {Huang}, X. and {Diego}, J.~M. and
	{Ford}, H. and {Lim}, J. and {Bauer}, F.~E. and {Infante}, L. and
	{Kelson}, D.~D. and {Molino}, A.},
    title = "{A Geometrically Supported z \~{} 10 Candidate Multiply Imaged by the Hubble Frontier Fields Cluster A2744}",
  journal = {\apjl},
archivePrefix = "arXiv",
   eprint = {1407.3769},
     year = 2014,
    month = sep,
   volume = 793,
      eid = {L12},
    pages = {L12},
      doi = {10.1088/2041-8205/793/1/L12},
   adsurl = {http://adsabs.harvard.edu/abs/2014ApJ...793L..12Z}
}

@ARTICLE{PostmanCLASHoverview,
   author = {{Postman}, M. and {Coe}, D. and {Ben{\'{\i}}tez}, N. and {Bradley}, L. and
	{Broadhurst}, T. and {Donahue}, M. and {Ford}, H. and {Graur}, O. and
	{Graves}, G. and {Jouvel}, S. and {Koekemoer}, A. and {Lemze}, D. and
	{Medezinski}, E. and {Molino}, A. and {Moustakas}, L. and {Ogaz}, S. and
	{Riess}, A. and {Rodney}, S. and {Rosati}, P. and {Umetsu}, K. and
	{Zheng}, W. and {Zitrin}, A. and {Bartelmann}, M. and {Bouwens}, R. and
	{Czakon}, N. and {Golwala}, S. and {Host}, O. and {Infante}, L. and
	{Jha}, S. and {Jimenez-Teja}, Y. and {Kelson}, D. and {Lahav}, O. and
	{Lazkoz}, R. and {Maoz}, D. and {McCully}, C. and {Melchior}, P. and
	{Meneghetti}, M. and {Merten}, J. and {Moustakas}, J. and {Nonino}, M. and
	{Patel}, B. and {Reg{\"o}s}, E. and {Sayers}, J. and {Seitz}, S. and
	{Van der Wel}, A.},
    title = "{The Cluster Lensing and Supernova Survey with Hubble: An Overview}",
  journal = {\apjs},
archivePrefix = "arXiv",
   eprint = {1106.3328},
 primaryClass = "astro-ph.CO",
     year = 2012,
    month = apr,
   volume = 199,
      eid = {25},
    pages = {25},
      doi = {10.1088/0067-0049/199/2/25},
   adsurl = {http://adsabs.harvard.edu/abs/2012ApJS..199...25P}
}

@ARTICLE{Coe2019RELICS,
       author = {{Coe}, Dan and {Salmon}, Brett and {Brada{\v{c}}}, Maru{\v{s}}a and
         {Bradley}, Larry D. and {Sharon}, Keren and {Zitrin}, Adi and
         {Acebron}, Ana and {Cerny}, Catherine and {Cibirka}, Nath{\'a}lia and
         {Strait}, Victoria and {Paterno-Mahler}, Rachel and
         {Mahler}, Guillaume and {Avila}, Roberto J. and {Ogaz}, Sara and
         {Huang}, Kuang-Han and {Pelliccia}, Debora and {Stark}, Daniel P. and
         {Mainali}, Ramesh and {Oesch}, Pascal A. and {Trenti}, Michele and
         {Carrasco}, Daniela and {Dawson}, William A. and {Rodney}, Steven A. and
         {Strolger}, Louis-Gregory and {Riess}, Adam G. and {Jones}, Christine and
         {Frye}, Brenda L. and {Czakon}, Nicole G. and {Umetsu}, Keiichi and
         {Vulcani}, Benedetta and {Graur}, Or and {Jha}, Saurabh W. and
         {Graham}, Melissa L. and {Molino}, Alberto and {Nonino}, Mario and
         {Hjorth}, Jens and {Selsing}, Jonatan and {Christensen}, Lise and
         {Kikuchihara}, Shotaro and {Ouchi}, Masami and {Oguri}, Masamune and
         {Welch}, Brian and {Lemaux}, Brian C. and {Andrade-Santos}, Felipe and
         {Hoag}, Austin T. and {Johnson}, Traci L. and {Peterson}, Avery and
         {Past}, Matthew and {Fox}, Carter and {Agulli}, Irene and
         {Livermore}, Rachael and {Ryan}, Russell E. and {Lam}, Daniel and
         {Sendra-Server}, Irene and {Toft}, Sune and {Lovisari}, Lorenzo and
         {Su}, Yuanyuan},
        title = "{RELICS: Reionization Lensing Cluster Survey}",
      journal = {\apj},
         year = 2019,
        month = oct,
       volume = {884},
       number = {1},
          eid = {85},
        pages = {85},
          doi = {10.3847/1538-4357/ab412b},
archivePrefix = {arXiv},
       eprint = {1903.02002},
 primaryClass = {astro-ph.GA},
       adsurl = {https://ui.adsabs.harvard.edu/abs/2019ApJ...884...85C}
}

@ARTICLE{SuyuEncore2026A&A...708A.291S,
       author = {{Suyu}, S.~H. and {Acebron}, A. and {Grillo}, C. and {Bergamini}, P. and {Caminha}, G.~B. and {Cha}, S. and {Diego}, J.~M. and {Ertl}, S. and {Foo}, N. and {Frye}, B.~L. and {Fudamoto}, Y. and {Granata}, G. and {Halkola}, A. and {Jee}, M.~J. and {Kamieneski}, P.~S. and {Koekemoer}, A.~M. and {Meena}, A.~K. and {Newman}, A.~B. and {Nishida}, S. and {Oguri}, M. and {Rosati}, P. and {Schuldt}, S. and {Zitrin}, A. and {Ca{\~n}ameras}, R. and {Hayes}, E.~E. and {Larison}, C. and {Mamuzic}, E. and {Millon}, M. and {Pierel}, J.~D.~R. and {Tortorelli}, L. and {Wang}, H.},
        title = "{Cosmology with supernova Encore in the strong lensing cluster MACS J0138-2155: Lens model comparison and H$_{0}$ measurement}",
      journal = {\aap},
         year = 2026,
        month = apr,
       volume = {708},
          eid = {A291},
        pages = {A291},
          doi = {10.1051/0004-6361/202557235},
archivePrefix = {arXiv},
       eprint = {2509.12319},
 primaryClass = {astro-ph.CO},
       adsurl = {https://ui.adsabs.harvard.edu/abs/2026A&A...708A.291S}
}

@ARTICLE{Richard2011,
       author = {{Richard}, Johan and {Kneib}, Jean-Paul and {Ebeling}, Harald and {Stark}, Daniel P. and {Egami}, Eiichi and {Fiedler}, Andrew K.},
        title = "{Discovery of a possibly old galaxy at z= 6.027, multiply imaged by the massive cluster Abell 383}",
      journal = {\mnras},
         year = 2011,
        month = jun,
       volume = {414},
       number = {1},
        pages = {L31-L35},
          doi = {10.1111/j.1745-3933.2011.01050.x},
archivePrefix = {arXiv},
       eprint = {1102.5092},
 primaryClass = {astro-ph.CO},
       adsurl = {https://ui.adsabs.harvard.edu/abs/2011MNRAS.414L..31R}
}

@ARTICLE{Bradac2006Bullet,
   author = {{Brada{\v c}}, M. and {Clowe}, D. and {Gonzalez}, A.~H. and
	{Marshall}, P. and {Forman}, W. and {Jones}, C. and {Markevitch}, M. and
	{Randall}, S. and {Schrabback}, T. and {Zaritsky}, D.},
    title = "{Strong and Weak Lensing United. III. Measuring the Mass Distribution of the Merging Galaxy Cluster 1ES 0657-558}",
  journal = {\apj},
   eprint = {arXiv:astro-ph/0608408},
     year = 2006,
    month = dec,
   volume = 652,
    pages = {937-947},
      doi = {10.1086/508601},
   adsurl = {http://adsabs.harvard.edu/abs/2006ApJ...652..937B}
}

@ARTICLE{Freedman2021ApJ...919...16Ftension,
       author = {{Freedman}, Wendy L.},
        title = "{Measurements of the Hubble Constant: Tensions in Perspective}",
      journal = {\apj},
         year = 2021,
        month = sep,
       volume = {919},
       number = {1},
          eid = {16},
        pages = {16},
          doi = {10.3847/1538-4357/ac0e95},
archivePrefix = {arXiv},
       eprint = {2106.15656},
 primaryClass = {astro-ph.CO},
       adsurl = {https://ui.adsabs.harvard.edu/abs/2021ApJ...919...16F}
}

@ARTICLE{Riess2021ApJ...908L...6Rtension,
       author = {{Riess}, Adam G. and {Casertano}, Stefano and {Yuan}, Wenlong and {Bowers}, J. Bradley and {Macri}, Lucas and {Zinn}, Joel C. and {Scolnic}, Dan},
        title = "{Cosmic Distances Calibrated to 1\% Precision with Gaia EDR3 Parallaxes and Hubble Space Telescope Photometry of 75 Milky Way Cepheids Confirm Tension with {\ensuremath{\Lambda}}CDM}",
      journal = {\apjl},
         year = 2021,
        month = feb,
       volume = {908},
       number = {1},
          eid = {L6},
        pages = {L6},
          doi = {10.3847/2041-8213/abdbaf},
archivePrefix = {arXiv},
       eprint = {2012.08534},
 primaryClass = {astro-ph.CO},
       adsurl = {https://ui.adsabs.harvard.edu/abs/2021ApJ...908L...6R}
}

@ARTICLE{Planck2020A&A...641A...6Pfinal,
       author = {{Planck Collaboration} and {Aghanim}, N. and {Akrami}, Y. and {Ashdown}, M. and {Aumont}, J. and {Baccigalupi}, C. and {Ballardini}, M. and {Banday}, A.~J. and {Barreiro}, R.~B. and {Bartolo}, N. and {Basak}, S. and {Battye}, R. and {Benabed}, K. and {Bernard}, J.-P. and {Bersanelli}, M. and {Bielewicz}, P. and {Bock}, J.~J. and {Bond}, J.~R. and {Borrill}, J. and {Bouchet}, F.~R. and {Boulanger}, F. and {Bucher}, M. and {Burigana}, C. and {Butler}, R.~C. and {Calabrese}, E. and {Cardoso}, J.-F. and {Carron}, J. and {Challinor}, A. and {Chiang}, H.~C. and {Chluba}, J. and {Colombo}, L.~P.~L. and {Combet}, C. and {Contreras}, D. and {Crill}, B.~P. and {Cuttaia}, F. and {de Bernardis}, P. and {de Zotti}, G. and {Delabrouille}, J. and {Delouis}, J.-M. and {Di Valentino}, E. and {Diego}, J.~M. and {Dor{\'e}}, O. and {Douspis}, M. and {Ducout}, A. and {Dupac}, X. and {Dusini}, S. and {Efstathiou}, G. and {Elsner}, F. and {En{\ss}lin}, T.~A. and {Eriksen}, H.~K. and {Fantaye}, Y. and {Farhang}, M. and {Fergusson}, J. and {Fernandez-Cobos}, R. and {Finelli}, F. and {Forastieri}, F. and {Frailis}, M. and {Fraisse}, A.~A. and {Franceschi}, E. and {Frolov}, A. and {Galeotta}, S. and {Galli}, S. and {Ganga}, K. and {G{\'e}nova-Santos}, R.~T. and {Gerbino}, M. and {Ghosh}, T. and {Gonz{\'a}lez-Nuevo}, J. and {G{\'o}rski}, K.~M. and {Gratton}, S. and {Gruppuso}, A. and {Gudmundsson}, J.~E. and {Hamann}, J. and {Handley}, W. and {Hansen}, F.~K. and {Herranz}, D. and {Hildebrandt}, S.~R. and {Hivon}, E. and {Huang}, Z. and {Jaffe}, A.~H. and {Jones}, W.~C. and {Karakci}, A. and {Keih{\"a}nen}, E. and {Keskitalo}, R. and {Kiiveri}, K. and {Kim}, J. and {Kisner}, T.~S. and {Knox}, L. and {Krachmalnicoff}, N. and {Kunz}, M. and {Kurki-Suonio}, H. and {Lagache}, G. and {Lamarre}, J.-M. and {Lasenby}, A. and {Lattanzi}, M. and {Lawrence}, C.~R. and {Le Jeune}, M. and {Lemos}, P. and {Lesgourgues}, J. and {Levrier}, F. and {Lewis}, A. and {Liguori}, M. and {Lilje}, P.~B. and {Lilley}, M. and {Lindholm}, V. and {L{\'o}pez-Caniego}, M. and {Lubin}, P.~M. and {Ma}, Y.-Z. and {Mac{\'\i}as-P{\'e}rez}, J.~F. and {Maggio}, G. and {Maino}, D. and {Mandolesi}, N. and {Mangilli}, A. and {Marcos-Caballero}, A. and {Maris}, M. and {Martin}, P.~G. and {Martinelli}, M. and {Mart{\'\i}nez-Gonz{\'a}lez}, E. and {Matarrese}, S. and {Mauri}, N. and {McEwen}, J.~D. and {Meinhold}, P.~R. and {Melchiorri}, A. and {Mennella}, A. and {Migliaccio}, M. and {Millea}, M. and {Mitra}, S. and {Miville-Desch{\^e}nes}, M.-A. and {Molinari}, D. and {Montier}, L. and {Morgante}, G. and {Moss}, A. and {Natoli}, P. and {N{\o}rgaard-Nielsen}, H.~U. and {Pagano}, L. and {Paoletti}, D. and {Partridge}, B. and {Patanchon}, G. and {Peiris}, H.~V. and {Perrotta}, F. and {Pettorino}, V. and {Piacentini}, F. and {Polastri}, L. and {Polenta}, G. and {Puget}, J.-L. and {Rachen}, J.~P. and {Reinecke}, M. and {Remazeilles}, M. and {Renzi}, A. and {Rocha}, G. and {Rosset}, C. and {Roudier}, G. and {Rubi{\~n}o-Mart{\'\i}n}, J.~A. and {Ruiz-Granados}, B. and {Salvati}, L. and {Sandri}, M. and {Savelainen}, M. and {Scott}, D. and {Shellard}, E.~P.~S. and {Sirignano}, C. and {Sirri}, G. and {Spencer}, L.~D. and {Sunyaev}, R. and {Suur-Uski}, A.-S. and {Tauber}, J.~A. and {Tavagnacco}, D. and {Tenti}, M. and {Toffolatti}, L. and {Tomasi}, M. and {Trombetti}, T. and {Valenziano}, L. and {Valiviita}, J. and {Van Tent}, B. and {Vibert}, L. and {Vielva}, P. and {Villa}, F. and {Vittorio}, N. and {Wandelt}, B.~D. and {Wehus}, I.~K. and {White}, M. and {White}, S.~D.~M. and {Zacchei}, A. and {Zonca}, A.},
        title = "{Planck 2018 results. VI. Cosmological parameters}",
      journal = {\aap},
         year = 2020,
        month = sep,
       volume = {641},
          eid = {A6},
        pages = {A6},
          doi = {10.1051/0004-6361/201833910},
archivePrefix = {arXiv},
       eprint = {1807.06209},
 primaryClass = {astro-ph.CO},
       adsurl = {https://ui.adsabs.harvard.edu/abs/2020A&A...641A...6P}
}

@ARTICLE{Suyu2017MNRAS.468.2590SHolicow,
       author = {{Suyu}, S.~H. and {Bonvin}, V. and {Courbin}, F. and {Fassnacht}, C.~D. and {Rusu}, C.~E. and {Sluse}, D. and {Treu}, T. and {Wong}, K.~C. and {Auger}, M.~W. and {Ding}, X. and {Hilbert}, S. and {Marshall}, P.~J. and {Rumbaugh}, N. and {Sonnenfeld}, A. and {Tewes}, M. and {Tihhonova}, O. and {Agnello}, A. and {Blandford}, R.~D. and {Chen}, G.~C.-F. and {Collett}, T. and {Koopmans}, L.~V.~E. and {Liao}, K. and {Meylan}, G. and {Spiniello}, C.},
        title = "{H0LiCOW - I. H$_{0}$ Lenses in COSMOGRAIL's Wellspring: program overview}",
      journal = {\mnras},
         year = 2017,
        month = jul,
       volume = {468},
       number = {3},
        pages = {2590-2604},
          doi = {10.1093/mnras/stx483},
archivePrefix = {arXiv},
       eprint = {1607.00017},
 primaryClass = {astro-ph.CO},
       adsurl = {https://ui.adsabs.harvard.edu/abs/2017MNRAS.468.2590S}
}

@ARTICLE{Harikane2025Ap&SS.370...85HReview,
       author = {{Harikane}, Yuichi},
        title = "{Early galaxies and supermassive black holes discovered by the James webb space telescope}",
      journal = {\apss},
         year = 2025,
        month = aug,
       volume = {370},
       number = {8},
          eid = {85},
        pages = {85},
          doi = {10.1007/s10509-025-04467-y},
       adsurl = {https://ui.adsabs.harvard.edu/abs/2025Ap&SS.370...85H}
}

@ARTICLE{Pierel2024ApJ...967L..37PEncore,
       author = {{Pierel}, J.~D.~R. and {Newman}, A.~B. and {Dhawan}, S. and {Gu}, M. and {Joshi}, B.~A. and {Li}, T. and {Schuldt}, S. and {Strolger}, L.~G. and {Suyu}, S.~H. and {Caminha}, G.~B. and {Cohen}, S.~H. and {Diego}, J.~M. and {D{\'S}ilva}, J.~C.~J. and {Ertl}, S. and {Frye}, B.~L. and {Granata}, G. and {Grillo}, C. and {Koekemoer}, A.~M. and {Li}, J. and {Robotham}, A. and {Summers}, J. and {Treu}, T. and {Windhorst}, R.~A. and {Zitrin}, A. and {Agarwal}, S. and {Agrawal}, A. and {Arendse}, N. and {Belli}, S. and {Burns}, C. and {Ca{\~n}ameras}, R. and {Chakrabarti}, S. and {Chen}, W. and {Collett}, T.~E. and {Coulter}, D.~A. and {Ellis}, R.~S. and {Engesser}, M. and {Foo}, N. and {Fox}, O.~D. and {Gall}, C. and {Garuda}, N. and {Gezari}, S. and {Gomez}, S. and {Glazebrook}, K. and {Hjorth}, J. and {Huang}, X. and {Jha}, S.~W. and {Kamieneski}, P.~S. and {Kelly}, P. and {Larison}, C. and {Moustakas}, L.~A. and {Pascale}, M. and {P{\'e}rez-Fournon}, I. and {Petrushevska}, T. and {Poidevin}, F. and {Rest}, A. and {Shahbandeh}, M. and {Shajib}, A.~J. and {Siebert}, M. and {Storfer}, C. and {Talbot}, M. and {Wang}, Q. and {Wevers}, T. and {Zenati}, Y.},
        title = "{Lensed Type Ia Supernova ``Encore'' at z = 2: The First Instance of Two Multiply Imaged Supernovae in the Same Host Galaxy}",
      journal = {\apjl},
         year = 2024,
        month = jun,
       volume = {967},
       number = {2},
          eid = {L37},
        pages = {L37},
          doi = {10.3847/2041-8213/ad4648},
archivePrefix = {arXiv},
       eprint = {2404.02139},
 primaryClass = {astro-ph.CO},
       adsurl = {https://ui.adsabs.harvard.edu/abs/2024ApJ...967L..37P}
}

@ARTICLE{Allingham2026arXiv260214074A_Eos,
       author = {{Allingham}, Joseph F.~V. and {Zitrin}, Adi and {Kokorev}, Vasily and {Yanagisawa}, Hiroto and {Diego}, Jose M. and {Furtak}, Lukas J. and {Asada}, Yoshihisa and {Coe}, Dan and {Coulter}, David A. and {Fujimoto}, Seiji and {Larison}, Conor and {Oguri}, Masamune and {Pierel}, Justin D.~R. and {Sun}, Fengwu and {Bradac}, Marusa and {Dayal}, Pratika and {Lopes}, Paulo A.~A. and {Meena}, Ashish K. and {Pascale}, Massimo and {Akins}, Hollis B. and {Bauer}, Franz E. and {Bradley}, Larry D. and {Brammer}, Gabriel and {Chisholm}, John and {Desprez}, Guillaume and {Fei}, Qinyue and {Ferguson}, Henry C. and {Finkelstein}, Steven L. and {Frye}, Brenda and {Golubchik}, Miriam and {Inayoshi}, Kohei and {Koekemoer}, Anton M. and {Lucas}, Ray A. and {Magdis}, Georgios E. and {Martis}, Nicholas S. and {Pan}, Richard and {Richard}, Johan and {Ricotti}, Massimo and {Rihtarsic}, Gregor and {Robbins}, Luke and {Sheu}, William and {Welch}, Brian and {Willott}, Chris and {Windhorst}, Rogier A.},
        title = "{VENUS: Strong-lensing model of MACS J1931.8-2635 -- revealing the farthest multiply imaged supernova}",
      journal = {arXiv e-prints},
         year = 2026,
        month = feb,
          eid = {arXiv:2602.14074},
        pages = {arXiv:2602.14074},
          doi = {10.48550/arXiv.2602.14074},
archivePrefix = {arXiv},
       eprint = {2602.14074},
 primaryClass = {astro-ph.CO},
       adsurl = {https://ui.adsabs.harvard.edu/abs/2026arXiv260214074A}
}

@ARTICLE{Coulter2026arXiv260104156C_SNEos,
       author = {{Coulter}, David A. and {Larison}, Conor and {Pierel}, Justin D.~R. and {Fujimoto}, Seiji and {Kokorev}, Vasily and {Allingham}, Joseph F.~V. and {Moriya}, Takashi J. and {Siebert}, Matthew and {Asada}, Yoshihisa and {Bezanson}, Rachel and {Brada{\v{c}}}, Maru{\v{s}}a and {Brammer}, Gabriel and {Chisholm}, John and {Coe}, Dan and {Dayal}, Pratika and {Engesser}, Michael and {Finkelstein}, Steven L. and {Fox}, Ori D. and {Furtak}, Lukas J. and {Koekemoer}, Anton M. and {Moore}, Thomas and {Nakane}, Minami and {Ouchi}, Masami and {Pan}, Richard and {Quimby}, Robert and {Rest}, Armin and {Richard}, Johan and {Robbins}, Luke and {Strolger}, Louis-Gregory and {Sun}, Fengwu and {Treu}, Tommaso and {Yanagisawa}, Hiroto and {Abdurro'uf} and {Agrawal}, Aadya and {Amor{\'\i}n}, Ricardo and {Anderson}, Joseph P. and {Angulo}, Rodrigo and {Atek}, Hakim and {Bauer}, Franz E. and {Bradley}, Larry D. and {Bromm}, Volker and {Bronikowski}, Mateusz and {Conselice}, Christopher J. and {DeCoursey}, Christa and {DerKacy}, James M. and {Desprez}, Guillaume and {Dhawan}, Suhail and {Diego}, Jose M. and {Egami}, Eiichi and {Faisst}, Andreas and {Frye}, Brenda and {Gomez}, Sebastian and {Gonz{\'a}lez-Otero}, Mauro and {Griggio}, Massimo and {Harikane}, Yuichi and {Inayoshi}, Kohei and {Jha}, Saurabh W. and {Jim{\'e}nez-Teja}, Yolanda and {Kartaltepe}, Jeyhan S. and {Kelly}, Patrick L. and {Kwok}, Lindsey A. and {Lane}, Zachary G. and {Li}, Xiaolong and {Lobbe}, Ivo and {Lopes}, Paulo A.~A. and {Lucas}, Ray A. and {Magdis}, Georgios E. and {Martis}, Nicholas S. and {Matthee}, Jorryt and {Meena}, Ashish K. and {Naidu}, Rohan P. and {Noirot}, Ga{\"e}l and {Oguri}, Masamune and {Padilla Gonzalez}, Estefania and {Pascale}, Massimo and {Petrushevska}, Tanja and {Ricotti}, Massimo and {Schaerer}, Daniel and {Schuldt}, Stefan and {Shahbandeh}, Melissa and {Sheu}, William and {Shukawa}, Koji and {Tsujita}, Akiyoshi and {Vanzella}, Eros and {Wang}, Qinan and {Weaver}, John and {Williams}, Robert and {Windhorst}, Rogier and {Xu}, Yi and {Zenati}, Yossef and {Zitrin}, Adi},
        title = "{A spectroscopically confirmed, strongly lensed, metal-poor Type II supernova at z = 5.13}",
      journal = {arXiv e-prints},
         year = 2026,
        month = jan,
          eid = {arXiv:2601.04156},
        pages = {arXiv:2601.04156},
          doi = {10.48550/arXiv.2601.04156},
archivePrefix = {arXiv},
       eprint = {2601.04156},
 primaryClass = {astro-ph.HE},
       adsurl = {https://ui.adsabs.harvard.edu/abs/2026arXiv260104156C}
}

@ARTICLE{Frye2024ApJ...961..171FSN_Hope,
       author = {{Frye}, Brenda L. and {Pascale}, Massimo and {Pierel}, Justin and {Chen}, Wenlei and {Foo}, Nicholas and {Leimbach}, Reagen and {Garuda}, Nikhil and {Cohen}, Seth H. and {Kamieneski}, Patrick S. and {Windhorst}, Rogier A. and {Koekemoer}, Anton M. and {Kelly}, Pat and {Summers}, Jake and {Engesser}, Michael and {Liu}, Daizhong and {Furtak}, Lukas J. and {Polletta}, Maria del Carmen and {Harrington}, Kevin C. and {Willner}, S.~P. and {Diego}, Jose M. and {Jansen}, Rolf A. and {Coe}, Dan and {Conselice}, Christopher J. and {Dai}, Liang and {Dole}, Herv{\'e} and {D'Silva}, Jordan C.~J. and {Driver}, Simon P. and {Grogin}, Norman A. and {Marshall}, Madeline A. and {Meena}, Ashish K. and {Nonino}, Mario and {Ortiz}, Rafael and {Pirzkal}, Nor and {Robotham}, Aaron and {Ryan}, Russell E. and {Strolger}, Lou and {Tompkins}, Scott and {Willmer}, Christopher N.~A. and {Yan}, Haojing and {Yun}, Min S. and {Zitrin}, Adi},
        title = "{The JWST Discovery of the Triply Imaged Type Ia ``Supernova H0pe'' and Observations of the Galaxy Cluster PLCK G165.7+67.0}",
      journal = {\apj},
         year = 2024,
        month = feb,
       volume = {961},
       number = {2},
          eid = {171},
        pages = {171},
          doi = {10.3847/1538-4357/ad1034},
archivePrefix = {arXiv},
       eprint = {2309.07326},
 primaryClass = {astro-ph.GA},
       adsurl = {https://ui.adsabs.harvard.edu/abs/2024ApJ...961..171F}
}

@ARTICLE{Harikane2025ApJ...980..138H,
       author = {{Harikane}, Yuichi and {Inoue}, Akio K. and {Ellis}, Richard S. and {Ouchi}, Masami and {Nakazato}, Yurina and {Yoshida}, Naoki and {Ono}, Yoshiaki and {Sun}, Fengwu and {Sato}, Riku A. and {Ferrami}, Giovanni and {Fujimoto}, Seiji and {Kashikawa}, Nobunari and {McLeod}, Derek J. and {P{\'e}rez-Gonz{\'a}lez}, Pablo G. and {Sawicki}, Marcin and {Sugahara}, Yuma and {Xu}, Yi and {Yamanaka}, Satoshi and {Carnall}, Adam C. and {Cullen}, Fergus and {Dunlop}, James S. and {Egami}, Eiichi and {Grogin}, Norman and {Isobe}, Yuki and {Koekemoer}, Anton M. and {Laporte}, Nicolas and {Lee}, Chien-Hsiu and {Magee}, Dan and {Matsuo}, Hiroshi and {Matsuoka}, Yoshiki and {Mawatari}, Ken and {Nakajima}, Kimihiko and {Nakane}, Minami and {Tamura}, Yoichi and {Umeda}, Hiroya and {Yanagisawa}, Hiroto},
        title = "{JWST, ALMA, and Keck Spectroscopic Constraints on the UV Luminosity Functions at z {\ensuremath{\sim}} 7-14: Clumpiness and Compactness of the Brightest Galaxies in the Early Universe}",
      journal = {\apj},
         year = 2025,
        month = feb,
       volume = {980},
       number = {1},
          eid = {138},
        pages = {138},
          doi = {10.3847/1538-4357/ad9b2c},
archivePrefix = {arXiv},
       eprint = {2406.18352},
 primaryClass = {astro-ph.GA},
       adsurl = {https://ui.adsabs.harvard.edu/abs/2025ApJ...980..138H}
}

@ARTICLE{Chemerynska2026MNRAS.546f2267C,
       author = {{Chemerynska}, Iryna and {Atek}, Hakim and {Furtak}, Lukas J. and {Chisholm}, John and {Endsley}, Ryan and {Kokorev}, Vasily and {Rosdahl}, Joki and {Blaizot}, Jeremy and {Adamo}, Angela and {Bouwens}, Rychard and {Fujimoto}, Seiji and {Korber}, Damien and {Mason}, Charlotte and {McQuinn}, Kristen B.~W. and {Mu{\~n}oz}, Julian B. and {Natarajan}, Priyamvada and {Nelson}, Erica and {Oesch}, Pascal A. and {Pan}, Richard and {Richard}, Johan and {Saldana-Lopez}, Alberto and {Schaerer}, Daniel and {Volonteri}, Marta and {Zitrin}, Adi and {Berg}, Danielle A. and {Claeyssens}, Ad{\'e}la{\"\i}de and {Dessauges-Zavadsky}, Miroslava and {Jecmen}, Michelle and {Labb{\'e}}, Ivo and {Naidu}, Rohan and {Trebitsch}, Maxime},
        title = "{The first GLIMPSE of the faint galaxy population at Cosmic Dawn with JWST: The evolution of the ultraviolet luminosity function across z {\ensuremath{\sim}} 9{\ensuremath{-}}15}",
      journal = {\mnras},
         year = 2026,
        month = feb,
       volume = {546},
       number = {2},
          eid = {staf2267},
        pages = {staf2267},
          doi = {10.1093/mnras/staf2267},
archivePrefix = {arXiv},
       eprint = {2509.24881},
 primaryClass = {astro-ph.GA},
       adsurl = {https://ui.adsabs.harvard.edu/abs/2026MNRAS.546f2267C}
}

@ARTICLE{Wong2020MNRAS.498.1420Wtension,
       author = {{Wong}, Kenneth C. and {Suyu}, Sherry H. and {Chen}, Geoff C.-F. and {Rusu}, Cristian E. and {Millon}, Martin and {Sluse}, Dominique and {Bonvin}, Vivien and {Fassnacht}, Christopher D. and {Taubenberger}, Stefan and {Auger}, Matthew W. and {Birrer}, Simon and {Chan}, James H.~H. and {Courbin}, Frederic and {Hilbert}, Stefan and {Tihhonova}, Olga and {Treu}, Tommaso and {Agnello}, Adriano and {Ding}, Xuheng and {Jee}, Inh and {Komatsu}, Eiichiro and {Shajib}, Anowar J. and {Sonnenfeld}, Alessandro and {Blandford}, Roger D. and {Koopmans}, L{\'e}on V.~E. and {Marshall}, Philip J. and {Meylan}, Georges},
        title = "{H0LiCOW - XIII. A 2.4 per cent measurement of H$_{0}$ from lensed quasars: 5.3{\ensuremath{\sigma}} tension between early- and late-Universe probes}",
      journal = {\mnras},
         year = 2020,
        month = oct,
       volume = {498},
       number = {1},
        pages = {1420-1439},
          doi = {10.1093/mnras/stz3094},
archivePrefix = {arXiv},
       eprint = {1907.04869},
 primaryClass = {astro-ph.CO},
       adsurl = {https://ui.adsabs.harvard.edu/abs/2020MNRAS.498.1420W}
}

@ARTICLE{Voit2005RvMP...77..207V,
       author = {{Voit}, G. Mark},
        title = "{Tracing cosmic evolution with clusters of galaxies}",
      journal = {Reviews of Modern Physics},
         year = 2005,
        month = apr,
       volume = {77},
       number = {1},
        pages = {207-258},
          doi = {10.1103/RevModPhys.77.207},
archivePrefix = {arXiv},
       eprint = {astro-ph/0410173},
 primaryClass = {astro-ph},
       adsurl = {https://ui.adsabs.harvard.edu/abs/2005RvMP...77..207V}
}

@ARTICLE{Franx1997,
   author = {{Franx}, M. and {Illingworth}, G.~D. and {Kelson}, D.~D. and
	{van Dokkum}, P.~G. and {Tran}, K.-V.},
    title = "{A Pair of Lensed Galaxies at z=4.92 in the Field of CL 1358+62}",
  journal = {\apjl},
   eprint = {arXiv:astro-ph/9704090},
     year = 1997,
    month = sep,
   volume = 486,
    pages = {L75},
      doi = {10.1086/310844},
   adsurl = {http://adsabs.harvard.edu/abs/1997ApJ...486L..75F}
}

@ARTICLE{Clowe2006Bullet,
   author = {{Clowe}, D. and {Brada{\v c}}, M. and {Gonzalez}, A.~H. and
	{Markevitch}, M. and {Randall}, S.~W. and {Jones}, C. and {Zaritsky}, D.
	},
    title = "{A Direct Empirical Proof of the Existence of Dark Matter}",
  journal = {\apjl},
   eprint = {arXiv:astro-ph/0608407},
     year = 2006,
    month = sep,
   volume = 648,
    pages = {L109-L113},
      doi = {10.1086/508162},
   adsurl = {http://adsabs.harvard.edu/abs/2006ApJ...648L.109C}
}

@ARTICLE{Lotz2016HFF,
   author = {{Lotz}, J.~M. and {Koekemoer}, A. and {Coe}, D. and {Grogin}, N. and 
	{Capak}, P. and {Mack}, J. and {Anderson}, J. and {Avila}, R. and 
	{Barker}, E.~A. and {Borncamp}, D. and {Brammer}, G. and {Durbin}, M. and 
	{Gunning}, H. and {Hilbert}, B. and {Jenkner}, H. and {Khandrika}, H. and 
	{Levay}, Z. and {Lucas}, R.~A. and {MacKenty}, J. and {Ogaz}, S. and 
	{Porterfield}, B. and {Reid}, N. and {Robberto}, M. and {Royle}, P. and 
	{Smith}, L.~J. and {Storrie-Lombardi}, L.~J. and {Sunnquist}, B. and 
	{Surace}, J. and {Taylor}, D.~C. and {Williams}, R. and {Bullock}, J. and 
	{Dickinson}, M. and {Finkelstein}, S. and {Natarajan}, P. and 
	{Richard}, J. and {Robertson}, B. and {Tumlinson}, J. and {Zitrin}, A. and 
	{Flanagan}, K. and {Sembach}, K. and {Soifer}, B.~T. and {Mountain}, M.
	},
    title = "{The Frontier Fields: Survey Design and Initial Results}",
  journal = {\apj},
archivePrefix = "arXiv",
   eprint = {1605.06567},
     year = 2017,
    month = mar,
   volume = 837,
      eid = {97},
    pages = {97},
      doi = {10.3847/1538-4357/837/1/97},
   adsurl = {http://adsabs.harvard.edu/abs/2017ApJ...837...97L}
}

@ARTICLE{Kneib2011review,
   author = {{Kneib}, J.-P. and {Natarajan}, P.},
    title = "{Cluster lenses}",
  journal = {\aapr},
archivePrefix = "arXiv",
   eprint = {1202.0185},
 primaryClass = "astro-ph.CO",
     year = 2011,
    month = nov,
   volume = 19,
    pages = {47},
      doi = {10.1007/s00159-011-0047-3},
   adsurl = {http://adsabs.harvard.edu/abs/2011A%26ARv..19...47K}
}

@ARTICLE{NarayanBartelmann1996Lectures,
   author = {{Narayan}, R. and {Bartelmann}, M.},
    title = "{Lectures on Gravitational Lensing}",
  journal = {ArXiv Astrophysics e-prints},
   eprint = {arXiv:astro-ph/9606001},
     year = 1996,
    month = jun,
   adsurl = {http://adsabs.harvard.edu/abs/1996astro.ph..6001N},
   volume = {astro-ph/9606001}
}

@ARTICLE{Markevitch2004ApJ...606..819M,
       author = {{Markevitch}, M. and {Gonzalez}, A.~H. and {Clowe}, D. and {Vikhlinin}, A. and {Forman}, W. and {Jones}, C. and {Murray}, S. and {Tucker}, W.},
        title = "{Direct Constraints on the Dark Matter Self-Interaction Cross Section from the Merging Galaxy Cluster 1E 0657-56}",
      journal = {\apj},
         year = 2004,
        month = may,
       volume = {606},
       number = {2},
        pages = {819-824},
          doi = {10.1086/383178},
archivePrefix = {arXiv},
       eprint = {astro-ph/0309303},
 primaryClass = {astro-ph},
       adsurl = {https://ui.adsabs.harvard.edu/abs/2004ApJ...606..819M}
}

@ARTICLE{Sereno2018ApJ...860L...4S,
       author = {{Sereno}, Mauro and {Umetsu}, Keiichi and {Ettori}, Stefano and {Sayers}, Jack and {Chiu}, I.-Non and {Meneghetti}, Massimo and {Vega-Ferrero}, Jes{\'u}s and {Zitrin}, Adi},
        title = "{CLUMP-3D: Testing {\ensuremath{\Lambda}}CDM with Galaxy Cluster Shapes}",
      journal = {\apjl},
         year = 2018,
        month = jun,
       volume = {860},
       number = {1},
          eid = {L4},
        pages = {L4},
          doi = {10.3847/2041-8213/aac6d9},
archivePrefix = {arXiv},
       eprint = {1804.00667},
 primaryClass = {astro-ph.CO},
       adsurl = {https://ui.adsabs.harvard.edu/abs/2018ApJ...860L...4S}
}

@ARTICLE{Sigel2018ApJ...861...71S,
       author = {{Siegel}, Seth R. and {Sayers}, Jack and {Mahdavi}, Andisheh and {Donahue}, Megan and {Merten}, Julian and {Zitrin}, Adi and {Meneghetti}, Massimo and {Umetsu}, Keiichi and {Czakon}, Nicole G. and {Golwala}, Sunil R. and {Postman}, Marc and {Koch}, Patrick M. and {Koekemoer}, Anton M. and {Lin}, Kai-Yang and {Melchior}, Peter and {Molnar}, Sandor M. and {Moustakas}, Leonidas and {Mroczkowski}, Tony K. and {Pierpaoli}, Elena and {Shitanishi}, Jennifer},
        title = "{Constraints on the Mass, Concentration, and Nonthermal Pressure Support of Six CLASH Clusters from a Joint Analysis of X-Ray, SZ, and Lensing Data}",
      journal = {\apj},
         year = 2018,
        month = jul,
       volume = {861},
       number = {1},
          eid = {71},
        pages = {71},
          doi = {10.3847/1538-4357/aac5f8},
archivePrefix = {arXiv},
       eprint = {1612.05377},
 primaryClass = {astro-ph.CO},
       adsurl = {https://ui.adsabs.harvard.edu/abs/2018ApJ...861...71S}
}

@ARTICLE{ZhangRXC221LRD2025arXiv251205180Z,
       author = {{Zhang}, Zijian and {Li}, Mingyu and {Oguri}, Masamune and {Lin}, Xiaojing and {Inayoshi}, Kohei and {Cerny}, Catherine and {Coe}, Dan and {Diego}, Jose M. and {Fujimoto}, Seiji and {Jiang}, Linhua and {Mahler}, Guillaume and {Matthee}, Jorryt and {Naidu}, Rohan P. and {Sharon}, Keren and {Shen}, Yue and {Zitrin}, Adi and {Abdurro'uf} and {Akins}, Hollis and {Allingham}, Joseph F.~V. and {Amor{\'\i}n}, Ricardo and {Asada}, Yoshihisa and {Atek}, Hakim and {Bauer}, Franz E. and {Brada{\v{c}}}, Maru{\v{s}}a and {Bradley}, Larry D. and {Cai}, Zheng and {Cantalupo}, Sebastiano and {Conselice}, Christopher and {Dai}, Liang and {Dayal}, Pratika and {Egami}, Eiichi and {Eisenstein}, Daniel J. and {Faisst}, Andreas L. and {Fan}, Xiaohui and {Fei}, Qinyue and {Frye}, Brenda L. and {Fudamoto}, Yoshinobu and {Furtak}, Lukas J. and {Golubchik}, Miriam and {Gonz{\'a}lez-Otero}, Mauro and {Harikane}, Yuichi and {Hsiao}, Tiger Yu-Yang and {Jim{\'e}nez-Teja}, Yolanda and {Kartaltepe}, Jeyhan S. and {Kiyota}, Tomokazu and {Koekemoer}, Anton M. and {Kohno}, Kotaro and {Kokorev}, Vasily and {Kumari}, Nimisha and {Labbe}, Ivo and {Lagos}, Claudia D.~P. and {Larison}, Conor and {Liang}, Yongming and {Lucas}, Ray A. and {Lyu}, Jianwei and {Martis}, Nicholas S. and {Magdis}, Georgios E. and {Messa}, Matteo and {Nakane}, Minami and {Noirot}, Ga{\"e}l and {Ortiz}, III, Rafael and {Ouchi}, Masami and {Pierel}, Justin D.~R. and {Postman}, Marc and {Reddy}, Naveen and {Ricotti}, Massimo and {Schaerer}, Daniel and {Schneider}, Raffaella and {Steidel}, Charles C. and {Tee}, Wei Leong and {Tripodi}, Roberta and {Trussler}, James A.~A. and {Umeda}, Hiroya and {Valentino}, Francesco and {Vanzella}, Eros and {Wang}, Feige and {Windhorst}, Rogier and {Wu}, Yunjing and {Wu}, Zihao and {Yanagisawa}, Hiroto and {Yang}, Jinyi and {Sun}, Fengwu},
        title = "{Little red dot variability over a century reveals black hole envelope via a giant Einstein cross}",
      journal = {arXiv e-prints},
         year = 2025,
        month = dec,
          eid = {arXiv:2512.05180},
        pages = {arXiv:2512.05180},
          doi = {10.48550/arXiv.2512.05180},
archivePrefix = {arXiv},
       eprint = {2512.05180},
 primaryClass = {astro-ph.GA},
       adsurl = {https://ui.adsabs.harvard.edu/abs/2025arXiv251205180Z}
}

@ARTICLE{Livermore2017ApJ...835..113L,
       author = {{Livermore}, R.~C. and {Finkelstein}, S.~L. and {Lotz}, J.~M.},
        title = "{Directly Observing the Galaxies Likely Responsible for Reionization}",
      journal = {\apj},
         year = 2017,
        month = feb,
       volume = {835},
       number = {2},
          eid = {113},
        pages = {113},
          doi = {10.3847/1538-4357/835/2/113},
archivePrefix = {arXiv},
       eprint = {1604.06799},
 primaryClass = {astro-ph.GA},
       adsurl = {https://ui.adsabs.harvard.edu/abs/2017ApJ...835..113L}
}

@ARTICLE{Bouwens2017ApJ...843..129Bturnover,
       author = {{Bouwens}, R.~J. and {Oesch}, P.~A. and {Illingworth}, G.~D. and {Ellis}, R.~S. and {Stefanon}, M.},
        title = "{The z {\ensuremath{\sim}} 6 Luminosity Function Fainter than -15 mag from the Hubble Frontier Fields: The Impact of Magnification Uncertainties}",
      journal = {\apj},
         year = 2017,
        month = jul,
       volume = {843},
       number = {2},
          eid = {129},
        pages = {129},
          doi = {10.3847/1538-4357/aa70a4},
archivePrefix = {arXiv},
       eprint = {1610.00283},
 primaryClass = {astro-ph.GA},
       adsurl = {https://ui.adsabs.harvard.edu/abs/2017ApJ...843..129B}
}

@ARTICLE{Zhang2025arXiv251205180ZRXC221LRDs,
       author = {{Zhang}, Zijian and {Li}, Mingyu and {Oguri}, Masamune and {Lin}, Xiaojing and {Inayoshi}, Kohei and {Cerny}, Catherine and {Coe}, Dan and {Diego}, Jose M. and {Fujimoto}, Seiji and {Jiang}, Linhua and {Mahler}, Guillaume and {Matthee}, Jorryt and {Naidu}, Rohan P. and {Sharon}, Keren and {Shen}, Yue and {Zitrin}, Adi and {Abdurro'uf} and {Akins}, Hollis and {Allingham}, Joseph F.~V. and {Amor{\'\i}n}, Ricardo and {Asada}, Yoshihisa and {Atek}, Hakim and {Bauer}, Franz E. and {Brada{\v{c}}}, Maru{\v{s}}a and {Bradley}, Larry D. and {Cai}, Zheng and {Cantalupo}, Sebastiano and {Conselice}, Christopher and {Dai}, Liang and {Dayal}, Pratika and {Egami}, Eiichi and {Eisenstein}, Daniel J. and {Faisst}, Andreas L. and {Fan}, Xiaohui and {Fei}, Qinyue and {Frye}, Brenda L. and {Fudamoto}, Yoshinobu and {Furtak}, Lukas J. and {Golubchik}, Miriam and {Gonz{\'a}lez-Otero}, Mauro and {Harikane}, Yuichi and {Hsiao}, Tiger Yu-Yang and {Jim{\'e}nez-Teja}, Yolanda and {Kartaltepe}, Jeyhan S. and {Kiyota}, Tomokazu and {Koekemoer}, Anton M. and {Kohno}, Kotaro and {Kokorev}, Vasily and {Kumari}, Nimisha and {Labbe}, Ivo and {Lagos}, Claudia D.~P. and {Larison}, Conor and {Liang}, Yongming and {Lucas}, Ray A. and {Lyu}, Jianwei and {Martis}, Nicholas S. and {Magdis}, Georgios E. and {Messa}, Matteo and {Nakane}, Minami and {Noirot}, Ga{\"e}l and {Ortiz}, III, Rafael and {Ouchi}, Masami and {Pierel}, Justin D.~R. and {Postman}, Marc and {Reddy}, Naveen and {Ricotti}, Massimo and {Schaerer}, Daniel and {Schneider}, Raffaella and {Steidel}, Charles C. and {Tee}, Wei Leong and {Tripodi}, Roberta and {Trussler}, James A.~A. and {Umeda}, Hiroya and {Valentino}, Francesco and {Vanzella}, Eros and {Wang}, Feige and {Windhorst}, Rogier and {Wu}, Yunjing and {Wu}, Zihao and {Yanagisawa}, Hiroto and {Yang}, Jinyi and {Sun}, Fengwu},
        title = "{Little red dot variability over a century reveals black hole envelope via a giant Einstein cross}",
      journal = {arXiv e-prints},
         year = 2025,
        month = dec,
          eid = {arXiv:2512.05180},
        pages = {arXiv:2512.05180},
          doi = {10.48550/arXiv.2512.05180},
archivePrefix = {arXiv},
       eprint = {2512.05180},
 primaryClass = {astro-ph.GA},
       adsurl = {https://ui.adsabs.harvard.edu/abs/2025arXiv251205180Z}
}

@ARTICLE{Yanagisawa2026arXiv260106015Y_doubleLRD,
       author = {{Yanagisawa}, Hiroto and {Ouchi}, Masami and {Golubchik}, Miriam and {Oguri}, Masamune and {Fujimoto}, Seiji and {Kokorev}, Vasily and {Brammer}, Gabriel and {Sun}, Fengwu and {Nakane}, Minami and {Harikane}, Yuichi and {Umeda}, Hiroya and {Akins}, Hollis B. and {Atek}, Hakim and {Bauer}, Franz E. and {Brada{\v{c}}}, Maru{\v{s}}a and {Chisholm}, John and {Coe}, Dan and {Diego}, Jose M. and {Ferguson}, Henry C. and {Finkelstein}, Steven L. and {Furtak}, Lukas J. and {Inayoshi}, Kohei and {Koekemoer}, Anton M. and {Matthee}, Jorryt and {Naidu}, Rohan P. and {Ono}, Yoshiaki and {Pan}, Richard and {Richard}, Johan and {Robbins}, Luke and {Willott}, Chris and {Zitrin}, Adi and {Amor{\'\i}n}, Ricardo O. and {Bradley}, Larry D. and {Bromm}, Volker and {Conselice}, Christopher J. and {Dayal}, Pratika and {Kartaltepe}, Jeyhan S. and {Lopes}, Paulo A.~A. and {Lucas}, Ray A. and {Magdis}, Georgios E. and {Martis}, Nicholas S. and {Papovich}, Casey and {Schaerer}, Daniel and {Valentino}, Francesco and {Vanzella}, Eros and {Allingham}, Joseph F.~V. and {Grogin}, Norman A. and {Gonz{\'a}lez-Otero}, Mauro and {Ricotti}, Massimo and {Windhorst}, Rogier A.},
        title = "{VENUS: Two Faint Little Red Dots Separated by $\sim70\,\mathrm{pc}$ Hidden in a Single Lensed Galaxy at $z\sim7$}",
      journal = {arXiv e-prints},
         year = 2026,
        month = jan,
          eid = {arXiv:2601.06015},
        pages = {arXiv:2601.06015},
          doi = {10.48550/arXiv.2601.06015},
archivePrefix = {arXiv},
       eprint = {2601.06015},
 primaryClass = {astro-ph.GA},
       adsurl = {https://ui.adsabs.harvard.edu/abs/2026arXiv260106015Y}
}

@ARTICLE{Golubchik2025arXiv251202117G_LRD,
       author = {{Golubchik}, Miriam and {Furtak}, Lukas J. and {Allingham}, Joseph F.~V. and {Zitrin}, Adi and {Akins}, Hollis B. and {Kokorev}, Vasily and {Fujimoto}, Seiji and {Abdurro'uf} and {Amor{\'\i}n}, Ricardo O. and {Bauer}, Franz E. and {Bezanson}, Rachel and {Brada{\v{c}}}, Marusa and {Bradley}, Larry D. and {Brammer}, Gabriel B. and {Chisholm}, John and {Coe}, Dan and {Conselice}, Christopher J. and {Dayal}, Pratika and {Dessauges-Zavadsky}, Miroslava and {Diego}, Jose M. and {Faisst}, Andreas L. and {Fei}, Qinyue and {Ferguson}, Henry C. and {Finkelstein}, Steven L. and {Frye}, Brenda L. and {Gonz{\'a}lez-Otero}, Mauro and {Greene}, Jenny E. and {Harikane}, Yuichi and {Hsiao}, Tiger Yu-Yang and {Inayoshi}, Kohei and {Jim{\'e}nez-Teja}, Yolanda and {Knudsen}, Kirsten and {Koekemoer}, Anton M. and {Labb{\'e}}, Ivo and {Lucas}, Ray A. and {Magdis}, Georgios E. and {Matthee}, Jorryt and {Messa}, Matteo and {Naidu}, Rohan P. and {Nakane}, Minami and {Noirot}, Ga{\"e}l and {Pan}, Richard and {Papovich}, Casey and {Richard}, Johan and {Ricotti}, Massimo and {Robbins}, Luke and {Stark}, Daniel P. and {Sun}, Fengwu and {Treu}, Tommaso and {Tripodi}, Roberta and {Vanzella}, Eros and {Willott}, Chris and {Windhorst}, Rogier A.},
        title = "{VENUS: When Red meets Blue -- A multiply imaged Little Red Dot with an apparent blue companion behind the galaxy cluster Abell 383}",
      journal = {arXiv e-prints},
         year = 2025,
        month = dec,
          eid = {arXiv:2512.02117},
        pages = {arXiv:2512.02117},
          doi = {10.48550/arXiv.2512.02117},
archivePrefix = {arXiv},
       eprint = {2512.02117},
 primaryClass = {astro-ph.GA},
       adsurl = {https://ui.adsabs.harvard.edu/abs/2025arXiv251202117G}
}

@ARTICLE{Baggen2026arXiv260202702B,
       author = {{Baggen}, Josephine F.~W. and {Scoggins}, Matthew T. and {van Dokkum}, Pieter and {Haiman}, Zolt{\'a}n and {Torralba}, Alberto and {Matthee}, Jorryt},
        title = "{Connecting the Dots: UV-Bright Companions of Little Red Dots as Lyman-Werner Sources Enabling Direct Collapse Black Hole Formation}",
      journal = {arXiv e-prints},
         year = 2026,
        month = feb,
          eid = {arXiv:2602.02702},
        pages = {arXiv:2602.02702},
          doi = {10.48550/arXiv.2602.02702},
archivePrefix = {arXiv},
       eprint = {2602.02702},
 primaryClass = {astro-ph.GA},
       adsurl = {https://ui.adsabs.harvard.edu/abs/2026arXiv260202702B}
}

@ARTICLE{YanagisawaLRDpair2026arXiv260106015Y,
       author = {{Yanagisawa}, Hiroto and {Ouchi}, Masami and {Golubchik}, Miriam and {Oguri}, Masamune and {Fujimoto}, Seiji and {Kokorev}, Vasily and {Brammer}, Gabriel and {Sun}, Fengwu and {Nakane}, Minami and {Harikane}, Yuichi and {Umeda}, Hiroya and {Akins}, Hollis B. and {Atek}, Hakim and {Bauer}, Franz E. and {Brada{\v{c}}}, Maru{\v{s}}a and {Chisholm}, John and {Coe}, Dan and {Diego}, Jose M. and {Ferguson}, Henry C. and {Finkelstein}, Steven L. and {Furtak}, Lukas J. and {Inayoshi}, Kohei and {Koekemoer}, Anton M. and {Matthee}, Jorryt and {Naidu}, Rohan P. and {Ono}, Yoshiaki and {Pan}, Richard and {Richard}, Johan and {Robbins}, Luke and {Willott}, Chris and {Zitrin}, Adi and {Amor{\'\i}n}, Ricardo O. and {Bradley}, Larry D. and {Bromm}, Volker and {Conselice}, Christopher J. and {Dayal}, Pratika and {Kartaltepe}, Jeyhan S. and {Lopes}, Paulo A.~A. and {Lucas}, Ray A. and {Magdis}, Georgios E. and {Martis}, Nicholas S. and {Papovich}, Casey and {Schaerer}, Daniel and {Valentino}, Francesco and {Vanzella}, Eros and {Allingham}, Joseph F.~V. and {Grogin}, Norman A. and {Gonz{\'a}lez-Otero}, Mauro and {Ricotti}, Massimo and {Windhorst}, Rogier A.},
        title = "{VENUS: Two Faint Little Red Dots Separated by $\sim70\,\mathrm{pc}$ Hidden in a Single Lensed Galaxy at $z\sim7$}",
      journal = {arXiv e-prints},
         year = 2026,
        month = jan,
          eid = {arXiv:2601.06015},
        pages = {arXiv:2601.06015},
          doi = {10.48550/arXiv.2601.06015},
archivePrefix = {arXiv},
       eprint = {2601.06015},
 primaryClass = {astro-ph.GA},
       adsurl = {https://ui.adsabs.harvard.edu/abs/2026arXiv260106015Y}
}

@ARTICLE{GouldLoeb1992ApJ...396..104G,
       author = {{Gould}, Andrew and {Loeb}, Abraham},
        title = "{Discovering Planetary Systems through Gravitational Microlenses}",
      journal = {\apj},
         year = 1992,
        month = sep,
       volume = {396},
        pages = {104},
          doi = {10.1086/171700},
       adsurl = {https://ui.adsabs.harvard.edu/abs/1992ApJ...396..104G}
}

@ARTICLE{NewmanBHmass2025arXiv250317478N,
       author = {{Newman}, Andrew B. and {Gu}, Meng and {Belli}, Sirio and {Ellis}, Richard S. and {Gangula}, Sai and {Greene}, Jenny E. and {Walsh}, Jonelle L. and {Suyu}, Sherry H. and {Ertl}, Sebastian and {Caminha}, Gabriel and {Granata}, Giovanni and {Grillo}, Claudio and {Schuldt}, Stefan and {Barone}, Tania M. and {Bird}, Simeon and {Glazebrook}, Karl and {Jafariyazani}, Marziye and {Kriek}, Mariska and {Matthews}, Allison and {Morishita}, Takahiro and {Nanayakkara}, Themiya and {Pierel}, Justin D.~R. and {Acebr\textbackslash'on}, Ana and {Bergamini}, Pietro and {Cha}, Sangjun and {Diego}, Jose M. and {Foo}, Nicholas and {Frye}, Brenda and {Fudamoto}, Yoshinobu and {Jee}, M. James and {Kamieneski}, Patrick S. and {Koekemoer}, Anton M. and {Meena}, Asish K. and {Nishida}, Shun and {Oguri}, Masamune and {Rosati}, Piero and {Zitrin}, Adi},
        title = "{A stellar dynamical mass measure of an inactive black hole in the distant universe}",
      journal = {arXiv e-prints},
         year = 2025,
        month = mar,
          eid = {arXiv:2503.17478},
        pages = {arXiv:2503.17478},
          doi = {10.48550/arXiv.2503.17478},
archivePrefix = {arXiv},
       eprint = {2503.17478},
 primaryClass = {astro-ph.GA},
       adsurl = {https://ui.adsabs.harvard.edu/abs/2025arXiv250317478N}
}

@ARTICLE{Claeyssens2023MNRAS.520.2180C,
       author = {{Claeyssens}, Ad{\'e}la{\"\i}de and {Adamo}, Angela and {Richard}, Johan and {Mahler}, Guillaume and {Messa}, Matteo and {Dessauges-Zavadsky}, Miroslava},
        title = "{Star formation at the smallest scales: a JWST study of the clump populations in SMACS0723}",
      journal = {\mnras},
         year = 2023,
        month = apr,
       volume = {520},
       number = {2},
        pages = {2180-2203},
          doi = {10.1093/mnras/stac3791},
archivePrefix = {arXiv},
       eprint = {2208.10450},
 primaryClass = {astro-ph.GA},
       adsurl = {https://ui.adsabs.harvard.edu/abs/2023MNRAS.520.2180C}
}

@ARTICLE{Jain2025A&A...703A..96J,
       author = {{Jain}, Rashi and {Wadadekar}, Yogesh},
        title = "{A grand-design spiral galaxy 1.5 billion years after the Big Bang with JWST}",
      journal = {\aap},
         year = 2025,
        month = nov,
       volume = {703},
          eid = {A96},
        pages = {A96},
          doi = {10.1051/0004-6361/202451689},
archivePrefix = {arXiv},
       eprint = {2412.04834},
 primaryClass = {astro-ph.GA},
       adsurl = {https://ui.adsabs.harvard.edu/abs/2025A&A...703A..96J}
}

@ARTICLE{Bradac2025ApJ...995L..74Bz11SF,
       author = {{Brada{\v{c}}}, Maru{\v{s}}a and {Jude{\v{z}}}, Jon and {Willott}, Chris and {Rihtar{\v{s}}ic}, Gregor and {Martis}, Nicholas S. and {Harshan}, Anishya and {Felicioni}, Giordano and {Asada}, Yoshihisa and {Desprez}, Guillaume and {Clowe}, Douglas and {Gonzalez}, Anthony H. and {Jones}, Christine and {Lemaux}, Brian C. and {Markevitch}, Maxim and {Markov}, Vladan and {Mowla}, Lamiya and {Noirot}, Ga{\"e}l and {Peter}, Annika H.~G. and {Robertson}, Andrew and {Sarrouh}, Ghassan T.~E. and {Sawicki}, Marcin and {Schrabback}, Tim and {Tripodi}, Roberta},
        title = "{Star Formation under a Cosmic Microscope: Highly Magnified z = 11 Galaxy behind the Bullet Cluster}",
      journal = {\apjl},
         year = 2025,
        month = dec,
       volume = {995},
       number = {2},
          eid = {L74},
        pages = {L74},
          doi = {10.3847/2041-8213/ae27d2},
archivePrefix = {arXiv},
       eprint = {2509.20446},
 primaryClass = {astro-ph.GA},
       adsurl = {https://ui.adsabs.harvard.edu/abs/2025ApJ...995L..74B}
}

@ARTICLE{Adamo2024Natur.632..513A,
       author = {{Adamo}, Angela and {Bradley}, Larry D. and {Vanzella}, Eros and {Claeyssens}, Ad{\'e}la{\"\i}de and {Welch}, Brian and {Diego}, Jose M. and {Mahler}, Guillaume and {Oguri}, Masamune and {Sharon}, Keren and {Abdurro'uf} and {Hsiao}, Tiger Yu-Yang and {Xu}, Xinfeng and {Messa}, Matteo and {Lassen}, Augusto E. and {Zackrisson}, Erik and {Brammer}, Gabriel and {Coe}, Dan and {Kokorev}, Vasily and {Ricotti}, Massimo and {Zitrin}, Adi and {Fujimoto}, Seiji and {Inoue}, Akio K. and {Resseguier}, Tom and {Rigby}, Jane R. and {Jim{\'e}nez-Teja}, Yolanda and {Windhorst}, Rogier A. and {Hashimoto}, Takuya and {Tamura}, Yoichi},
        title = "{Bound star clusters observed in a lensed galaxy 460 Myr after the Big Bang}",
      journal = {\nat},
         year = 2024,
        month = aug,
       volume = {632},
       number = {8025},
        pages = {513-516},
          doi = {10.1038/s41586-024-07703-7},
archivePrefix = {arXiv},
       eprint = {2401.03224},
 primaryClass = {astro-ph.GA},
       adsurl = {https://ui.adsabs.harvard.edu/abs/2024Natur.632..513A}
}

@ARTICLE{Atek2025arXiv251107542AOverview,
       author = {{Atek}, Hakim and {Chisholm}, John and {Kokorev}, Vasily and {Endsley}, Ryan and {Pan}, Richard and {Furtak}, Lukas and {Chemerynska}, Iryna and {Richard}, Johan and {Claeyssens}, Ad{\'e}la{\"\i}de and {Oesch}, Pascal and {Fujimoto}, Seiji and {Naidu}, Rohan and {Korber}, Damien and {Schaerer}, Daniel and {Blaizot}, Jeremy and {Rosdahl}, Joki and {Adamo}, Angela and {Asada}, Yoshihisa and {Basu}, Arghyadeep and {Beauchesne}, Benjamin and {Berg}, Danielle and {Bezanson}, Rachel and {Bouwens}, Rychard and {Brammer}, Gabriel and {Dessauges-Zavadsky}, Miroslava and {Ellien}, Ama{\"e}l and {Ezziati}, Meriam and {Fei}, Qinyue and {Goovaerts}, Ilias and {Heurtier}, Sylvain and {Hsiao}, Tiger Yu-Yang and {Jecmen}, Michelle and {Khullar}, Gourav and {Kneib}, Jean-Paul and {Labb{\'e}}, Ivo and {Leclercq}, Floriane and {Marques-Chaves}, Rui and {Mason}, Charlotte and {McQuinn}, Kristen B.~W. and {Mu{\~n}oz}, Julian B. and {Natarajan}, Priyamvada and {Saldana-Lopez}, Alberto and {Stephenson}, Mabel G. and {Trebitsch}, Maxime and {Volonteri}, Marta and {Weibel}, Andrea and {Zitrin}, Adi},
        title = "{JWST's GLIMPSE: an overview of the deepest probe of early galaxy formation and cosmic reionization}",
      journal = {arXiv e-prints},
         year = 2025,
        month = nov,
          eid = {arXiv:2511.07542},
        pages = {arXiv:2511.07542},
          doi = {10.48550/arXiv.2511.07542},
archivePrefix = {arXiv},
       eprint = {2511.07542},
 primaryClass = {astro-ph.GA},
       adsurl = {https://ui.adsabs.harvard.edu/abs/2025arXiv251107542A}
}

@ARTICLE{Nakane2025arXiv251114483N,
       author = {{Nakane}, Minami and {Kokorev}, Vasily and {Fujimoto}, Seiji and {Ouchi}, Masami and {McLeod}, Derek J. and {Golubchik}, Miriam and {Oguri}, Masamune and {Zitrin}, Adi and {Bondestam}, Cecilia and {Donnan}, Callum T. and {Brammer}, Gabriel and {Finkelstein}, Steven L. and {Willott}, Chris and {Rihtarsic}, Gregor and {Desprez}, Guillaume and {Adamo}, Angela and {Vanzella}, Eros and {Brada{\v{c}}}, Maru{\v{s}}a and {Messa}, Matteo and {Yanagisawa}, Hiroto and {Sun}, Fengwu and {Ferguson}, Henry C. and {Lucas}, Ray A. and {Coe}, Dan and {Richard}, Johan and {Abdurro'uf} and {Akins}, Hollis B. and {Allingham}, Joseph F.~V. and {Amor{\'\i}n}, Ricardo O. and {Asada}, Yoshihisa and {Atek}, Hakim and {Bezanson}, Rachel and {Bradley}, Larry D. and {Chisholm}, John and {Conselice}, Christopher J. and {Dayal}, Pratika and {Dessauges-Zavadsky}, Miroslava and {Diego}, Jose M. and {Faisst}, Andreas L. and {Fei}, Qinyue and {Frye}, Brenda L. and {Fudamoto}, Yoshinobu and {Furtak}, Lukas J. and {Harikane}, Yuichi and {Hsiao}, Tiger Yu-Yang and {Jim{\'e}nez-Teja}, Yolanda and {Kartaltepe}, Jeyhan S. and {Kiyota}, Tomokazu and {Koekemoer}, Anton M. and {Lagos}, Claudia del P. and {Magdis}, Georgios E. and {Meena}, Ashish Kumar and {Mowla}, Lamiya and {Noirot}, Ga{\"e}l and {Oesch}, Pascal A. and {Ono}, Yoshiaki and {Ortiz}, III, Rafael and {Pan}, Richard and {Papovich}, Casey and {Pierel}, Justin D. and {Ricotti}, Massimo and {Robbins}, Luke and {Schaerer}, Daniel and {Schneider}, Raffaella and {Treu}, Tommaso and {Valentino}, Francesco and {Windhorst}, Rogier A. and {Bauer}, Franz E. and {Bromm}, Volker and {Egami}, Eiichi and {Gonz{\'a}lez-Otero}, Mauro and {Kohno}, Kotaro and {Labbe}, Ivo and {Matthee}, Jorryt and {Mun}, Marcie and {Naidu}, Rohan P. and {Tripodi}, Roberta},
        title = "{VENUS: A Strongly Lensed Clumpy Galaxy at $z\sim11-12$ behind the Galaxy Cluster MACS J0257.1-2325}",
      journal = {arXiv e-prints},
         year = 2025,
        month = nov,
          eid = {arXiv:2511.14483},
        pages = {arXiv:2511.14483},
          doi = {10.48550/arXiv.2511.14483},
archivePrefix = {arXiv},
       eprint = {2511.14483},
 primaryClass = {astro-ph.GA},
       adsurl = {https://ui.adsabs.harvard.edu/abs/2025arXiv251114483N}
}

@ARTICLE{Hsiao2023ApJ...949L..34Hz11,
       author = {{Hsiao}, Tiger Yu-Yang and {Coe}, Dan and {Abdurro'uf} and {Whitler}, Lily and {Jung}, Intae and {Khullar}, Gourav and {Meena}, Ashish Kumar and {Dayal}, Pratika and {Barrow}, Kirk S.~S. and {Santos-Olmsted}, Lillian and {Casselman}, Adam and {Vanzella}, Eros and {Nonino}, Mario and {Jim{\'e}nez-Teja}, Yolanda and {Oguri}, Masamune and {Stark}, Daniel P. and {Furtak}, Lukas J. and {Zitrin}, Adi and {Adamo}, Angela and {Brammer}, Gabriel and {Bradley}, Larry and {Diego}, Jose M. and {Zackrisson}, Erik and {Finkelstein}, Steven L. and {Windhorst}, Rogier A. and {Bhatawdekar}, Rachana and {Hutchison}, Taylor A. and {Broadhurst}, Tom and {Dimauro}, Paola and {Andrade-Santos}, Felipe and {Eldridge}, Jan J. and {Acebron}, Ana and {Avila}, Roberto J. and {Bayliss}, Matthew B. and {Ben{\'\i}tez}, Alex and {Binggeli}, Christian and {Bolan}, Patricia and {Brada{\v{c}}}, Maru{\v{s}}a and {Carnall}, Adam C. and {Conselice}, Christopher J. and {Donahue}, Megan and {Frye}, Brenda and {Fujimoto}, Seiji and {Henry}, Alaina and {James}, Bethan L. and {Kassin}, Susan A. and {Kewley}, Lisa and {Larson}, Rebecca L. and {Lauer}, Tod and {Law}, David and {Mahler}, Guillaume and {Mainali}, Ramesh and {McCandliss}, Stephan and {Nicholls}, David and {Pirzkal}, Norbert and {Postman}, Marc and {Rigby}, Jane R. and {Ryan}, Russell and {Senchyna}, Peter and {Sharon}, Keren and {Shimizu}, Ikko and {Strait}, Victoria and {Tang}, Mengtao and {Trenti}, Michele and {Vikaeus}, Anton and {Welch}, Brian},
        title = "{JWST Reveals a Possible z {$\sim$} 11 Galaxy Merger in Triply Lensed MACS0647-JD}",
      journal = {\apjl},
         year = 2023,
        month = jun,
       volume = {949},
       number = {2},
          eid = {L34},
        pages = {L34},
          doi = {10.3847/2041-8213/acc94b},
archivePrefix = {arXiv},
       eprint = {2210.14123},
 primaryClass = {astro-ph.GA},
       adsurl = {https://ui.adsabs.harvard.edu/abs/2023ApJ...949L..34H}
}

@ARTICLE{Mowla2022ApJ...937L..35M,
       author = {{Mowla}, Lamiya and {Iyer}, Kartheik G. and {Desprez}, Guillaume and {Estrada-Carpenter}, Vicente and {Martis}, Nicholas S. and {Noirot}, Ga{\"e}l and {Sarrouh}, Ghassan T. and {Strait}, Victoria and {Asada}, Yoshihisa and {Abraham}, Roberto G. and {Brammer}, Gabriel and {Sawicki}, Marcin and {Willott}, Chris J. and {Bradac}, Marusa and {Doyon}, Ren{\'e} and {Muzzin}, Adam and {Pacifici}, Camilla and {Ravindranath}, Swara and {Zabl}, Johannes},
        title = "{The Sparkler: Evolved High-redshift Globular Cluster Candidates Captured by JWST}",
      journal = {\apjl},
         year = 2022,
        month = oct,
       volume = {937},
       number = {2},
          eid = {L35},
        pages = {L35},
          doi = {10.3847/2041-8213/ac90ca},
archivePrefix = {arXiv},
       eprint = {2208.02233},
 primaryClass = {astro-ph.GA},
       adsurl = {https://ui.adsabs.harvard.edu/abs/2022ApJ...937L..35M}
}

@ARTICLE{MaoPaczynski1991ApJ...374L..37M,
       author = {{Mao}, Shude and {Paczynski}, Bohdan},
        title = "{Gravitational Microlensing by Double Stars and Planetary Systems}",
      journal = {\apjl},
         year = 1991,
        month = jun,
       volume = {374},
        pages = {L37},
          doi = {10.1086/186066},
       adsurl = {https://ui.adsabs.harvard.edu/abs/1991ApJ...374L..37M}
}

@ARTICLE{LSST2019ApJ...873..111I,
       author = {{Ivezi{\'c}}, {\v{Z}}eljko and {Kahn}, Steven M. and {Tyson}, J. Anthony and {Abel}, Bob and {Acosta}, Emily and {Allsman}, Robyn and {Alonso}, David and {AlSayyad}, Yusra and {Anderson}, Scott F. and {Andrew}, John and {Angel}, James Roger P. and {Angeli}, George Z. and {Ansari}, Reza and {Antilogus}, Pierre and {Araujo}, Constanza and {Armstrong}, Robert and {Arndt}, Kirk T. and {Astier}, Pierre and {Aubourg}, {\'E}ric and {Auza}, Nicole and {Axelrod}, Tim S. and {Bard}, Deborah J. and {Barr}, Jeff D. and {Barrau}, Aurelian and {Bartlett}, James G. and {Bauer}, Amanda E. and {Bauman}, Brian J. and {Baumont}, Sylvain and {Bechtol}, Ellen and {Bechtol}, Keith and {Becker}, Andrew C. and {Becla}, Jacek and {Beldica}, Cristina and {Bellavia}, Steve and {Bianco}, Federica B. and {Biswas}, Rahul and {Blanc}, Guillaume and {Blazek}, Jonathan and {Blandford}, Roger D. and {Bloom}, Josh S. and {Bogart}, Joanne and {Bond}, Tim W. and {Booth}, Michael T. and {Borgland}, Anders W. and {Borne}, Kirk and {Bosch}, James F. and {Boutigny}, Dominique and {Brackett}, Craig A. and {Bradshaw}, Andrew and {Brandt}, William Nielsen and {Brown}, Michael E. and {Bullock}, James S. and {Burchat}, Patricia and {Burke}, David L. and {Cagnoli}, Gianpietro and {Calabrese}, Daniel and {Callahan}, Shawn and {Callen}, Alice L. and {Carlin}, Jeffrey L. and {Carlson}, Erin L. and {Chandrasekharan}, Srinivasan and {Charles-Emerson}, Glenaver and {Chesley}, Steve and {Cheu}, Elliott C. and {Chiang}, Hsin-Fang and {Chiang}, James and {Chirino}, Carol and {Chow}, Derek and {Ciardi}, David R. and {Claver}, Charles F. and {Cohen-Tanugi}, Johann and {Cockrum}, Joseph J. and {Coles}, Rebecca and {Connolly}, Andrew J. and {Cook}, Kem H. and {Cooray}, Asantha and {Covey}, Kevin R. and {Cribbs}, Chris and {Cui}, Wei and {Cutri}, Roc and {Daly}, Philip N. and {Daniel}, Scott F. and {Daruich}, Felipe and {Daubard}, Guillaume and {Daues}, Greg and {Dawson}, William and {Delgado}, Francisco and {Dellapenna}, Alfred and {de Peyster}, Robert and {de Val-Borro}, Miguel and {Digel}, Seth W. and {Doherty}, Peter and {Dubois}, Richard and {Dubois-Felsmann}, Gregory P. and {Durech}, Josef and {Economou}, Frossie and {Eifler}, Tim and {Eracleous}, Michael and {Emmons}, Benjamin L. and {Fausti Neto}, Angelo and {Ferguson}, Henry and {Figueroa}, Enrique and {Fisher-Levine}, Merlin and {Focke}, Warren and {Foss}, Michael D. and {Frank}, James and {Freemon}, Michael D. and {Gangler}, Emmanuel and {Gawiser}, Eric and {Geary}, John C. and {Gee}, Perry and {Geha}, Marla and {Gessner}, Charles J.~B. and {Gibson}, Robert R. and {Gilmore}, D. Kirk and {Glanzman}, Thomas and {Glick}, William and {Goldina}, Tatiana and {Goldstein}, Daniel A. and {Goodenow}, Iain and {Graham}, Melissa L. and {Gressler}, William J. and {Gris}, Philippe and {Guy}, Leanne P. and {Guyonnet}, Augustin and {Haller}, Gunther and {Harris}, Ron and {Hascall}, Patrick A. and {Haupt}, Justine and {Hernandez}, Fabio and {Herrmann}, Sven and {Hileman}, Edward and {Hoblitt}, Joshua and {Hodgson}, John A. and {Hogan}, Craig and {Howard}, James D. and {Huang}, Dajun and {Huffer}, Michael E. and {Ingraham}, Patrick and {Innes}, Walter R. and {Jacoby}, Suzanne H. and {Jain}, Bhuvnesh and {Jammes}, Fabrice and {Jee}, M. James and {Jenness}, Tim and {Jernigan}, Garrett and {Jevremovi{\'c}}, Darko and {Johns}, Kenneth and {Johnson}, Anthony S. and {Johnson}, Margaret W.~G. and {Jones}, R. Lynne and {Juramy-Gilles}, Claire and {Juri{\'c}}, Mario and {Kalirai}, Jason S. and {Kallivayalil}, Nitya J. and {Kalmbach}, Bryce and {Kantor}, Jeffrey P. and {Karst}, Pierre and {Kasliwal}, Mansi M. and {Kelly}, Heather and {Kessler}, Richard and {Kinnison}, Veronica and {Kirkby}, David and {Knox}, Lloyd and {Kotov}, Ivan V. and {Krabbendam}, Victor L. and {Krughoff}, K. Simon and {Kub{\'a}nek}, Petr and {Kuczewski}, John and {Kulkarni}, Shri and {Ku}, John and {Kurita}, Nadine R. and {Lage}, Craig S. and {Lambert}, Ron and {Lange}, Travis and {Langton}, J. Brian and {Le Guillou}, Laurent and {Levine}, Deborah and {Liang}, Ming and {Lim}, Kian-Tat and {Lintott}, Chris J. and {Long}, Kevin E. and {Lopez}, Margaux and {Lotz}, Paul J. and {Lupton}, Robert H. and {Lust}, Nate B. and {MacArthur}, Lauren A. and {Mahabal}, Ashish and {Mandelbaum}, Rachel and {Markiewicz}, Thomas W. and {Marsh}, Darren S. and {Marshall}, Philip J. and {Marshall}, Stuart and {May}, Morgan and {McKercher}, Robert and {McQueen}, Michelle and {Meyers}, Joshua and {Migliore}, Myriam and {Miller}, Michelle and {Mills}, David J.},
        title = "{LSST: From Science Drivers to Reference Design and Anticipated Data Products}",
      journal = {\apj},
         year = 2019,
        month = mar,
       volume = {873},
       number = {2},
          eid = {111},
        pages = {111},
          doi = {10.3847/1538-4357/ab042c},
archivePrefix = {arXiv},
       eprint = {0805.2366},
 primaryClass = {astro-ph},
       adsurl = {https://ui.adsabs.harvard.edu/abs/2019ApJ...873..111I}
}

@ARTICLE{Merritt2006AJ....132.2685M,
       author = {{Merritt}, David and {Graham}, Alister W. and {Moore}, Ben and {Diemand}, J{\"u}rg and {Terzi{\'c}}, Bal{\v{s}}a},
        title = "{Empirical Models for Dark Matter Halos. I. Nonparametric Construction of Density Profiles and Comparison with Parametric Models}",
      journal = {\aj},
         year = 2006,
        month = dec,
       volume = {132},
       number = {6},
        pages = {2685-2700},
          doi = {10.1086/508988},
archivePrefix = {arXiv},
       eprint = {astro-ph/0509417},
 primaryClass = {astro-ph},
       adsurl = {https://ui.adsabs.harvard.edu/abs/2006AJ....132.2685M}
}

@ARTICLE{Jullo2007Lenstool,
   author = {{Jullo}, E. and {Kneib}, J.-P. and {Limousin}, M. and {El{\'{\i}}asd{\'o}ttir}, {\'A}. and
	{Marshall}, P.~J. and {Verdugo}, T.},
    title = "{A Bayesian approach to strong lensing modelling of galaxy clusters}",
  journal = {New Journal of Physics},
archivePrefix = "arXiv",
   eprint = {0706.0048},
     year = 2007,
    month = dec,
   volume = 9,
    pages = {447},
      doi = {10.1088/1367-2630/9/12/447},
   adsurl = {http://adsabs.harvard.edu/abs/2007NJPh....9..447J}
}

@ARTICLE{Kneib1993Lenstool,
   author = {{Kneib}, J.~P. and {Mellier}, Y. and {Fort}, B. and {Mathez}, G.
	},
    title = "{The Distribution of Dark Matter in Distant Cluster Lenses - Modelling A:370}",
  journal = {\aap},
     year = 1993,
    month = jun,
   volume = 273,
    pages = {367},
   adsurl = {http://adsabs.harvard.edu/abs/1993A%26A...273..367K}
}

@ARTICLE{Diego2005Nonparam,
   author = {{Diego}, J.~M. and {Protopapas}, P. and {Sandvik}, H.~B. and
	{Tegmark}, M.},
    title = "{Non-parametric inversion of strong lensing systems}",
  journal = {\mnras},
   eprint = {arXiv:astro-ph/0408418},
     year = 2005,
    month = jun,
   volume = 360,
    pages = {477-491},
      doi = {10.1111/j.1365-2966.2005.09021.x},
   adsurl = {http://adsabs.harvard.edu/abs/2005MNRAS.360..477D}
}

@ARTICLE{Refsdal1964MNRAS,
   author = {{Refsdal}, S.},
    title = "{On the possibility of determining Hubble's parameter and the masses of galaxies from the gravitational lens effect}",
  journal = {\mnras},
     year = 1964,
   volume = 128,
    pages = {307},
   adsurl = {http://adsabs.harvard.edu/abs/1964MNRAS.128..307R}
}
\bibliographystyle{tfnlm.bst}

\end{document}